\documentclass[twocolumn,twoside]{article}

\usepackage{a4}
\usepackage{amssymb}
\usepackage{amsmath}
\usepackage[numbers,sort&compress]{natbib}
\usepackage{graphicx}
\usepackage{blindtext}

\begin{document}

\title{\bf {\Large Laboratory modeling of jets from young stars
using plasma focus facilities}} 

\date{{\normalsize\textit{P N Lebedev Physical Institute, Russian Academy of Sciences,
Leninskii prosp. 53, 119991, Moscow, Russian Federation,\\
Moscow Institute of Physics and Technology
(National Research University), Institutskii per. 9,
141701 Dolgoprudny, Moscow region, Russian Federation \\
National Research Center Kurchatov Institute,
pl. Akademika Kurchatova 1, 123182 Moscow, Russian Federation \\
Sternberg Astronomical Institute,
Lomonosov Moscow State University,
Universitetskii prosp. 13, 119234 Moscow, Russian Federation}} \\[1ex]
{\small \textit{Usp.\ Fiz.\ Nauk} \textbf{193}, 
 345--381 (2023) 
[in Russian]\\
English translation: \textit{Physics -- Uspekhi}, \textbf{66} 327--359 (2023)}
\\{\small Translated by M Zh Shmatikov}
}

\author{V S Beskin, V I Krauz, S A Lamzin}

\maketitle

{\bf {\underline {Abstract.}} Jets from young stars are used as an example to review how laboratory modeling enables advancement in understanding the main physical processes responsible for the formation and stability of these amazing objects. The discussion focuses on the options for modeling jet emissions in a laboratory experiment at the PF-3 facility at the National Research Center Kurchatov Institute. Many properties of the flows obtained using this experimental setup are consistent with the main features of jets from young stars.}


\setcounter{secnumdepth}{3}
\setcounter{tocdepth}{2}

\tableofcontents

\section{Introduction}

Immense progress in the development of information technology could not but affect the structure of scientific research.
While in the 1960s and 1970s the rapid development of astronomy was associated with a technological breakthrough, which resulted in the mastering of ever new ranges of the electromagnetic spectrum, now the key point is the possibility of the wide coverage of the objects under study, which previously was unattainable. This refers not only to the concurrent exploration of cosmic sources at various frequencies (now, only those research applications look worthy in which the authors are not limited to a single spectral band, but propose different spectral range observations, e.g. in the radio, optical, and X-ray). Due to the transition to large
collaborations, research teams increasingly frequently include, along with observers, theorists who carry out basic theoretical research (though, lately, using numerical rather than analytical methods).

On the other hand, laboratory modeling of astrophysical processes is now increasingly important. The range of areas in which significant progress has been made in laboratory modeling is very wide. Such areas as the study of the equations of state, nonlinear hydrodynamics, radiative dynamics, and even modeling of the quantum properties of black holes [1] should be noted. Interesting results have been obtained in the study of the properties of carbonaceous and siliceous dust grains for molecules important for astrophysics,
which can play the role of progenitors of the formation of planetesimals [2, 3], in the study of dusty ion-sound shock waves, which are widely represented in the near-Earth plasma and in the Universe [4, 5], in modeling plasma-dust processes near the surface of the Moon [6], etc. Laboratory experiments that study the conversion of magnetic energy into the energy of high-energy plasma flows, accelerated particles, and radiation in various wavelength ranges, i.e., the processes of magnetic reconnection during the formation and destruction of current sheets, carried out at the Prokhorov Institute of General Physics of the Russian Academy of Sciences using
the TC-3D facility [7, 8], make it possible to reproduce many astrophysical phenomena and provide fundamental possibilities for predicting burst-type phenomena.

Significant progress has also been achieved due to the emergence of a whole set of new facilities with a high energy density developed as part of the program of inertial confinement fusion (ICF), in particular, state-of-the-art laser [9] and Z-pinch systems [10]. The new installations made it possible to simulate many astrophysical processes. The use of superpower lasers enabled simulation of the interaction of radiation with matter in superstrong electromagnetic fields of neutron stars [11, 12]

A vast number of studies have been carried out at the Institute of Laser Physics of the Siberian Branch of the Russian Academy of Sciences at the KI-1 facility. In experiments that explored the formation of collisionless shock waves (CSWs) in the background plasma, laser plasma bunches were injected across the magnetic field with a maximum energy of up to 100 J per unit solid angle and a fairly high degree of ion magnetization [13]. For the first time, under laboratory conditions, it was possible to detect intense deceleration by the background of a super-Alfven laser plasma flow and the formation in it of a strong perturbation with the properties of a subcritical CSW propagating perpendicular to the magnetic field vector. New results have been obtained regarding the
formation of an extended (up to $\sim 0.5$ m) jet in a vacuum in a magnetic field (up to 300 G) by means of the injection of laser plasma bunches across this field [14].

Experiments on plasma expansion into an external magnetic field were also carried out at the Institute of Applied Physics (IAP) of the Russian Academy of Sciences on a laboratory bench for studying laser-plasma interaction, which was created on the basis of the PEARL laser complex (PEtawatt pARametric Laser) (https://pearl.iapras.ru) [15]. A high-speed dense plasma flow was formed by thermal ablation of a substance from the surface of a solid target under the action of high-power laser radiation. A comparison of the results of a laboratory experiment with examples of typical astrophysical flows in the system of the polar AM Herculis, the intermediate polar EX Hydrae, and the `hot Jupiter' WASP-12b showed the conceptual possibility of laboratory modeling of accretion flows. The experimental study of the expansion of a laser plasma in a strong external magnetic field (with an induction of 135 kG) at various sizes of the plasma formation region on the surface of a solid target has shown that, for sizes of the plasma formation region smaller than the classical plasma stagnation radius, an almost identical topology of plasma flows is observed, which is characterized by the formation of a thin plasma sheet oriented along the external magnetic field [16]. Laser and Z-pinch systems are also widely used to simulate jets from compact astrophysical objects [17, 18]. This issue is covered in more detail in Section 2.3. 

In this review, we use the modeling of jets from young stars at the PF-3 plasma focus facility operated by the National Research Center Kurchatov Institute [19-21] as an example to show that such a wide range of research really enables significant advancement in understanding the physical processes which are responsible for the formation and relative stability of these amazing objects. We do not provide here any definitive answers but only discuss how a laboratory experiment, despite significant differences in many key parameters (energy characteristics, dimensions), makes it possible to obtain new, sometimes unique, information about the physical processes responsible for the observed activity of jets.

\section{Jets from young stellar objects}

\subsection{Astrophysical aspect}

{\bf 2.1.1. Star formation process.} We start by outlining the main stages of the star formation process, referring to books [22, 23] for details. 

(1) Stars are formed from interstellar matter primarily inside giant molecular clouds --- condensations of interstellar gas with a characteristic size of the order of several tens of parsecs (1 pc $\approx 3 \times 10^{18}$ cm) and a mass of $\sim 10^{4}$--$10^{6} M_{\odot}$. A small but extremely important addition to the gas is dust particles with sizes not exceeding 0.1 mm, which absorb visible
light and shield the molecular cloud matter from the radiation of surrounding stars.

(2) Molecular clouds, which feature an inhomogeneous structure, and condensations with a mass of about \mbox{100 $M_{\odot}$} called prestellar cores, are the `nurseries' of stars. The lifetime of prestellar cores does not exceed $10^{5}$ years, after which they
either are destroyed or begin to shrink, breaking up into separate stellar-mass fragments.

(3) Initially, these fragments (protostellar clouds) contract almost in the free collapse regime, but after about $10^{4}$ years, the inner regions become opaque to their own radiation, as a result of which they rapidly heat up. The gas pressure gradient in such areas stops the compression, and the cloud turns into a protostar an object consisting of a hydrostatically equilibrium core with a mass of about 1\% of the mass of the cloud, onto which the matter of the outer layers falls (accretes). The core temperature is $\sim 1000$--$3000$ K, and it radiates in the optical and near infrared (IR) ranges, but this radiation does not reach the external observer, since it is completely absorbed by dust particles in the shell and reemitted in the wavelength band $\lambda > 20$ $\mu$m; objects with the specified properties are called Class 0 Young Stellar Objects.

(4) Protostellar clouds have a nonzero angular momentum $J_{\rm cl}$; therefore, as the cloud contracts, a disk-like condensation is formed in its equatorial plane, inside which the matter falls in a spiral onto the forming hydrostatic core due to the viscosity between the layers. If the angular momentum $J_{\rm cl}$ exceeds a certain critical value, centrifugal forces tear the cloud into two or more parts, each of which further contracts on its own --- this is apparently the way binary and multiple stellar systems are formed. 

(5) Over time, the mass and temperature of the core increase, and the shell mass decreases. From an observational point of view, it is manifested in the protostar becoming a source of radiation at ever shorter IR wavelengths: at this stage, protostars are called Class I objects. At some point, the shell becomes transparent along the line of sight also for the optical radiation from the core, which in the first approximation looks like an ordinary star, and for the observer, the protostar turns into a young star.

(6) After approximately $3 \times 10^{5}$ years, almost the entire mass of the parent protostellar cloud is concentrated in the young star: the accretion disk retains about 1\% of the mass, but more than 90\% of the initial angular momentum. Disk accretion in young stars is not able to compensate for energy losses due to radiation from the surface, as a result of which such stars slowly shrink, and their central regions heat up. For objects with mass $M > 0.08 \, M_{\odot}$, after $3 \times 10^{7}(M/M_{\odot})^{-2}$ years, the temperature reaches a value sufficient to ensure that the intensity of nuclear reactions, in which hydrogen is converted into helium, can compensate losses due to radiation from the surface. From this time on, the young star becomes an 'adult' star of the so-called main sequence. For the Sun, for example, the stage of youth was about 1000 times shorter than the duration of its 'adult' life, in the middle of which it is now.

(7) Young stars, depending on their mass, are divided into T Tauri stars ($M < 2$--$3 \, M_{\odot}$) and Ae/Be Herbig stars. In their spectrum in the wavelength band up to about 2 $\mu$m, the radiation of the star dominates, while in the far IR band, the radiation of the accretion disk, or rather its dust heated by the radiation of the star and the heat released due to accretion, dominates. The disk gas partly accretes onto the star and partly evaporates from the disk surface in the form of so-called disk wind. The dust settles to the equatorial plane of the disk; the concentration of dust particles increases, and they begin to stick together during collisions, forming ever larger agglomerates. When (and if) dust clumps grow to a size of about 1 km, their further growth up to the formation of planets occurs due to the accretion of the surrounding gas. 

(8) Over several million years, the gas from the disk almost completely disappears; accretion stops, and a planetary system and/or `construction debris --- dust particles, asteroids, comets --- remains around the young star. Virtually nonaccreting disks are observed around T Tauri stars with weak lines(age over 3--10 Myr), while their more massive brothers, which evolve faster, exhibit almost nonaccreting disks even at the main sequence stage. Protostars and young stars are usually called by a generalized name --- young stellar
objects (YSOs).

(9) A fundamentally important role in the process of star formation is played by the magnetic field, which affects the redistribution of the angular momentum inside the protostellar cloud and the processes of its fragmentation and disk accretion onto the hydrostatic core [24]. Already in pre-stellar cores, the field induction is $B \sim 10$--$100 \, \mu$G [25], implying that in these objects the magnetic pressure is comparable to thermal and turbulent pressure.

\vspace{0.3cm}

{\bf 2.1.2. Jets of young stellar objects.} Astronomers noticed about 60 years ago that some emission lines in the spectra of classical T Tauri stars have absorption components on the short-wavelength side, which indicated the outflow of matter from these objects [26]. For almost a quarter of a century, it was believed that this phenomenon is associated with the coronal chromospheric activity of young stars and is similar to the solar wind, although much more intense [27]. The isotropy of the wind itself was not questioned.

However, the history of the discovery of jets, i.e., collimated outflows from young stellar objects, began much earlier, in the early 1950s, when compact diffuse nebulae with an unusual spectrum were discovered in star-forming regions, which moved across the sky at speeds as high as 100 km s$^{-1}$ [28--30]. V A Ambartsumyan coined the name Herbig--Haro (HH) objects for these nebulae after the names of the astronomers who discovered them (Fig. 1). It was found later that such nebulae are bright knots in weakly luminous jets of gas escaping from young stellar objects [31--33], in connection with which the term Herbig--Haro flow is now used more often, and a catalog number is assigned not to each newly discovered object but to the entire flow as a whole [34].

\begin{figure}
\begin{center}
\includegraphics[width=250pt]{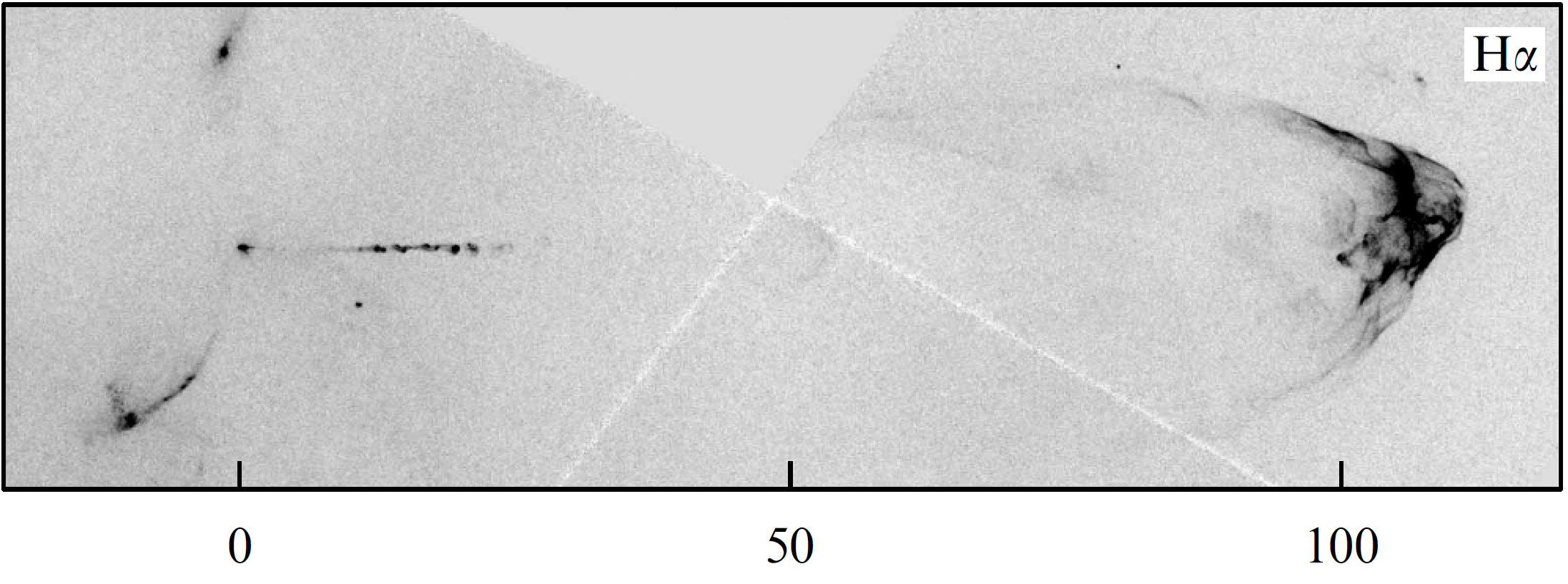}
\caption{\small Optical image of jet in the HH 34 system (see, for example, [31]).
Scale corresponds to 100 arc seconds (46,000 au = 0.22 pc). Jet details and
complex structure of the bow shock (object HH 34S) are clearly seen.}  
\label{fig01}
\end{center}
\end{figure}

By now, jets have been detected in several hundred young stars [35]. Approximately 2/3 of these stars have two oppositely directed jets, while in other cases the jet moving away from us is not observed, presumably due to the absorption of light in the circumstellar gas and dust envelope obscuring it. The observed length of the jets varies from 0.01 to 3 pc [36], and the degree of jet collimation (the ratio of the observed length to width) can reach 30, i.e., the full jet opening angle is 5--10$^{\circ}$. It should be stressed that this refers to so-called atomic jets, i.e., gas flows that radiate (and are observed) in the optical and near infrared spectral bands in the lines of atoms and ions, mainly in the lines of hydrogen and the so-called forbidden lines of elements heavier than helium.

Intensity ratios of these lines enabled determining that the characteristic gas number density in atomic jets ranges from $10^{2}$ to $10^{6}$ cm$^{-3}$; their temperature is close to $10^{4}$ K, but hydrogen is almost neutral: the degree of ionization is about 10\% [37]. In knots (Herbig-Haro objects per se), the density is higher due to which they look brighter (see Section 2.1.3 for details). Comparing images of HH objects obtained with a lapse of several years (see a selection of animations on the page https://sparky.rice.edu//~hartigan/movies.html [38]) and the Doppler shift of spectral lines shows that they move at a speed of about \mbox{300 km s$^{-1}$,} which correlates with the luminosity (and mass) of the central star, reaching 1000 km s$^{-1}$ for the most massive ($M > 10 \, M_{\odot}$) stars and gradually decreasing as distance from the central source increases [37]. The jets move relative to the colder interstellar medium in the supersonic regime with a Mach number exceeding 30. The loss rate of the mass carried away by atomic jets of young stars is about $10^{-9}$--$10^{-11} M_{\odot}$yr$^{-1}$, which is approximately 10--20\% of the accretion rate [31].

Usually, chains of HH objects form a straight line passing through the central star, perpendicular to the inner regions of the disk, and each object in the part of the jet approaching us (in the jet) corresponds to an HH object similar in parameters and distance from the star in the flow part moving away from us (in the counter jet). However, there are exceptions to this rule. For some young stars, for example, for RW Aur A, the gas velocity and the number of HH objects in the jet and counter jet differ by about a factor of two [37]. It is not yet clear whether this difference is related to the difference between conditions in the circumstellar medium on both sides of the accretion disk, or whether the asymmetry is due to the operation of the `central engine' [39]. In some young stars, jet chains form an arc, at the top of which there is a star. Most likely, this is due to the motion of the star relative to the circumstellar gas. Of more interest are situations when chains of HH objects bend, forming an S-shaped (spiral) curve (see, for example, the image of the jet and counterjet of the V1331 Cyg star in [40]). It is assumed that this effect is associated with the precession of the inner regions of the disk due to the presence of a companion star moving in a highly elongated orbit. The precession rate does not exceed several degrees per decade [41].

The most important discovery of recent years has been the detection of the rotation of jets of young stars, the history of which is also complicated. According to the data of studies [42--44] that date to the 2000 s, the characteristic velocities at a distance of 20--30 astronomical units (au) from the jet axis were 3--10 km s$^{-1}$. Later, these results, obtained at the resolution limit, were not confirmed [45, 46]. However, recent results from the ALMA (Atacama Large Millimeter Array) observatory (see [47] and references therein) undoubtedly indicate that jet rotation exists, albeit at a slightly slower rate.

In the late 1970s, extended (ranging from 0.1 to 10 pc) bipolar flows of cold (T $\sim$ 30 K) molecular gas were discovered in some protostars, moving away from the central source at a speed of about several tens of kilometers per second. Such a gas was discovered in the radio lines of the CO molecule [48], but now it is also observed in the lines of other molecules [49, 50]. The mass flux carried away by CO flows is proportional to the luminosity of the protostar. For massive ($M > 10\,M_{\odot}$) protostars with an age of the order of (1--3) $\times 10^{4}$ yr, it can be as high as $\sim 10^{-5}M_{\odot}$ yr$^{-1}$, amounting to, as in young stars, about 10\% of the accretion rate (see, for example, [41, Fig. 4]). The CO flows have a conical shape with different opening angles ranging from almost 180$^{\circ}$ to \mbox{20--30$^{\circ}$.} As the distance from the axis of the cone increases, the velocity of the gas in the flow decreases from $\sim$ 100 km s$^{-1}$ near the axis to $\sim$10 km s$^{-1}$ at the periphery (see [51, Fig. 3]).

Near the axis of many CO flows in the radio lines of excited molecules, chains of hotter (T $\sim$ 1000--3000 K) knots are observed, which trace the 'molecular' jet, just as Herbig-Haro objects in young stars trace atomic jets. Thus, high-speed (V $>$ 100 km s$^{-1}$) jets of protostars and young stars are like the core of a cocoon of gas escaping from a central source. In the case of young stars, the jet is surrounded primarily by a weakly collimated wind `blowing' from the disk surface, while in Class 0 protostars, the moving gas mainly consists of the protostellar cloud matter entrained by the jet [41]. 

It is important to note that the transverse size \mbox{($\sim$ 10--30 au),} the degree of collimation ($\sim$ 30), and the ratio of the outflow rate to the accretion rate ($\sim$ 0.1) for molecular jets of protostars and atomic jets of young stars are almost the same. Such jet parameters as the gas velocity, fluxes of mass ${\rm d}M/{\rm d}t$, and momentum $F_{\rm jet} = V{\rm d}M/{\rm d}t$ depend to a greater extent on the luminosity of the central object $L_{\rm ac}$ due to accretion rather than on its evolutionary stage [50]. In particular, for the entire range of ages of young stellar objects (from $10^{4}$ to $3 \times 10^{6}$ yr), the momentum flux is $F_{\rm jet} \sim 200 \, L_{\rm ac}/c$, where $L_{\rm ac}/c$ is the photon momentum flux, c is the speed of light, and this dependence spans almost five orders of $L_{\rm ac}$ [50]. From a quantitative point of view, the dependence under consideration indicates an insignificant role of radiation pressure in the acceleration of the substance of jets of young stellar objects.
 
However, even more important is the proportionality of the momentum flux $F_{\rm jet}$ precisely to the value of $L_{\rm ac}$, which indicates not only that the jet formation mechanism is inextricably linked with the accretion process but also the universality of this mechanism. Indeed, first, the parameters of the `central engine' for protostars and young stars differ greatly, and second, for protostars, the total luminosity $L^{*}$ almost coincides with $L_{\rm ac}$, while for young stars, as a rule, $L_{\rm ac} \ll L^{*}$. In connection with the above, it should be noted that those T Tauri stars with weak lines, in which the gas from the disk almost completely disappeared and only solid particles remained (from dust grains to planets), have no jets.

\vspace{0.3cm}

{\bf 2.1.3. Central engine.} The 'central engine,' i.e. the region in the jet base from which the energy required for its launch is released, in the situation under consideration consists of a central object (a protostar or a young star) and a surrounding accretion (protoplanetary) disk. By now, the regions of jet formation of young stars have been studied much better than those of protostars; therefore, we primarily focus on young starts that feature properties which are required for explaining their activity: rapid rotation and a strong global magnetic field.

Observations show that rotation periods of young stars range from 1.6 to 12 days [52]. Measurements of the magnetic field based on the Zeeman effect have shown that on the surface of T Tauri type stars it may reach several kG and have a fairly complex structure (see, for example, [53, 54]). It is convenient to represent the global field of these stars as a sum of multipoles, the magnetic moments of which are in general oriented with respect to the star rotation axis in a different way [55]. For example, the magnetic field of the BP Tau star measured using magnetic Doppler tomography can be approximated near the surface with an accuracy of the order of 10\% by dipole ($n = 1$) and octupole ($n = 3$) moments, which contain 50\% and 30\% of the magnetic energy, respectively [56]. On the surface of Herbig Ae/Be stars, the induction of the field is usually significantly less than 300 G and also has a multipole structure [57, 58]. In relation to this, it should be kept in mind that Herbig Ae/Be stars, unlike less massive T Tauri type stars, do not have external convective zones. Unfortunately, data on the axial rotation speed or magnetic field of hydrostatically equilibrium cores of protostars are very scant [59].

The parameters of the accretion disks of young stars may be quite different [60]: their masses range from several hundredths of a percent to several percent of the mass of the central star, and their external radii range from several to several hundred astronomical units. The radii of the protostar disks can be even larger, and, on their external regions, remnants of the protostellar cloud accrete. In the disk regions closest to the star, the dust evaporates (at distances from 0.01 to 1 au, depending on the luminosity of the star), while, at larger distances, the dust in the disk is either uniformly mixed with gas or settles to some extent on the central plane.

It has been firmly established that the activity of classic T Tauri stars is due to the magnetospheric accretion of the protoplanetary disk matter (see reviews [47, 61] and references therein). It is assumed in this concept that the magnetic field of the star destroys the accretion disk at distances of the order of several star radii, where the magnetic pressure becomes equal to the pressure of the accreting matter. The matter of the disk is frozen in some way into the force lines of the magnetic field of the star and slides down along them to the surface, accelerating to a velocity of $\sim 300$ km s$^{-1}$, i.e., to a value close to the free-fall velocity. The falling matter is decelerated in a shock wave, behind the front of which the gas is heated to a temperature of $\sim 10^{6}$ K and later cools down due to radiation in an ultraviolet (UV) and soft X-ray radiation range. This radiation heats the star atmosphere, creating on its surface a hot spot and an ionization zone ahead of the shock wave (SW) front. Three-dimensional magneto-hydrodynamic (MHD) calculations of accretion onto a young star with various configurations of the magnetic field [62] and calculations of the SW and hot spot radiation spectra [63, 64] provide an explanation for the main array of observational data, although there is not yet complete quantitative agreement between the theory and observations.

There are many reasons to believe that magnetospheric accretion is responsible for the activity of Herbig Ae/Be stars or at least some of them [58, 61]. Unfortunately, observational data that would allow making a conclusion regarding the role of the magnetic field in the accretion of the disk matter on hydrostatically equilibrium cores of protostars are virtually unavailable.

As mentioned in Section 2.1.2, an intense outflow of matter from the vicinity of young stars was discovered long ago. However, it was only in the early 1990s that it became clear that this process was associated not with coronal-chromospheric activity, as in our Sun, but with the presence of accretion disks in stars. We will not present numerous observational facts confirming this conclusion, but simply recall that the mass loss rate $({\rm d}M/{\rm d}t)_{\rm wind}$ clearly correlates with the accretion rate $({\rm d}M/{\rm d}t)_{\rm acc}$.

As it gradually became clear, matter outflows from the surface of protoplanetary disks, and this process occurs as a result of the action of various factors [65]. In disk regions far enough from the star, the outflow is most likely associated with the Blandford-Payne mechanism [66], which is facilitated by the heating of the upper layers of the disk atmosphere by ultraviolet and X-ray radiation from the star (see [67] and references therein). In the flow of matter from the disk, the so-called disk wind, low-velocity sections of the profiles of forbidden lines of the optical range with radial velocities of less than \mbox{30 km s$^{-1}$} and [O I] 63-$\mu$m and [Ne II] 12.8-$\mu$m IR-range lines are formed. As was noted above, the disk wind is weakly collimated. The issue of how effectively it removes the angular momentum from the disk, thereby contributing to accretion, remains open for now [68].

In areas of the disk at a distance of $\ll 10$ stellar radii ($R_{\ast} \sim 10^{11}$ cm) from the star, matter outflows as a result of the dynamic interaction of the disk with the star's magnetosphere. The presence of rotation and a regular magnetic field should inevitably involve electromagnetic processes, leading to a more active release of energy and, as a result, to an intense outflow of matter. We call such a wind magnetospheric, and it is probably this wind that is subsequently collimated into a jet. This wind forms the emission and absorption features of the lines observed in the spectra of young stars (see, for example, [61] and references therein). The size of the region (distance from the axis of rotation) in which the magnetospheric wind is formed depends on the models briefly discussed in Section 2.2.

Unfortunately, the angular resolution of modern optical telescopes, even for the closest young stars, does not enable studying the jet collimation region. For several classical T Tauri stars (HN Tau, UZ Tau E [69], RW Aur [37]), it was found that the wind gas is collimated into a jet $\sim 20$--40 au wide already at a distance of 10--50 au from a star. For the wind to transform from a region much smaller than 1 au into a conical jet with such parameters and an opening angle of 5--10$^{\circ}$, it must initially be quasi-spherical. It is curious that the size of the collimation region for protostars also turns out to be of the same order (see [50] and references therein).

The question remains open: what contribution to the formation of the jet is made by the wind `blowing' from the surface of a young star? This wind can, in principle, originate due to the acceleration of the matter of the star's atmosphere by the gas pressure gradient, as in the case of the Sun, or the pressure of Alfv{\' e}n waves excited by the accretion jet [70, 71]. Anyway, no observational confirmation of the existence of such a wind has been found so far, implying that the jet is formed primarily from the magnetospheric wind. We note, however, that the presence of a strong magnetic field and an accretion disk in a young star is apparently a necessary but not sufficient condition for the formation of a jet. For example, a BP Tau T Tauri star has all the features of magnetospheric accretion and intense wind [72]; nevertheless, the jet is missing.

\begin{figure}
\begin{center}
\includegraphics[width=250pt]{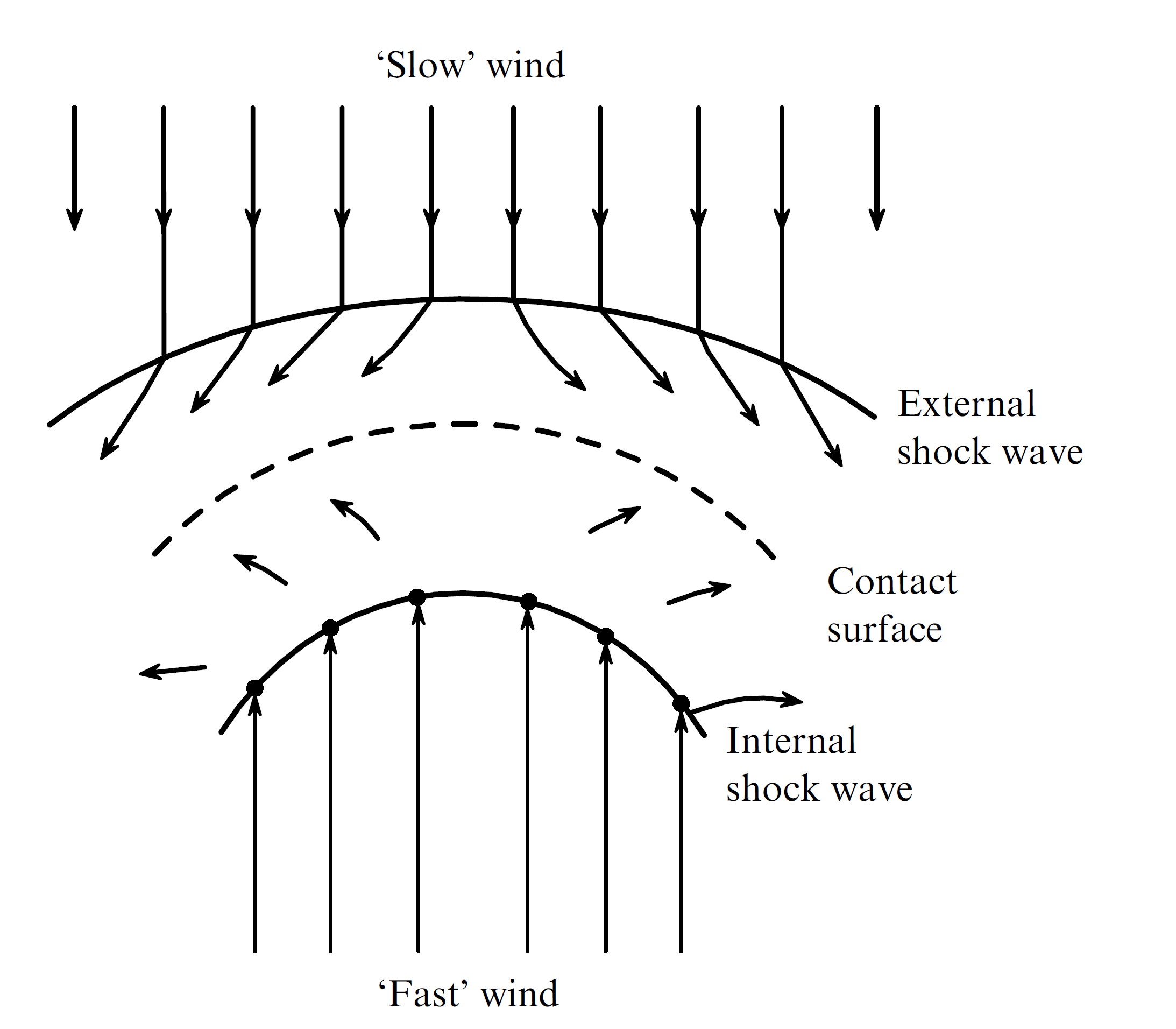}
\caption{\small Double structure of a shock wave emerging as a result of the
collision of a fast flow with a slower one. Arrows show the direction of
streamlines in the reference frame related to the front of the head shock
wave.}  
\label{fig02}
\end{center}
\end{figure}

\vspace{0.3cm}

{\bf 2.1.4. Jets and shock waves.} As was noted in Section 2.1.2, the jets of protostars and young stars move at velocities several tens of times higher than the speed of sound in the circumstellar medium. As is known, when a hypersonic flow collides with the surrounding interstellar gas, a double structure should appear (Fig. 2), consisting of a leading SW, which compresses and heats the interstellar gas, and a 'reverse' SW, which slows down the supersonic flow [73, 74]. Such a double structure emerges because, in the frame of reference related to the immobile external medium, not only the jet itself but also the SW arising due to the interaction of the jet with the external medium propagate at supersonic speed. Thus, in the reference frame where the front of the head SW is at rest, there are two colliding supersonic flows, each of which, before starting to interact with the other, must slow down to subsonic speed.

However, if in the region in front of the moving gas the longitudinal component of the magnetic field is so large that the Alfv{\' e}n velocity exceeds the sonic velocity, then the 'classical' SW with a density jump at the front does not arise. In this case (referred to as C-shock), the gas temperature, density, and velocity change smoothly when switching from a moving gas to a stationary one [75]. Such a situation is observed in some HH objects (see, for example, [76] and references therein).

Initially, it seemed natural to identify arc-shaped HH-objects observed `at the ends' of jets (see Fig. 1) with the bow shock. As regards the chain of HH objects located closer to the star, the following explanation for their appearance was proposed in 1990 [77], which later became generally accepted. It is assumed that, on average, wind speed $V_{\rm w}$ is much greater than the speed of sound in the jet matter $c_{\rm s} < 10$ km s$^{-1}$. If the wind speed at the base of the jet increases by $\Delta V$ on the order of several tens of km s$^{-1}$, a new portion of matter will collide with the gas ejected previously, generating an SW ($\Delta V > c_{\rm s}$!), which will propagate along the jet. Behind the SW front, the gas heats up to a temperature
\begin{equation}
T_{\rm sh} \approx 13 000 \left(\frac{\Delta V}{30 \,  {\rm km} \, {\rm s}^{-1}}\right)^2 {\rm K}. 
\end{equation} 
The subsequent cooling of the gas, accompanied by an increase in its density, occurs primarily due to radiation in the hydrogen and forbidden lines of oxygen, nitrogen, sulfur, and iron. HH-objects tracing the jets of young stars are precisely the compactions that appear in the cooling zone behind the front of 'internal' SWs of this kind. The spectrum of an HH object enables determination of both the density in the cooling zone and the temperature in the wake of the SW front, and thus the value of $\Delta V$; for example, the [O III] (5007 \r{A}) line has a noticeable intensity only at $\Delta V > 70$ km s$^{-1}$ [78]. The areas of lower density between the compactions are jet regions that consist of the gas ejected from the star at a velocity slightly less than the average value $V_{\rm w}$ [79].

It cannot be completely ruled out that, in some cases, HH objects arise as a result of the interaction between a high-speed jet and inhomogeneities of the interstellar medium and/or the wind surrounding the jet (see, for example, [80]). In turn, analysis of a large array of observational data shows that the main reason for the appearance of Herbig-Haro objects is the change in jet base velocity [41, 77]. However, the change in the speed of the magnetospheric wind is most likely due to the nonstationary nature of accretion onto the central star. A detailed discussion of this topic can be found in review [61]; we only point out that such a conclusion is well justified (see, for example, [39, 81, 82]).

Calculations show that, if the matter in pre-shock zone is partially neutral, the hydrogen atoms are excited by collisions even before their ionization, thereby causing intense radiation in the Balmer lines from the region immediately adjacent to the front. But [O I] 6300 \r{A} line, for example, is originated in the post-shock cooling zone, which is so far from the shock front that in some cases the angular resolution of modern telescopes makes it possible to observe its radiation separately from Balmer lines formation region [83, 84]. However, if \mbox{$\Delta V > 110$ km s$^{-1}$,} the flux of UV radiation from the cooling zone will completely ionize the matter ahead of the front [85].

Spectral diagnostics of HH objects also makes it possible to measure the magnetic field ahead of the SW front, or more precisely, its component $B_{t}$ parallel to the front (see [86] and references therein). The measurement method is based on the following: the $B_{t}$ component reduces the degree of gas compression in the cooling zone, thereby increasing the extension of the zone along the streamlines and changing the relative intensity of the spectral lines formed in it. Using this approach, it was found in [87, 88], for example, that, in the oncoming flow for two protostars $B_{t} \approx 20$--30 $\mu$G at a number density of $n_{H} \sim 100$--200 cm$^{-3}$, and in the jets of two young stars, it was obtained in [89, 90] that $B_{t} \approx  500 \, \mu$G at $n_{H} \approx 3 \times 10^{4}$ cm$^{-3}$. To avoid confusion, it should be noted that, in the case under consideration, the dynamic role of the magnetic field ahead of the SW front is small. However, behind the front, because the field is frozen into the plasma, $B \propto \rho$; therefore, $P_{\rm mag}/P_{\rm gas} \propto B^{2}/\rho T \propto \rho/T$. For this reason, in the post-shock cooling zone, as the gas temperature and its velocity ($\rho \propto 1/V$) decrease, the dynamic role of the magnetic field increases, which results in a change in the structure of the region.

We now return to the bow shock (see Fig. 1). As it turns out, HH-objects, which are arc-shaped and seem to be located at the `end' of the jet, move not through the interstellar medium but through the gas moving at the same or a slightly slower speed as the `internal' HH-objects. In other words, the region that was originally considered the region of collision of the jet with the circumstellar medium, strictly speaking, is not such. An example is the HH 34S object shown in Fig. 1, which, judging by its spectrum, features a shock wave velocity $\Delta V \sim 80$ km s$^{-1}$, while the object itself is moving away from the central star at a speed $V \sim 260$--320 km s$^{-1}$ [78]. This observation is consistent with the fact that the actual length of the HH 34 jet and counterjet is more than an order of magnitude greater than the distance from the central star to HH 34S. This is not surprising, given that not only young stars (about 1 Myr old) have jets, but so do their predecessors, protostars with an age starting from $\sim$ 10,000 yr. Thus, the jets from young stars propagate not through the interstellar medium but through a kind of channel created by a much more powerful gas outflow at an earlier stage of star formation.

On the other hand, the velocity $\Delta V$ of motion of the HH 34S object along the jet is approximately three times greater than that of the chains of HH objects located closer to the star (see Fig. 1). The gas number density in front of the HH 34S of $\sim$ 3 cm$^{-3}$ is comparable to that of interstellar gas in star-forming regions and is three orders of magnitude lower than that of HH objects in the chain [78]. The temperature (about 300 K) and the magnetic field induction ($B_{t} \sim 10 \mu$G) ahead of the front and in the interstellar gas also turn out to be of the same order [88]. Given these observations, it may be concluded that the HH 34S object, in contrast to the chain of HH objects closer to the star, can in a certain sense be considered as a bow shock similar to that observed in plasma focus setups during bunch propagation (see Section 3.5.2 and Fig. 16 below).

Since the HH 34S object is now under discussion, we note that it is one of the few examples when both direct and reverse SWs can be separately observed [87, 91]. Another interesting feature of HH 34S is the `lace' structure, which is also observed in many other objects. It is not yet clear what the reason is for the emergence of the `lace' structure: thermal instability at the SW front, inhomogeneous density of the gas and/or magnetic field ahead of the front, the appearance of turbulence, or a combination of the above factors [41, 88, 92, 93] (see also Section 3.5.4).

\subsection{Theoretical aspect}

{\bf 2.2.1. Formulation of the problem.} As discussed in Section 2.1.1, the formation of jets, as well as the formation of a disk wind, is associated with the need to most effectively remove the angular momentum that prevents the formation of a star. The similarity of the observational properties of young stars and active galactic nuclei (a rotating `central engine' and bipolar jets) suggests that the physical mechanisms of energy release, despite the significant differences among many parameters, may be similar for them. It should be recalled that one of the main differences is the nonrelativistic nature of the flow in jets from young stars.

As a result, an electromagnetic mechanism of energy release was proposed as a basic idea, which is implemented in a Faraday disk (see, for example, [94]). It is this mechanism of energy release that operates in the magnetospheres of radio pulsars and (with some important refinements [95]) in supermassive black holes. Its implementation requires the presence of a central rotating body with a strong regular magnetic field. All of these elements are also inherent in the `central engine' of young stellar objects.

Indeed, as is well known, the rotation of a body with characteristic size $R$ with an angular velocity $\Omega$ and a regular magnetic field $B_{0}$ results in the emergence on the surface of the body of a potential difference $U \sim ER$ between its pole and the equator:
\begin{equation}
U \sim \frac{\Omega R^2}{c} B_{0}.
\end{equation}
The energy release
\begin{equation}
W_{\rm tot} = I U    
\end{equation}
is determined by the magnitude of the electric current $I$ circulating in the magnetosphere of the 'central engine.' Determining the current $I$ (i.e., the internal and external load in the resulting system of currents) thus becomes the main task of the theory.

As can be seen, the main parameters of the problem are the size $R$ and the angular velocity $\Omega$ of the 'central engine' and the magnetic field $B_{0}$ on its surface. In addition, the key parameters include the ejection rate ${\dot M}_{\rm jet}$. Below, we define these parameters more strictly. Here, we make the following important remark: despite the rapid rotation of a young star, its angular velocity is always much lower than the Keplerian angular velocity of the inner regions of the accretion disk $\Omega_{\rm K} = \sqrt{GM/R^{3}}$. It is for this reason, as was specially emphasized above, that the properties of the jet ejection correlate with the accretion disk, rather than with the young star itself. Thus, as has now become quite clear [47], the `central engine' should be not the star itself but the magnetosphere and accretion disk surrounding it.

For the characteristic parameters of young stars (${\dot M}_{\rm acc} \sim 10^{-8} M_{\odot}$ yr$^{-1}$, $B_{0} \sim$ 1 kG), the inner radius of the accretion disk is only a few stellar radii:
\begin{eqnarray}
R_{\rm A} \approx 3 \, R
\left(\frac{M}{1.4 \, M_{\odot}}\right)^{-1/7} 
\left(\frac{R}{2 \,R_{\odot}}\right)^{5/7}
\left(\frac{B_{0}}{10^{3} \, {\rm G}}\right)^{4/7}
\nonumber \\
\times  \left(\frac{{\dot M}_{\rm acc}}{10^{-8}M_{\odot} \, {\rm yr}^{-1}}\right)^{-2/7}. 
\label{RAR}
\end{eqnarray}
Here,
\begin{equation}
R_{\rm A} = 
\left(\frac{B_{0}^{2}R^{6}}{2{\dot M}_{\rm acc}\sqrt{2 G M}}\right)^{2/7}   
\label{RA}
\end{equation}
is the so-called Alfv{\' e}n radius [96], at which the energy density of the star's dipole magnetic field becomes comparable to the
energy density of the accreting matter. Consequently, the angular velocity of the disk here turns out to be approximately an order of magnitude greater than that of the star:
\begin{eqnarray}
\frac{\Omega_{\rm K}(R_{\rm A})}{\Omega} \approx 20
\left(\frac{M}{1.4 \, M_{\odot}}\right)^{5/7} 
\left(\frac{R}{2 \, R_{\odot}}\right)^{-18/7} 
\nonumber \\
\times \left(\frac{B_{0}}{10^{3} \, {\rm G}}\right)^{-6/7} 
\left(\frac{{\dot M}_{\rm acc}}{10^{-8}M_{\odot} \, {\rm yr}^{-1}}\right)^{3/7}
\left(\frac{P_{\ast}}{10 \, {\rm days}}\right) 
\label{RARa}
\end{eqnarray}
($P_{\ast}$ is the rotation period of the star). All this suggests that the
characteristic value of the angular velocity should be taken as
the Keplerian angular velocity of the inner part of the accretion disk,
\begin{equation}
\Omega_{\rm A} = \sqrt{\frac{G M}{R_{\rm A}^3}},    
\label{OmegaK}
\end{equation}
and, as a characteristic dimension, the Alfv{\' e}n radius $R_{\rm A}$.

\vspace{0.3 cm}

{\bf 2.2.2 Method of the Grad-Shafranov equation.} We now recall the main relations underlying the MHD approach. Under the assumption of stationarity and axial symmetry of the flows of interest to us (which, as we have seen, is a reasonable approximation for our problem), the approach based on the method of the Grad-Shafranov equation [97--100] has become very helpful. The Grad-Shafranov method is attractive since, in the ideal MHD approximation, one can formulate a number of integrals of motion that are preserved on magnetic surfaces. Moreover, for the known structure of the magnetic field, these integrals of motion are sufficient to express all the main quantities characterizing the flow using fairly simple algebraic relations. As for the structure of the magnetic field itself, its definition is reduced to solving only one second-order equation in partial derivatives.

So, the main 'actor' is the scalar function $\Psi(r,z)$, which is a magnetic flux through a surface bounded by a circle centered on the axis of rotation, passing through a point with coordinates{\footnote{{For quasi-cylindrical jets, it is natural to use cylindrical coordinates.}}} $r, z$. The function $\Psi(r,z)$ is related to the poloidal magnetic field ${\bf B}_{\rm p}$ by the relation [95]
\begin{equation}
{\bf B}_{\rm p} = \frac{\nabla\Psi\times{\bf e}_{\varphi}}{2\pi r}.
\label{Bp}
\end{equation}
Due to the relation ${\bf B}\nabla \Psi = 0$, the condition $\Psi(r,z) =$ const just determines the structure of the magnetic field lines. Consequently, it is convenient to express the toroidal magnetic field $B_{\varphi}$ in terms of the total current $I(r, z)$ flowing inside the same magnetic tube:
\begin{equation}
B_{\varphi} = -\frac{2\,I}{c r}.
\label{Bphi}
\end{equation}
The minus sign in Eqn (9) was chosen so that, when $\bf{\Omega B} > $ 0, which is usually implicitly assumed, the value $I$ is positive. Then, $I$ is the current flowing near the axis towards the 'central engine.' It is clear that the opposite case $\bf{\Omega B} < $ 0 is not forbidden either, in which the current near the axis flows from the 'central engine.'

In what regards the electric field ${\bf E}$, due to the condition of axial symmetry and stationarity (according to which $E_{\varphi} = 0$), and the freezing condition \mbox{${\bf E} + {\bf v} \times {\bf B}/c = 0$} (according to which $E_{\parallel} = 0$), it can only be directed along the gradient $\nabla\Psi$. Therefore, it is convenient to represent the field ${\bf E}$ in the form
\begin{equation}
{\bf E} = -\frac{\Omega_{\rm F}}{2\pi c}\nabla\Psi,
\label{defE}
\end{equation}
where $\Omega_{\rm F}$ is some scalar function. Since, in the stationary case, $\nabla \times {\bf E} = 0$, we arrive at the relation
\begin{equation}
\Omega_{\rm F} = \Omega_{\rm F}(\Psi).
\label{Ferraro}
\end{equation}
The quantity $\Omega_{\rm F}(\Psi)$ is the first integral of motion, which ensures the equipotentiality of magnetic surfaces.

Furthermore, given the freezing-in condition, the hydrodynamic flow velocity v can be represented as
\begin{equation}
{\bf v} = 
\frac{\eta_{\rm n}}{\rho}{\bf B}
+\Omega_{\rm F} r \,{\bf e}_{\varphi},
\label{v-to-B}
\end{equation}
where the scalar function $\eta_{\rm n}$ is, by definition, the ratio of the particle flux to the magnetic field flux (here and below, the subscript n indicates the nonrelativistic nature of the flows under consideration). First of all, it can be seen that the value of $\Omega_{\rm F}$ introduced above really has the meaning of the angular velocity of rotation. Therefore, Eqn (11) is the isorotation law [101]. In addition, due to the relation $\nabla \eta_{\rm n} {\bf B} = 0$, the quantity $\eta_{\rm n}$ is also an integral of motion:
\begin{equation}
\eta_{\rm n} = \eta_{\rm n}(\Psi).
\label{etan}
\end{equation}

Two more integrals of motion are obtained by integrating two components of the total balance of forces. For example, it is known that its projection onto the direction of the velocity of medium ${\bf v}$ results in the conservation of the Bernoulli integral
\begin{equation}
E_{\rm n}(\Psi)=\frac{v^{2}}{2} + w + \frac{\Omega_{\rm F}I}{2\pi\eta_{\rm n} c} 
+ \varphi_{\rm g},
\label{intE}
\end{equation}
where $w$ is the enthalpy and $\varphi_{\rm g}$ is the gravitational potential. Consequently, the projection onto the $\varphi$-axis gives the conservation of the specific angular momentum:
\begin{equation}
L_{\rm n}(\Psi) = r v_{\varphi} + \frac{I}{2\pi\eta_{\rm n} c}.
\label{intL}
\end{equation}
Finally, the entropy $s = s(\Psi)$ becomes the fifth integral of motion in the approximation of ideal magnetohydrodynamics.

As a result, the five integrals presented above turn out to be sufficient to determine all other flow parameters using the algebraic relations
\begin{equation}
\rho = \frac{4\pi\eta_{\rm n}^{2}}{{\cal M}^2},
\label{rho}
\end{equation}
\begin{equation}
I = 2\pi\eta_{\rm n} c \, \frac{L_{\rm n}-\Omega_{\rm F}r^{2}}{1-{\cal M}^2},
\label{BE-I}
\end{equation}
\begin{equation}
v_{\varphi}=\frac{1}{r} \, 
\frac{\Omega_{\rm F}r^{2} - L_{\rm n}{\cal M}^2}{1-{\cal M}^2}.
\label{BE-v}
\end{equation}
Here, the quantity
\begin{equation}
{\cal M}^2=\frac{4\pi\eta_{\rm n}^{2}}{\rho}
\label{M2}
\end{equation}
is the square of the Alfv{\' e}n Mach number
\begin{equation}
{\cal M}^2 = \frac{v_{\rm p}^{2}}{v_{\rm A,p}^{2}}
=v_{\rm p}^{2}\frac{4\pi\rho}{B_{\rm p}^{2}}.
\label{calM}
\end{equation}
To find ${\cal M}^2$, Bernoulli equation (14) should be used, which, taking into consideration the definitions above, can be represented in the form
\begin{eqnarray}
\frac{{\cal M}^4}{64\pi^4\eta_{\rm n}^2}\left(\nabla\Psi\right)^2 
= 2r^{2}[E_{\rm n} - w(\rho, s) - \varphi_{\rm g}] 
\nonumber \\
-\frac{(\Omega_{\rm F}r^2 - L_{\rm n}{\cal M}^2)^2}{(1-{\cal M}^2)^2}
-2r^2\Omega_{\rm F}\frac{L_{\rm n}-\Omega_{\rm F}r^2}{1-{\cal M}^2}. 
\label{B-Me}
\end{eqnarray}

Since, according to (19), $\rho = \rho(\eta, {\cal M}^2)$, enthalpy $w(\rho, s)$ can be represented as a function of $\eta$, 
$s$, and ${\cal M}^2$, i.e., the first two integrals of motion and the Mach number. Therefore, Eqn (21) defines, albeit in an implicit form, the Alfv{\' e}n Mach number as a function of magnetic flux $\Psi$ and five integrals of motion:
\begin{equation}
{\cal M}^2 = 
{\cal M}^2(\nabla \Psi, E_{\rm n}, L_{\rm n}, \Omega_{\rm F}, \eta_{\rm n}, s).
\end{equation}

As to magnetic flux $\Psi(r,\theta)$ itself, to determine it, the remaining component of the equation of force balance should be used, which, in a compact form, can be represented as [95, 102]
\begin{eqnarray}
\frac{1}{16 \pi^3 \rho}
\nabla_{k}\left(\frac{1-{\cal M}^2}{r^{2}}
    \nabla^{k}\Psi\right)
+\frac{{\rm d}E_{\rm n}}{{\rm d}\Psi}
+\frac{\Omega_{\rm F}r^2 - L_{\rm n}}{1-{\cal M}^2} \,
\frac{{\rm d}\Omega_{\rm F}}{{\rm d}\Psi}
\nonumber \\
+\frac{1}{r^2} \,
\frac{{\cal M}^2 L_{\rm n} - \Omega_{\rm F} r^2}{1-{\cal M}^2}
\frac{{\rm d}L_{\rm n}}{{\rm d}\Psi}
+\left[2(E_{\rm n} - w(\rho, s) - \varphi_{\rm g}) \frac{\frac{}{}}{} \right.
\nonumber \\
\left.
+ \frac{1}{r^2} \,
\frac{\Omega_{\rm F}^2 r^4 -2\Omega_{\rm F}L_{\rm n} r^2
+ {\cal M}^2 L_{\rm n}^2}{1-{\cal M}^2}\right]
\frac{1}{\eta_{\rm n}}\frac{{\rm d}\eta_{\rm n}}{{\rm d}\Psi}
-\frac{T}{m_{\rm p}} \, \frac{{\rm d}s}{{\rm d}\Psi} = 0.
\nonumber
\end{eqnarray}
\begin{eqnarray}
\label{GS}
\end{eqnarray}
Thus, taking into account the Bernoulli integral represented as Eqn (21), Eqn (23) only contains one unknown function, the magnetic flux $\Psi(r,\theta)$. It is a generalization of the Grad-Shafranov equation for the case of nonzero flow velocities and, for this reason, is often referred to as the Grad-Shafranov equation. It is of special importance to stress once again that Eqn (23) originates from the equation of force balance in the direction perpendicular to the direction of plasma motion.

Finally, it should be recalled that, to study cylindrical configurations (a jet away from the 'central engine'), it turned out to be extremely convenient to reduce the problem to two ordinary first-order differential equations [102--104]. One of these equations is the Bernoulli equation (21)
\begin{eqnarray}
\frac{{\cal M}^4}{64\pi^4\eta_{\rm n}^2}
\left(\frac{{\rm d}\Psi}{{\rm d}r}\right)^2 
= 2r^{2}(E_{\rm n} - w)  
\nonumber \\
-\frac{(\Omega_{\rm F}r^2-L_{\rm n}{\cal M}^2)^2}{(1-{\cal M}^2)^2}
-2r^2\Omega_{\rm F}
\frac{L_{\rm n}-\Omega_{ \rm F}r^2}{1-{\cal M}^2}
\label{B-Me1}
\end{eqnarray}
for the flux function $\Psi(r)$, and the other, the equation for the Mach number squared ${\cal M}^2$, obtained by integrating the Grad-Shafranov equation, which becomes possible in the one-dimensional case [95]:
\begin{eqnarray}
&&\left[2(E_{\rm n}^{\prime} - w)
+\Omega_{\rm F}^2r^2- (1-{\cal M}^2)c_{\rm s}^2\right]
\frac{{\rm d}{\cal M}^2}{{\rm d}r}
=
\nonumber \\ 
&&\frac{{\cal M}^6}{1-{\cal M}^2}\frac{L_{\rm n}^2}{r^3}
-\frac{\Omega_{\rm F}^2r}{1-{\cal M}^2}{\cal M}^2(2{\cal M}^2-1)
\nonumber \\ 
&&+ {\cal M}^2\frac{{\rm d}\Psi}{{\rm d}r} 
\frac{{\rm d}E_{\rm n}^{\prime}}{{\rm d}\Psi}
+ {\cal M}^2r^2\Omega_{\rm F}
\frac{{\rm d}\Psi}{{\rm d}r}
\frac{{\rm d}\Omega_{\rm F}}{{\rm d}\Psi}
\nonumber \\ 
&& + \left[2(E_{\rm n}^{\prime} - w) + \Omega_{\rm F}^{2}r^{2} 
- 2(1-{\cal M}^2)c_s^2\right]
\frac{{\cal M}^2}{\eta_{\rm n}}
\frac{{\rm d}\Psi}{{\rm d}r}\frac{{\rm d}\eta_{\rm n}}{{\rm d}\Psi}
\nonumber \\ 
&&-{\cal M}^2\left[(1-{\cal M}^2)\frac{1}{\rho}
\left(\frac{\partial P}{\partial s}\right)_{\rho}+\frac{T}{m_{\rm
p}}\right]\frac{{\rm d}\Psi}{{\rm d}r} \frac{{\rm d}s}{{\rm d}\Psi}.
\label{GS-Me}
\end{eqnarray}
Here, $r$ is again the cylindrical radius, $E_{\rm n}^{\prime} = E_{\rm n}-\Omega_{\rm F}L_{\rm n}$, and $T$ is temperature (in energy units); we naturally disregarded the effect of the gravitational field.

\vspace{0.3 cm}

{\bf 2.2.3 Main properties of nonrelativistic jets.} The method of the Grad-Shafranov equation enabled significant progress in the
formalization of the entire problem of the activity of compact astrophysical objects [66, 97--99, 102--106]. Below, we only consider the results which pertain to the main topic of this review.

We start with the observation that, in any theory, dimensionless parameters are of importance. One such parameter is the so-called magnetization parameter $\sigma_{\rm n}$, equal by definition to the ratio of the electromagnetic field flow and the flow of particle energy at the jet base{\footnote{To avoid confusion, we note that the parameter $\sigma_{\rm n}$ is equivalent to the relativistic Michel magnetization parameter $\sigma_{\rm M}$ rather than the current magnetization value $\sigma$ widely used at present.}}. The quantities defined above can be used to represent $\sigma_{\rm n}$ in the form
\begin{equation}
\sigma_{\rm n} = \frac{\Omega_{\rm F}^{2}\Psi_{\rm tot}^2}
{8\pi^2 v_{\rm in}^{3}{\dot M}_{\rm jet}}.
\label{sigma}
\end{equation}
Here and below, the subscript 'in' corresponds to quantities near the `central engine,' and we set $\eta_{\rm n} = {\dot M}_{\rm jet}/\Psi_{\rm tot}$, where $\Psi_{\rm tot} \approx \pi B_{0} R^2$ is the total magnetic flux in a jet. Below, we are naturally interested in the case $\sigma_{\rm n} > 1$, when the energy flow near the jet base is fully determined by the electromagnetic field flow. It should be noted that the condition of strong magnetization $\sigma_{\rm n} > 1$ can be represented in the form of the condition of fast rotation of the 'central engine' $\Omega_{\rm F} > \Omega_{\rm cr}$, where
\begin{equation}
\Omega_{\rm cr} = \left(
\frac{v_{\rm in}^3\eta_{\rm n}}{\Psi_0}\right)^{1/2}
= \frac{v_{\rm in}}{R_{\rm in}}\left(
\frac{4\pi\rho_{\rm in}v_{ \rm in}^2}{B_{ \rm in}^2}\right)^{1/2}.
\label{ocr}
\end{equation}

We shall see that it is convenient to consider a ratio of the accretion and ejection rates as another important parameter [66]:
\begin{equation}
\lambda = \frac{{\dot M}_{\rm acc}}{{\dot M}_{\rm jet}}.
\label{B-lambda}
\end{equation}
The point is that the property of the angular velocity of a star to stay virtually invariable (and, hence, the accreting matter fully transfers its angular momentum to a jet), which was described above, can be used to obtain
\begin{equation}
{\dot M}_{\rm jet}L_{\rm n} = {\dot M}_{\rm acc} \Omega(R_{\rm in}) R_{\rm in}^2,
\label{dotMM}
\end{equation}
where $R_{\rm in}$ is the distance from the axis of the specific field line on the disk surface. As a result, using Eqn (17), according to which on the Alfv{\' e}n surface ${\cal M} = 1$ the relation
\begin{equation}
L_{\rm n} = \Omega_{\rm F} r_{\rm A}^2,
\label{LA}
\end{equation}
should hold, we finally arrive at [107]
\begin{equation}
\lambda = \frac{r_{\rm A}^2}{R_{\rm in}^2}.
\label{B-lambdaaa}
\end{equation}
Here, $r_{\rm A}$ is the distance from the axis to the Alfv{\' e}n surface that corresponds to the given field line.

We now discuss the innermost regions at the jet base, in which the transverse size of the flow increases from the dimensions of the 
'central engine' to jet dimensions, i.e., by a factor of several hundred or even thousands. As a result of this expansion, thermal effects (which, as we keep in mind, did not play a decisive role in the `central engine' itself) due to adiabatic cooling apparently become negligible. Therefore, below, we primarily use the cold plasma approximation ($s = 0$), noting thermal effects only in cases when they cannot be discarded.

However, in any case, the key point is that, due to a significant expansion of the outflowing matter, the flow inevitably should pass the Alfv{\' e}n and the fast magnetosonic surface (in the nonrelativistic case, they are located not far from each other, touching each other on the rotation axis.) Thus, the flow is inevitably transonic. This basic feature leads to an entire 'chain' of very important conclusions.

First of all, the Alfv{\' e}n and the fast magnetosonic surfaces are critical surfaces, i.e., their existence results in additional restrictions on flow parameters [108]. For the Alfv{\' e}n surface, this is clear already from Eqns (17) and (18), which require that corresponding numerators vanish if the condition ${\cal M}^2 = 1$ holds. In particular, for a quasi-monopole magnetic field, the following explicit formula can be derived:
\begin{equation}
L_{\rm n}(\Psi) = i_{0} \, \frac{\Omega_{\rm F}(\Psi)}{4 \pi^2 c \eta_{\rm n}} \, \Psi,
\label{Ln}
\end{equation}
which we need below. The quantity $i_{0}$ introduced here for convenience is a dimensionless longitudinal current $I$, i.e., $I = i_{0}I_{\rm GJ}$, where $I_{\rm GJ}$ is the so-called Goldreich-Julian current [109]
\begin{equation}
I_{\rm GJ}  = \pi R^2 c \rho_{\rm GJ}
\label{IGJ}
\end{equation}
and the quantity
\begin{equation}
\rho_{\rm GJ}  = \frac{\Omega B_{0}}{2 \pi c}
\label{IGJbis}
\end{equation}
is the module of Goldreich-Julian charge density (for this estimate (and below),  $\Omega$ is the angular velocity of the rotation of the 'central engine'). For strongly magnetized nonrelativistic flows $\sigma_{\rm n} \gg 1$ [95, 102],
\begin{equation}
i_0 \approx \frac{c}{v_{\rm in}}\left(
\frac{\Omega}{\Omega_{\rm cr}}\right)^{-2/3} = \frac{c}{v_{\rm in}}\sigma_{\rm n}^{-1/3},
\label{i00}
\end{equation}
while for slow rotation,
\begin{equation}
i_0 \approx \frac{c}{v_{\rm in}}.
\label{i_0n}
\end{equation} 

For a fast magnetosonic surface, the critical condition arises due to the appearance in the Grad-Shafranov equation (23) of the expression $\nabla {\cal M}^2 = N/D$, the denominator of which,
\begin{equation}
D = \frac{1 - {\cal M}^2}{{\cal M}^2} + \frac{B_{\varphi}^2}{{\cal M}^2B_{\rm p}^2}
- \frac{c_{\rm s}^2}{v_{\rm p}^2} \frac{1 - {\cal M}^2}{{\cal M}^2},
\label{B-D}
\end{equation}
vanishes, which is easy to verify, just under the condition that the poloidal flow velocity vp and the speed of the fast magnetosonic wave $v_{\rm fm}$ are equal.

The next conclusion we can draw here is that a second-order equation with four integrals of motion and two critical surfaces (it should be kept in mind that we set $s = 0$) requires four boundary conditions on the surface of the 'central engine' $r \sim R$ [95]. For further analysis, it is convenient to choose the magnetic field $B_{0}$, the angular velocity of the 'central engine' $\Omega$, the ejection rate ${\dot M}_{\rm jet}$, and the flow velocity on the surface{\footnote{At a nonzero temperature, the outflow rate should be determined from the critical condition on the slow magnetosonic surface, which is absent in the case of a cold plasma.}} $v_{\rm in}$.

Furthermore, as noted, Eqn (31) shows that the Alfv{\' e}n scale $r_{\rm A}$ is a characteristic scale of our problem. On the other hand, in the nonrelativistic case, $r_{\rm A}$ can be estimated using the formula for the radius of a fast magnetosonic surface, which gives
\begin{equation}
r_{\rm F} \approx \left(\frac{\Psi_{\rm tot}^2}{\Omega {\dot M}_{\rm jet}}\right)^{1/3}.
\label{B-rF}
\end{equation}
If we now choose as the angular velocity the Keplerian velocity at the Alfv{\' e}n radius $R_{\rm A}$ (5) (which should be
distinguished from the radius of the Alfv{\' e}n surface $r_{\rm A}$!), for the characteristic values of young stars we eventually obtain
\begin{eqnarray}
r_{\rm F} \approx 0.4 \, {\rm au}
\left(\frac{B_{0}}{10^{3} \, {\rm G}}\right)^{20/21} 
\left(\frac{M}{1.4 \, M_{\odot}}\right)^{-5/21} 
\nonumber \\
\times \left(\frac{R}{2 \,R_{\odot}}\right)^{46/21} 
\left(\frac{{\dot M}_{\rm jet}}{10^{-9}M_{\odot} \, {\rm yr}^{-1}}\right)^{-10/21}.
\label{B-rFnum}
\end{eqnarray}

It can be seen that on a logarithmic scale the singular surfaces are located just between the size of the 'central engine' and the transverse dimensions near the jet base. Therefore, as was noted, on a scale of several tens of astronomical units, the supersonic flow should inevitably interact with the surrounding medium, resulting in an efficient heating of electrons to a temperature of several
million degrees and, consequently, to X-ray radiation [110].

Another basic feature related to the presence of singular surfaces is that, similarly to the way the critical condition determines the accretion rate in the Bondi solution, the critical surface on the fast magnetosonic surface sets the value of the longitudinal current $I$ that circulates in the magnetosphere [105, 111, 112] and, consequently, determines the total energy release of the 'central engine.'

As a result, the total electromagnetic energy losses $W_{\rm tot} = (c/4\pi)\int [{\bf E} \times {\bf B}] {\rm d}s$, which can be represented in the form
\begin{equation}
W_{\rm tot} \approx i_{0} \left(\frac{\Omega R}{c}\right)^{2} B_{0}^{2} R^{2} c,
\label{Wtot}
\end{equation}
for a rapidly rotating 'central engine,' $\Omega > \Omega_{\rm cr}$, can be described by an amazingly simple formula in terms of observable quantities (see, for example, [102]):
\begin{equation}
W_{\rm tot} \approx \Omega^{4/3} \Psi_{\rm tot}^{4/3} {\dot M}^{1/3},
\label{exta1}
\end{equation}
i.e., in terms of the total magnetic flux $\Psi_{\rm tot} = \pi R^2 B_0$, angular rotation velocity $\Omega$, and the rate of mass loss in a jet${\dot M}$. For the parameters characteristic of young stars,
\begin{eqnarray}
W_{\rm tot} \sim 10^{36} \,\, {\rm erg} \,\,  {\rm s}^{-1} 
\left(\frac{P}{10 \, {\rm days}}\right)^{4/3}
\left(\frac{B_{0}}{10^{3} \, {\rm G}} \right)^{4/3}
\nonumber \\
\times
\left(\frac{R}{2 \, R_{\odot}}\right)^{8/3}
\left(\frac{\dot M}{10^{-9}\, M_{\odot} \, {\rm yr}^{-1}}\right)^{1/3}.
\label{exta3}
\end{eqnarray}

The value of $W_{\rm tot}$ is actually close to the energy losses characteristic of young stellar objects. It is of interest that, if total energy losses $W_{\rm tot}$ are known, it possible to immediately estimate the total longitudinal current circulating in the `central engine.' Indeed, comparing Eqns (33) and (40), we immediately obtain
\begin{equation}
I \approx i_0 c^{-1/2}W_{\rm tot}^{1/2}.
\label{i_0}
\end{equation}

Finally, it should be noted that the transonic nature of astrophysical flows shows that care must be taken in analyzing their properties based on the so-called magnetic tower model [113], which is often mentioned both in connection with theoretical studies and in the analysis of laboratory experiments. In this subsonic cylindrical model, the jet width does not change significantly with distance from the 'central engine,' as a result of which the longitudinal current is determined not by the conditions on critical
surfaces but by the degree of swirling of the toroidal magnetic field, i.e., dissipative processes.

As regards assertions related to the region of quasi-cylindrical flow, the following should be noted. First of all, a nonrelativistic cylindrical flow always contains an internal scale [95, 102]
\begin{equation}
r_{\rm core}  = \frac{v_{\rm in}}{\Omega_{\rm F}}.
\label{core}
\end{equation}
At distances $r$ from the axis that are much greater than $r_{\rm core}$, both the velocity and the density of the outflowing matter decrease sharply. For the characteristic parameters of young stars,
\begin{equation}
r_{\rm core}  = 0.3 \, {\rm au} 
\left(\frac{v_{\rm in}}{300 \, {\rm km} \, {\rm s}^{-1}}\right) 
\left(\frac{P}{10 \, {\rm days}}\right),
\label{corea}
\end{equation}
i.e., this scale is still beyond the resolution of modern telescopes.

On the other hand, an analysis of Eqn (25) showed in [97, 98] that, at large distances from the jet axis, $r \gg r_{\rm core}$ (and in the cold plasma approximation), it reduces simply to the zero derivative
\begin{equation}
\frac{{\rm d}}{{\rm d} r} 
\left(\frac{\eta_{\rm n}\Omega_{\rm F}r^{2}}{{\cal M}^2}\right) = 0,
\label{II}
\end{equation}
i.e., to the conservation of the quantity
\begin{equation}
I_{\rm c} = \frac{2 \pi c\eta_{\rm n}\Omega_{\rm F}r^{2}}{{\cal M}^2}.
\label{III}
\end{equation}
However, according to Eqn (17), for supersonic nonrelativistic flows, the value of $I_{\rm c}$ coincides with the total current flowing within the central magnetic tube. Consequently, we arrive at the most important conclusion: in the cylindrical flow region, the longitudinal current $j_{\parallel}$ should be concentrated in a small region $r \approx r_{\rm core}$. Thus, the reverse current should only return in the region of the so-called cocoon (the peripheral region of the jet), where the cold plasma approximation becomes inapplicable. Hence, the conservation of the quantity $I_{\rm c}$  enables an easy determination of the relationship between the transverse size of the cocoon $r_{\rm coc}$ and the external pressure $P_{\rm ext}$, which follows from the equilibrium condition  $P_{\rm ext} = B_{\varphi}^2(r)/8\pi$:
\begin{equation}
r_{\rm coc} \approx \left(\frac{I_{\rm c}^2}{2 \pi c^2 P_{\rm ext}}\right)^{1/2}.
\label{rcocoon}
\end{equation}
It should be emphasized here that the assertion about the constancy of the total current $I(r)$ only refers to nonrelativistic flows; in the relativistic case, $I(r)$ can vary significantly within the jet. Therefore, the longitudinal current $I_{\rm c}$ (47) should not coincide with the current $i_0 I_{\rm GJ}$ (43) determined by us for the 'central engine,' since, in the full MHD version, the current $I$, in contrast to the current in the force-free approximation, is not an integral of motion.

On the other hand, using relation (47) and definitions (9) and (19), outside the dense core, $r > r_{\rm core}$, we obtain
\begin{equation}
B_{\varphi}(r) = B_{\rm p}(r_{\rm core}) \left(\frac{r}{r_{\rm core}}\right)^{-1}.
\label{Bphmore}
\end{equation}
Consequently, at the very boundary of the core, the toroidal and longitudinal magnetic fields should be close to each other. We emphasize that this result is a consequence of a fine adjustment of the magnitude of the longitudinal current, which, keep in mind, is associated with the condition of smooth passage of the flow through singular surfaces. As regards the poloidal magnetic field $B_{z}$, it also turns out to be concentrated within the limits of the central compaction. Moreover, as shown in [114], outside the central compaction, the poloidal magnetic field decreases as
\begin{equation}
B_{z}(r) = B_{\rm p}(r_{\rm core}) \left(\frac{r}{r_{\rm core}}\right)^{-\alpha},
\label{Bphmorea}
\end{equation}
where $\alpha \approx 2$. Finally, taking into account Eqn (49), we can represent the transverse size of the cocoon in a convenient
form:
\begin{equation}
r_{\rm coc} \approx \left(\frac{B_{\varphi}^2(r_{\rm core})}{8 \pi P_{\rm ext}}\right)^{1/2} \, r_{\rm core}. 
\label{rcocoonbis}
\end{equation}

Furthermore, as it turns out, the above-mentioned critical conditions, among other things, imply that the flow in the supersonic region cannot be strongly magnetized, i.e., the plasma energy flux should be of the order of the energy flux of the electromagnetic field, even if the flow at the jet base was strongly magnetized ($\sigma_{\rm n} \gg 1$). Using the asymptotic form of Eqn (17) for the current $I$ at ${\cal M}^2 \gg 1$ ($I = 2 \pi \eta_{\rm n}\Omega_{\rm F}r^2/{\cal M}^2$) and the definition of the Goldreich-Julian current $I_{\rm GJ} = \pi r^2 c \rho_{\rm GJ}$, we find that, in the region $r < r_{\rm core}$, the condition $i_{0} = I/I_{\rm GJ} = c/v_{\rm in}$ should be satisfied. This formula, which coincides with Eqn (36) for the case of slow rotation, can be
considered a condition of approximate equality of the energy fluxes of particle and electromagnetic field. The difference from the value (35) should not be surprising, since the current I is not an invariant for MHD flows.

Another basic property that follows from the MHD theory discussed here is the rotation of the jet. As can be seen from exact formula (18), the angular velocity of rotation of particles $\omega = v_{\varphi}/r$ should not coincide with the angular velocity $\Omega_{\rm F}$, since, in the approximation considered here, the plasma motion is not only rotation with an angular velocity $\Omega_{\rm F}$ but also a sliding along the magnetic field lines [95]. Therefore, due to the strong toroidal magnetic field, the angular velocity of plasma rotation should differ from the angular velocity $\Omega_{\rm F}$. In particular, for a supersonic flow ${\cal M} > 1$, the angular velocity of plasma rotation $\omega$ within the central core can be represented as
\begin{equation}
\omega   \approx  \sigma_{\rm n}^{-1/3}{\cal M}^{-2} \, \Omega_{\rm F}.
\label{7_1}
\end{equation}
Equation (52) can be easily derived from the estimate $\omega \approx L_{\rm n}/r^2$ that follows from Eqn (18) for ${\cal M} > 1$, and from the explicit expression for $L_{n}(\Psi)$ (32) that follows from the condition of smooth passage through singular surfaces. Consequently, for $r > r_{\rm core}$ , we have
\begin{equation}
\omega   \approx  \sigma_{\rm n}^{-1/3}{\cal M}^{-2} 
\left(\frac{r}{r_{\rm core}}\right)^{-\alpha} \, \Omega_{\rm F},
\label{7_2}
\end{equation}
so, at $\alpha \approx 2$, the rotational speed outside the central core should decrease approximately as $v_{\varphi} \propto r^{-1}$.

Finally, one more remarkable result should be pointed out that connects the above assertions. As noted, the Grad-Shafranov equation is an equation for the balance of forces in the direction perpendicular to the flow velocity. Therefore, in the nonrelativistic case, the Grad-Shafranov equation cannot be anything but the balance of the centrifugal force and the Lorentz force, which in the supersonic regime has the form
\begin{equation}
\frac{\rho v_{\rm p}^2}{R_{\rm c}} = 
\frac{j_{\parallel}B_{\varphi}}{c}.
\label{GSGS}
\end{equation}
Here, $R_{\rm c}$ is the curvature radius of the magnetic surface, and we only presented the terms that are leading for a highly magnetized flow.

Equation (54) enables us to draw a number of important conclusions. First of all, it can be easily checked that, for the characteristic currents flowing in the magnetosphere,
\begin{equation}
\frac{R_{\rm c}}{r} \approx 
\frac{v_{\rm A}^2(r)}{\Omega^2r_{\rm A}^2} \, \frac{r^2}{r_{\rm A}^2} 
\sim \left(\frac{r}{r_{\rm A}}\right)^{-2},
\label{Rcr}
\end{equation}
so that, at the Alfv{\' e}n radius $r = r_{\rm A}$, the curvature radius $R_{\rm c}$ becomes comparable to $r$; hence, the nonrelativistic flow inevitably begins to collimate along the rotation axis. In other words, the longitudinal current $I$ (which, as should be kept in mind, is determined from the condition of passage of singular surfaces) turns out to be large enough for the pinching effect (parallel currents are attracted) to lead to collimation. Actually, this is the main attractive feature of this approach, which made it possible to explain the observed collimation of nonrelativistic jets using a simple and physically transparent model{\footnote{It should be kept in mind that, for relativistic jets, inherent collimation turns out to be inefficient.}}.

On the other hand, as can be seen from Eqn (54), the sign of the curvature radius depends on the direction of the longitudinal current $j_{\parallel}$. In other words, in the reverse current region, it is not collimation but decollimation of the flow that should occur; at small distances $r \ll r_{\rm A}$, the decollimation was actually reproduced both analytically [115] and numerically [105]. Thus, at large distances, this should lead to the emergence of an empty region, i.e., an area not filled with a poloidal magnetic field.

However, nature is known abhors a vacuum. Therefore, as shown numerically [105], instead of a region not occupied by a poloidal magnetic field, a region emerges where longitudinal current density is zero. Mathematically, this result is related to the following: since the radius of curvature $R_{\rm c}$ cannot be less than the characteristic size $r$, at distances from the axis that are much larger than the Alfv{\' e}n radius, $r \gg r_{\rm A}$, the left side of Eqn (54) can be neglected, and then we arrive at relation $j_{\parallel} = 0$, which we have already derived by analyzing the cylindrical asymptotic form of the Grad-Shafranov equation.

Concluding this section, it is necessary to note two important circumstances that can significantly limit the scope of the above analytical results. The first is related to the very possibility of using the MHD approach for a medium that is not completely ionized. The second circumstance concerns another basic assumption about the existence of a regular magnetic field that penetrates the accretion disk. The point is that this assumption disagrees with one of the main provisions of the theory of accretion disks, according to which accretion becomes possible due to the viscosity caused by strong turbulence. To clarify the latter assertion, some
comments on the results of numerical simulation should be made.

Of course, the numerical simulation of jets from young stellar objects, which has been rapidly developing in recent years, is not limited to only the problem of turbulence in accretion disks. Work is also underway to simulate the generation of a magnetic field and its interaction with an accretion disk, the collimation and stability of the jets themselves, and the structure of the region of interaction of a jet with the environment. A detailed discussion of each of these areas of research would require a separate review.
Therefore, we again confine ourselves to a few remarks directly related to the main topic of the review.

As regards the first issue (namely, the applicability of the MHD approach to an incompletely ionized medium), here, ambipolar diffusion comes to the rescue [47]. As a result, despite the low degree of ionization, neutral and ionized particles will be bound by collisions, so the substance will move primarily along the magnetic field lines. Thus, the results of the MHD approximation are generally adequate.

The second issue concerns the mechanism for maintaining turbulence in accretion disks, which is necessary for the very existence of accretion. As is well known, in recent years, magneto-rotational instability has been considered to be the most promising mechanism [116]. Moreover, in the first studies with numerical simulations of the formation of outflow from accretion disks, the development of turbulence led to the complete disappearance of the regular component of the magnetic field (see, for example, [117]). However, it was
shown later that, in the presence of a significant vertical magnetic flow and taking into account ambipolar diffusion in combination with ohmic dissipation, the magneto-rotational instability is suppressed, and a powerful magneto-centrifugal wind is generated [47, 118, 119], as expected in an analytical approach. Thus, the above model of the loss of angular momentum from the surface of an accretion disk certainly is justified.

However, the issue of the properties of a turbulent disk is not directly related to laboratory studies that simulate jets from young stars. In any case, as shown in Section 2.2.3, in most laboratory experiments, the nature of the mechanism for the launch of a collimated plasma flow is completely different. Therefore, we discuss here neither the process of collimation itself nor the structure of the magnetic field in the region of the central engine, since, despite the large number of studies devoted to them [120--126], they are not directly related to the laboratory experiment discussed in this review (see Section 3).

As for the internal structure of jets already formed, as discussed in Sections 2.2.1 and 2.2.2 (see also [127, 128]), according to both observations and one-dimensional analytical models, jets have a dense core; moreover, their longitudinal velocity is higher in the central regions of the jet, decreasing with increasing distance from the axis. A similar structure has also been confirmed in 3D numerical simulations for both nonrelativistic [123] and relativistic [129--131] jets.

Finally, many studies are devoted to the numerical simulation of active regions of interaction between a jet and interstellar gas. As early as the 1980s--1990s, the emergence of two shock waves during the interaction of a supersonic jet with the environment was successfully simulated, and the role of radiation processes in general was clarified [132--134]. Later, for a consistent analysis of the processes of heating and radiation in shock waves, all the main processes of ionization and recombination were included in the calculations [135]. The complex multicomponent structure of 'head parts' has also been reproduced [93, 136, 137], and even the
interaction of a jet with a side wind has been simulated [138] (see also review [49]). A significant number of studies on
numerical simulation are also related to the analysis of the results obtained with experimental setups (see, for example, Refs [139, 140]). Moreover, in all numerical experiments, the magnetic field really played a decisive role, making it possible to reproduce the main morphological features of the observed flows.

Thus, at present, the theory of jets from young stars makes it possible to formulate a fairly reliable model that can explain
the main features of the observed flows. It is based on a physically transparent electrodynamic idea, which has been successfully tested for other compact objects, such as active galactic nuclei, radio pulsars, and micro-quasars. However, as already noted, such key predictions as the presence of a dense central core and the absence of a longitudinal current outside it have not been confirmed. These and other unanswered questions required an even wider scope of research. Here, the role of the missing link was played precisely by the laboratory experiment.

\subsection{Laboratory aspect}

The idea to turn to the laboratory experiment as the most important area of research into astrophysical processes was already suggested more than 20 years ago [17, 141]. Indeed, although the characteristic lengths and time scales of laboratory experiments are about 20 orders of magnitude smaller than those of real astrophysical objects, they can be scaled for astrophysical situations to the extent that both are described by the laws of ideal MHD. This is due to the fact that, in cases where dissipative processes do not play a decisive role, the ideal MHD equations do not have their intrinsic scale and can describe both laboratory and astrophysical flows. However, when dissipative processes turn out to be significant, the issue of similarity requires a more thorough analysis [15].

The relocation of studies of astrophysical jets to the laboratory has a number of unquestionable advantages. First of all, flow parameters can be easily varied in laboratory plasma, which is extremely important for testing the predictions of theoretical models. In particular, theoretical models provide an important test for the reliability of the description of astrophysical jets in the MHD approximation. Furthermore, the time frame of laboratory experiments is small, so the dynamics of ongoing processes can be easily
monitored, while tracking the dynamics of real astrophysical jets can take several years or even decades. In addition, laboratory experiments can in principle be fully diagnosed, while the diagnosis of real astrophysical jets is limited, so many important characteristics, such as the structure of the internal magnetic field and the density profile, are poorly known. Finally, laboratory experiments are relatively inexpensive compared to today's ground-based and space-based telescopes, owing to which a great deal of information needed to clarify the underlying physical processes can be obtained using modest resources. Laboratory experiments can also be used to verify the MHD software used to describe real astrophysical jets.

As noted in the Introduction, progress in the laboratory modeling of astrophysical jets was achieved due to the studies performed as part of the inertial controlled thermonuclear fusion program, which led to the rapid development of Z-pinch systems and high-power lasers, i.e., state-of-the-art installations with a high energy density. Table 1 displays the main characteristics of plasma jets obtained at a number of experimental facilities, where laboratory simulations of astrophysical jets were studied. It can be seen that all kinds of equipment have been used, including high-power pulsed lasers, installations with high energy power, in particular, fast
Z-pinches, and plasma accelerators. Moreover, the scale of the laboratory experiments themselves differ greatly.

\begin{table*}
\caption{Main features of plasma jets obtained at experimental facilities where they are simulated (PF is the plasma focus and PG is the plasma gun).}
%{\footnotesize
\vspace{0.2cm}
\centering
{\begin{tabular}{|c|c|c|c|c|c|c|c|c|}
\hline
Name  & Facility & Length& Radius&  Time & $V$ & $B$  & $n$  & $T$  \\ 
 & type & cm &  cm &  $\mu$s & km s$^{-1}$ & G  & cm$^{-3}$   & eV   \\ 
\hline
Typical YSO  & -- & $10^{17}$ &$10^{15}$ & $10^{9}$ c& $\sim 100$& $\sim 10^{-4}$ & $\sim 10^{4}$ & $\sim$ 1 \\
\hline
PF-3  & PF & 10--100& 1--10 & $\sim 10$ & $\sim 100$& $10^{3}$--$10^{4}$  &$10^{16}$ & $1$--$5$\\
PF-1000 & PF & 10--100& 1--10 & $\sim 10$ & $\sim 100$& $10^{3}$--$10^4$  &$(1$--$6)\times 10^{17}$ & 3--5\\
Phoenix  & PF & 10--100& 1--10 & $\sim 10$ & $\sim 100$& $10^{3}$--$10^4$  & -- & --\\
  \hline
MAGPIE & Pinch &1--3.5 & 0.1 & 0.1--0.5 & 200& $\sim 10^{5}$& $10^{18}$-- $10^{19}$ & 50--100 \\
COBRA & Pinch & 1--5 &  0.1--1 & $\sim$ 0.1 & $\sim$ 100  & (1--2)$\times 10^4$ &  $10^{18}$-- $10^{19}$ & 15 \\
NOVA & Laser & 0.1 & 0.01 & 0.001 & 60 & -- & $10^{20}$-- $10^{21}$ & $\sim$ 100\\
OMEGA & Laser & 0.7 & 0.05 & 0.001 & 400 & 2 $\times 10^{6}$ & $\sim 5 \times 10^{19}$ & $\sim$ 300 \\
LULI & Laser & 1 & 0.05 & 0.001 & $90$--$200$ & $4 \times 10^{5}$ & (1--3)$ \times 10^{18}$& $>$ 70 \\
Neodim & Laser & 0.1--1 & 0.1 & 0.1--0.01 & $\sim 100$ & $\sim 10^8$ & $\sim 2 \times 10^9$ & 1 \\
PEARL & Laser & 0.1--1 & 0.1--1 & 0.1  & $100$--$500$ & $1 \times 10^{5}$ & $10^{17}$--$10^{18}$& 30 \\
KI-1 Superjet & Laser & 70 & 30 & 10 & $100$ & $300$ & -- & 10 \\
Caltech & PG & 25 & 1 & 15 & 60 & (0.2--1)$ \times 10^3$ &  $10^{14}$ & 5--20 \\
LabJet & PG&  100 & 1--5 & 10--100 & 20--100 & $5 \times 10^3$& $10^{16}$ & 1--10 \\
 \hline
 \end{tabular} 
\label{ta0}
}
%}
\end{table*}

Some of the first experiments of this kind were carried out with laser devices. For example, in the scheme of indirect irradiation at the Nova facility at the Livermore National Laboratory (USA) [142, 143], laser radiation with a duration of 1 ns and an energy of 20 kJ was fed into a closed volume (the so-called hohlraum). The interaction with the walls of this volume created a powerful pulse of X-ray radiation, leading to the ablation of specially prepared targets and the generation of a plasma flow. In other setups, the laser
beam was focused directly on the target, which ensured the release of very high power on its surface. Several laser beams directed at conical targets were often used. This approach was used in experiments at the facilities Nova [144], Omega at the University of Rochester (USA) [145], the Vulcan laser system at the Rutherford--Appleton Laboratory (UK) [146], etc.

Experiments with laser jets [147, 148] were primarily aimed at studying hydrodynamic instabilities in a jet interacting with localized dense obstacles [149], which lead to jet deflection similar to that observed in the Herbig-Haro HH 110 object. A detailed comparison of the experimental results with numerical simulation showed their good agreement. In a recent experiment [150], the interaction between direct and reflected shock waves (Mach stem) was studied, which is relevant to the observation of similar structures in Herbig-Haro objects. Such collisions, which are expected during the interaction of inhomogeneous jets with the surrounding gas, can result in the formation of shock waves normal to the flow [38].

In experiments carried out at the {\' E}cole Polytechnique (France) using the LULI facility{\footnote{Laboratoire d'Utilization des Lasers Intenses (LULI): a scientific research laboratory specializing in the study of plasmas generated by laser-matter interaction at high intensities and their applications.}} [151], a 0.5-ns laser pulse delivers energy up to 50 J to a massive (plastic) target in a spot 0.75 mm in diameter. This leads to an explosive ejection of the target material, which freely expands at a large angle, like the wind of young stellar objects. To collimate this flow, a strong longitudinal magnetic field of the order of 200 kG is applied. Studies in the time interval of 20 ns, according to the scaling laws (see Section 3.4.2), correspond to six years in the astrophysical environment, which makes it possible to simulate the morphology of the jet when it propagates over a distance of more than 600 au. Recently, similar experiments have started at the Central Research Institute of Mechanical Engineering (TsNIIMASh) using a Neodim picosecond laser facility with an energy of up to 10 J [152]. The focusing system provides a concentration of at least 40\% of the laser beam energy onto a spot with a diameter of 15 mm and a peak intensity of  $2 \times 10^{18}$ W cm$^{-2}$. It has been shown that the formation and development of a jet in a laboratory laser experiment is a complex physical phenomenon that includes a large number of various physical processes. A comparison of the results of a laboratory experiment and numerical calculations of magnetized jets in Ref. [152] confirms that ring structures can be formed whose characteristics depend on the magnitude of the magnetic field.

Since the beginning of the 2000s, experiments with an astrophysical aspect have also been carried out at so-called high electrical pulse power facilities; in other words, at facilities where a very large current flows in a short time (typical value of the order of 100 ns) through a specially prepared load. The main principles of operation of such installations are described in review [10].

The loads used in such experiments are very diverse. A classic example is wire arrays --- thin ($\sim $10 $\mu$m) wires of various materials (usually aluminum or tungsten) stretched along the perimeter parallel to the axis between the annular anode and cathode, through which current is passed. The flow of a high-amperage current --- from 1 MA at the COBRA (COrnell Beam Research Accelerator, Cornell University, USA) to 26 MA at the ZR (Z-Refurbished) facility (Sandia National Laboratories, USA) --- causes ablation of the wire substance. Under the action of Amp{\' e}re forces ${\bf j}\times{\bf B}$, it is driven to the axis of the system, resulting in plasma pinching. More details about the physics of processes leading to pinching of multiwire arrays can be found in review [153]. Plasma flows with a magnetic field can be used to model various astrophysical processes. Of particular interest from the point of view of modeling jets are conical wire arrays, in which plasma flows converging toward the axis and having a pronounced axial component of the momentum can be used to simulate the emergence and propagation of strongly radiating jets with a weak magnetic field.

The first experiment of this kind carried out at the MAGPIE facility at Imperial College (Great Britain) is described in [154]. The jet in this experiment had a characteristic longitudinal size of 1.5 cm with a radius of 1 mm; the time scale was several hundred nanoseconds. The jet collimation was shown to be affected by radiation cooling (by changing the material of the array wires), and the interaction of the jet with various obstacles was studied.

Notably, this experimental scheme was used to study the interaction of stellar outflows with gas clouds created by the injection of an argon jet into the region where the plasma flow propagates [155] and with the plasma wind [156, 157]. In the latter case, plastic foil was placed some distance from the axis at an angle to it. Extreme UV radiation from the pinch and individual wires led to foil ablation and the appearance of a plasma flow crossing the jet path, which made it possible to study the deflection (bending) of supersonic jets by the pressure of the crosswind flow. The experimental results were compared with those of numerical simulations, and the same computer code was used to simulate astrophysical systems with extended initial conditions [158]. Both experiments and astrophysical modeling show that the jet can be deflected by a significant angle (up to 30$^{\circ}$) without breaking. The interaction between the jet and the crosswind also leads to the disruption of the initial laminar flow caused by the onset of Kelvin-Helmholtz instability.

Another characteristic type of load is a thin foil stretched between two concentric electrodes. In this case, the density of the current flowing through the foil is maximal near the central electrode. Due to nonuniform heating of the foil over the radius, the most intense ablation of the foil substance occurs in the center near the cathode, which facilitates the formation of a narrowly directed plasma flow. A similar scheme was also used to study the interaction of the plasma flow with the environment using various pulsed gas injectors. However, even in the absence of additional injection, there is always a 'halo' surrounding the jet in the experiment due to the ablation of neighboring, less intensely evaporating regions of the foil.

Another modification of the load in the experiments under discussion is a combination of the two loads described above: instead of a foil, wires are used that are radially stretched between two concentric electrodes. In this case, the initial stage of the discharge is similar to the discharge with a cylindrical wire liner. However, the ablation rate is highest near the axis, and at some point the wire material near the central electrode completely passes into the plasma state. This leads to the formation of a magnetic bubble, i.e., a jet with a strong magnetic field{\footnote{Although this flow is often referred to as the `magnetic tower,' this term is
not quite correct, since the flow is supersonic.}}. As a result, it was possible to simulate many processes characteristic of real jet emissions, for example, the interaction of a supersonic radiation-cooled plasma jet with the environment [159].

Finally, a method based on the technology of a planar coaxial plasma gun with a longitudinal magnetic field was used at the California Institute of Technology (Caltech) (USA) [160, 161] and at the LabJet facility at the University of Washington [162]. The design of the coaxial gun at the Caltech facility is geometrically very simple: it consists of a disk surrounded by a coplanar ring and a magnetic field coil located immediately behind the disk. The disk and the ring, separated by a vacuum gap 6 mm wide, each have eight holes
through which the working gas is supplied by means of a pulse valve. After applying voltage to the electrodes, a breakdown occurs between the holes along the lines of the magnetic field. So-called spider legs are formed, which, under the influence of the magnetic field pressure, increase in height and shrink towards the center, forming a plasma column, which is what is called a jet. The jet is formed in this case in accordance with the `magnetic tower' model: under the effect of magnetic field pressure, the column propagates in space at a speed of 50--100 km s$^{-1}$ to distances of $\sim$ 10 cm with characteristic lifetimes of the order of 5 $\mu$s. Similar results were also obtained in experiments using the LabJet setup.

Thus, the parameters of experiments carried out at various facilities differ by a factor of two to three, but this difference is negligible in comparison with their difference of about 20 orders of magnitude from the parameters of real astrophysical jets.

As noted above, the most important element of the theory of both relativistic and nonrelativistic jets is their rotation. As for the laboratory experiment, the rotation of plasma ejections was realized for first time with the MAGPIE setup [163] using a wire array with wires twisted relative to the direction of the axis. With the COBRA setup, in a scheme with a radial foil and a longitudinal field of 10 kG, an azimuthal rotation of ions at the boundary of an aluminum jet with a velocity of 15--20 km s$^{-1}$ was detected [164]. Recently, the rotation of a plasma ejection was successfully simulated at the LabJet facility by applying a radial electric
field at the jet base [162]. It should be kept in mind that, according to Eqn (10), rotation and the radial electric field are
closely related.

Finally, several experiments [165, 166] were devoted to studying the interaction of the jet and the environment under conditions of very strong radiative cooling. Small-scale formations caused by the instability associated with cooling were found to rapidly develop in the region of the shock front, and the characteristic dimensions of such formations decrease with increasing radiative cooling [167]. These experiments constitute the first study of thermal instabilities in their nonlinear stage of development, which can have observable consequences, for example, for emission lines at high and low degrees of ionization.

As a result, laboratory experiments made it possible to reproduce such features inherent in astrophysical jets as the efficient conversion of magnetic energy into the kinetic energy of the flow; a high degree of jet collimation ($<$ 10$^{\circ}$) with sufficiently strong radiation cooling; enhanced collimation for secondary jets, since the magnetic fields trapped in the earlier ejected plasma contribute to the collimation of later jets; the generation of an X-ray pulse with each new plasma jet (this is due to the jets being compressed towards the axis by a magnetic field); and, finally, the development of current MHD instabilities leading to density (100\%) and elocity (3\%) variability.

In particular, laboratory experiments have shown that the kink mode instability (one of the main unstable modes, which has been actively studied for many years in connection with the problem of the stability of jets) does not destroy the MHD jet, despite the dominant toroidal field. On the contrary, the nonlinear saturation of unstable modes leads to the division of the jet into a chain of dense knots, which propagate with different velocities along the jet axis. In particular, the similarity of the episodic behavior of jets in laboratory experiments with that of the structures observed in XZ Tau and DG Tau jets is discussed in [168, 169]. Finally, the
stability and possible observational features of various configurations of magnetic jets have recently been studied by means of numerical simulations using the AstroBEAR code [170].

\section{Plasma focus as a promising area of laboratory research}

\subsection{Principle of operation} 

Recently, significant progress has also been made in the laboratory modeling of Herbig-Haro flows at 'plasma focus' type facilities. PF installations, being one version of systems that use the pinch effect (compression of a plasma with a current under the action of its own magnetic field generated by this current), are also known as sources of intense plasma flows that are widely used in various fields. In the mid-1950s, the phenomenon of plasma focusing near an anode was discovered in two devices, the types of which were apparently different, by \mbox{N V Filippov} at the National Research Center Kurchatov Institute in the study of a Z-pinch with conducting walls [171, 172] and J Mather (USA) in the study of coaxial plasma accelerators with a positive polarity of the central electrode [173]. In subsequent years, such systems gained popularity worldwide, primarily due to the high neutron yield achieved with them. For several decades, until the advent of high-power laser systems, the PF provided a record-setting neutron yield obtained as a result of a fusion reaction. Combined with the relative simplicity of the PF installation design, this led to the rapid development of PF research in many laboratories and universities around the world.

\begin{figure*}
\begin{center}
\includegraphics[width=520pt]{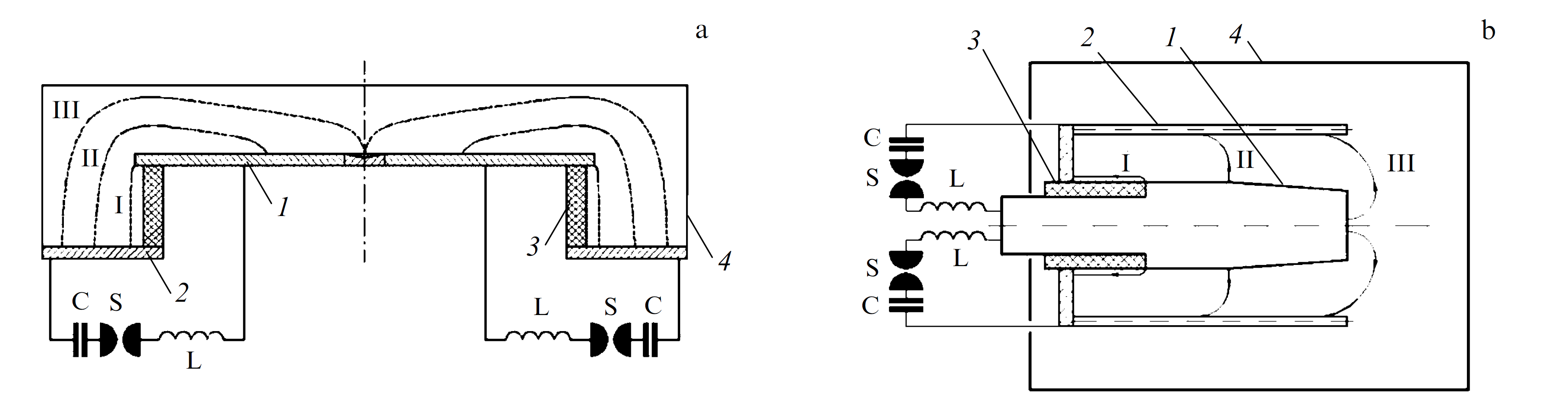}
\caption{\small Schematic diagrams of the Filippov (a) and Mather (b) plasma focus: 
$1$ -- anode, $2$ -- cathode, $3$ -- insulator, $4$ -- vacuum chamber, C -- power
supply, L -- external inductance, S -- discharger. I is the breakdown stage, 
II is the acceleration stage, III is the stage of PF formation.}  
\label{fig03}
\end{center}
\end{figure*}

Despite the obvious differences in the design of installations and in the direction of research, the resulting plasma parameters at installations with Filippov electrode geometry (Filippov plasma focus; FPF) and Mather electrode geometry (Mather plasma focus; MPF) turned out to be surprisingly similar. The schematic diagram of the FPF and MPF devices is extremely simple (Fig. 3): anode $1$, separated from cathode $2$ by tubular insulator $3$ in vacuum chamber $4$. In the FPF, the vacuum chamber often also serves as the cathode. The main distinguishing feature of these installation designs is the ratio of electrode sizes. In Filippov's modification (the so-called flat system), the transverse dimensions of the discharge device, determined by the anode diameter $d$, are much larger than the longitudinal dimension $l$, actually set by the insulator height: $l/d \ll 1$. In the Mather geometry, the coaxial accelerator length is usually greater than the anode diameter: $l/d > 1$. The vacuum chamber, after preliminary evacuation, is filled with gas (hydrogen, deuterium, or a heavier gas, depending on the problem under study) usually at a pressure of 0.1--10 Torr. Sometimes, pulsed gas injection is also used. The capacitor storage (C) is most often used as a power source.

PF-type systems are actually an inductive energy storage device, in which the electrical energy previously stored in a capacitor bank is transformed into the magnetic energy of the current at the stage when the current sheath is formed and moves along the electrodes. At the initial stage of the discharge (stage I), the current is skinned along insulator $3$ and a current-plasma sheath (CPS) is formed. Under the action of ponderomotive forces, the formed CPS is torn off from the insulator. In Filippov-type installations, the formation of a current sheath with a duration of about 1 ms ends with the sheath coming out to the anode edge and transforming into the radial compression process toward the axis (stage II).

In the Mather geometry, the CPS first accelerates along the electrodes and, only after reaching the anode end and the emergence of the longitudinal component of the discharge current, is compressed towards the axis under the action of its own magnetic field. This stage of CPS motion (stage II), developing at supersonic speed and accompanied by raking and ionization of neutral gas, is well described by the 'snowplow' model. During stage II, which is the longest, the discharge current increases, reaching a maximum in a few microseconds, depending on the anode size and circuit parameters, and the energy of the capacitor bank is transformed into the magnetic energy of the discharge current flowing through the sheath.

The main task at stage II is to achieve the maximum possible discharge current by the time the CPS converges on the axis (the so-called matching procedure). The CPS convergence is completed by plasma compression to a density of $\sim 10^{19}$ cm$^{-3}$ and its heating (stage III) (see Figs 3 and 4). The processes occurring at stage III with characteristic times of $\sim$ 100 ns are in many respects similar those in the classical Z-pinch. A distinctive feature is the noncylindrical shape of the CPS (an alternative name for the plasma focus is the noncylindrical Z-pinch).

At the dense pinch stage (actually, a 'dense plasma focus' (DPF), a current density of $> 10^{7}$ A cm$^{-2}$ is achieved, which leads to a buildup of strong current instabilities, the appearance of anomalous turbulent resistance, and a sharp interruption of the current. In fact, we are dealing with an efficient plasma circuit breaker, and the energy stored in the magnetic field of the pinch is put into a 'load': anomalous plasma heating, the generation of charged particle beams, intense neutron and X-ray radiation, and plasma flows occur. Due to the space-time dynamics of the current sheath, a rapid surge of the electric power occurs. The pinching process is accompanied by a decrease in the discharge current and, consequently, the emergence of a sharp peak (dip) at its derivative (Fig. 4c). The characteristic time of processes in the DPF stage ranges from several nanoseconds to several hundreds of nanoseconds. More details on the physics of the formation of a plasma focus and accompanying radiation can be found in reviews [174--177].

Plasma flows in PF facilities have been discovered at the early stages of research [172], but, due to the original focus on thermonuclear topics, their study was not given sufficient attention. Now, plasma flows are applied in practice increasingly widely. In particular, they are used to study the interaction of a plasma with a surface, to modify structural materials, and to create materials with new properties, including the production of nanocoatings and other applications (see, for example, [178--181]). PF installations have also been used to model astrophysical processes [182--186]. For example, in studies [185, 186], the plasma flows generated in a PF discharge were used to simulate the interaction of the solar wind with Earth's magnetosphere. It was studies [185, 186] that were subsequently used as the basis for the development of methods for modeling astrophysical jets of young stellar
objects.

\begin{figure*}
\begin{center}
\includegraphics[width=500pt]{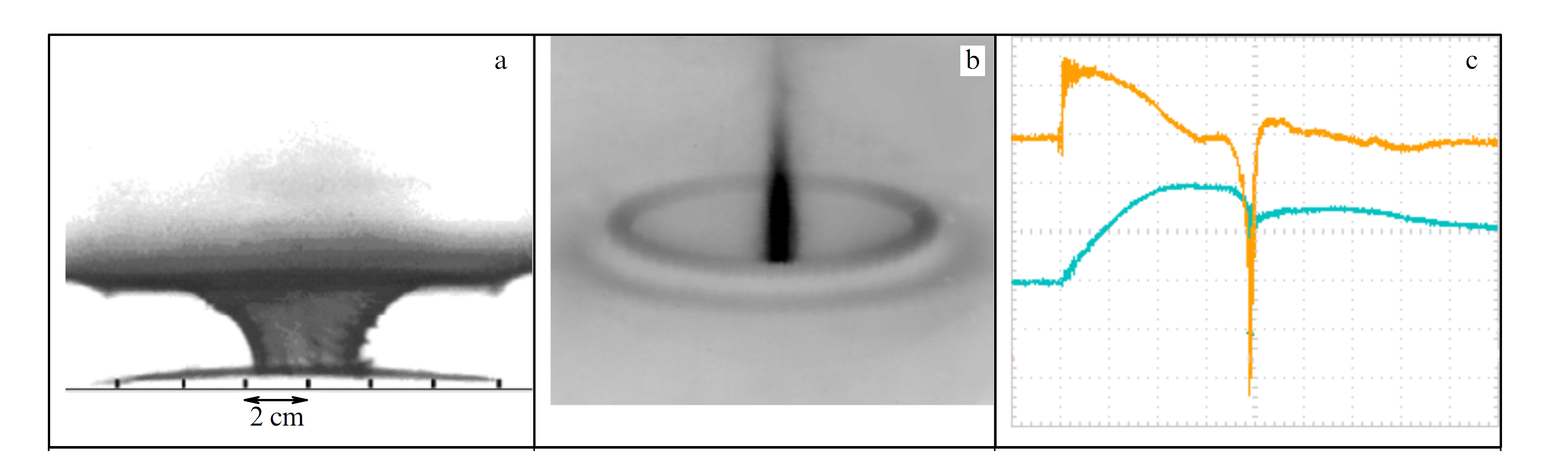}
\caption{\small (a) Image of a CPS converging toward the axis in the visible range. 
(b) Image of the pinch in the soft X-ray radiation range. (c) Oscillograms of
discharge current (lower curve) and its derivative (upper curve). Current amplitude
is 2 MA, and time scale is 5 ms per cell. PF-3 facility.}  
\label{fig04}
\end{center}
\end{figure*}

\subsection{Experimental setups}

In 2012-2014, experiments on the simulation of astrophysical jets started at the PF-3 facility at the National Research Center Kurchatov Institute [187]. The PF-3 setup is a plasma focus with a Filippov-type configuration of electrodes (Fig. 5). The PF-3 discharge system consists of a copper anode disk with a diameter of \mbox{92 cm} and a thickness of 2.5 cm and a cathode, which simultaneously is the lower flange of the vacuum chamber, with a diameter of 250 cm. At a diameter of 115 cm, 48 rods are mounted in the form of a squirrel cage that form a side wall of the return conductor. The top cover of the return conductor is a perforated duralumin disk. The interelectrode gap between the anode and the top cover is 10 cm. The electrodes are separated by a glass-ceramic
insulator 90 cm in diameter and 25 cm in height. The total capacity of the power source is 9.2 mF; the maximum charging voltage is 25 kV, and the maximum stored energy is 2.8 MJ. With the current geometry of the discharge chamber, which is determined by the insulator size, the facility operates in regimes optimized for obtaining a high degree of plasma compression at voltage $U_{0} =$  8--14 kV and current $I =$ 2--3 MA. The characteristic rise time of the current to the maximum value is on the order of 10 ms. After preliminary evacuation, the chamber is filled with a working gas (hydrogen, deuterium, helium, neon, argon, or a mixture of these gases, depending on the experimental problem) under a pressure of several Torr.

To study the propagation of plasma flows over considerable distances, the PF-3 facility was upgraded [21]. A three-section diagnostic flight chamber with the necessary set of diagnostic windows was designed and manufactured, which makes it possible to measure plasma flow parameters at distances of up to 100 cm (Fig. 6). The chamber consists of three sections 300 mm in length and 210 mm in diameter.
Each section of the flight chamber has a set of diagnostic windows evenly spaced along the perimeter in the central cross sections of each section. The center of the observation regions is located at distances of 35, 65, and 95 cm relative to the anode plane. The flight chamber is filled with the same working gas as the discharge chamber. Thus, taking into consideration the standard diagnostic windows of the setup discharge chamber, it is possible to study the parameters of the plasma flow as it propagates in the background gas at a distance of up to 100 cm from the anode plane, near which the flow generation region is located.

The experiments used a wide range of diagnostic tools, including

-- a Rogowski coil and magnetic coils for recording the total discharge current and its derivative;

-- a unit for diagnosing pinch emission parameters;

-- a unit for spectral-technique studies of plasma density and temperature;

-- a diagnostic unit for measuring plasma flow velocity, including high-speed photo recorders and optical collimators;

-- frame cameras based on EP-16 type electron-optical converters (EOCs) with electrostatic image focusing for studying plasma flow shape and dynamics;

-- SFER-6 and K-008 high-speed optical streak cameras;

-- a ballistic pendulum-calorimeter for measuring plasma flow momentum and energy;

-- magneto-optical probes for studying the distribution of magnetic fields captured by the flow.

\begin{figure*}
\begin{center}
\includegraphics[width=500pt]{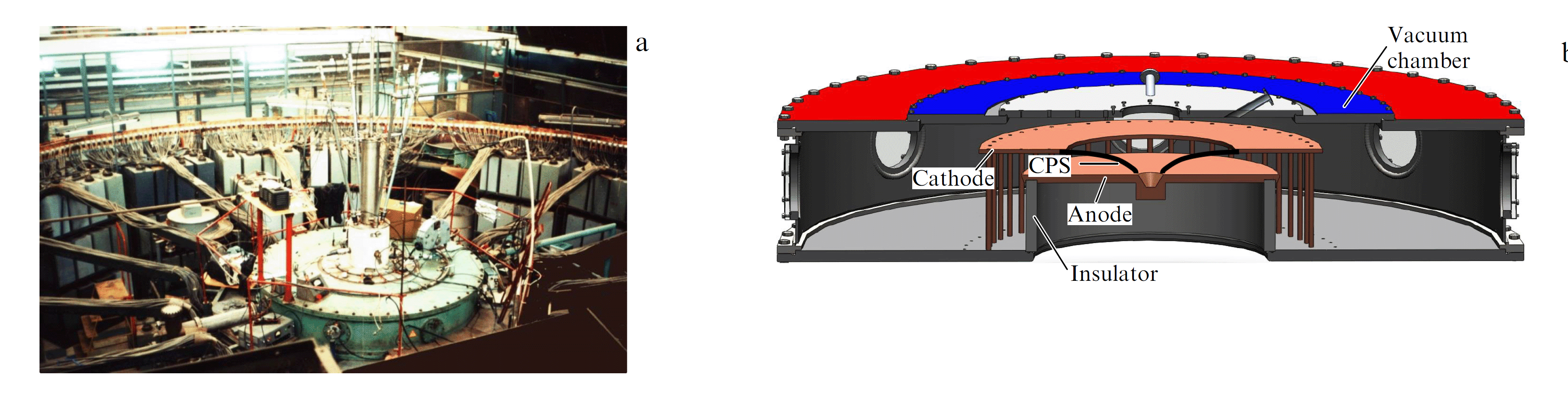}
\caption{\small PF-3 facility (a) and a cross section of the facility chamber (b).}  
\label{fig05}
\end{center}
\end{figure*}

\begin{figure*}
\begin{center}
\includegraphics[width=500pt]{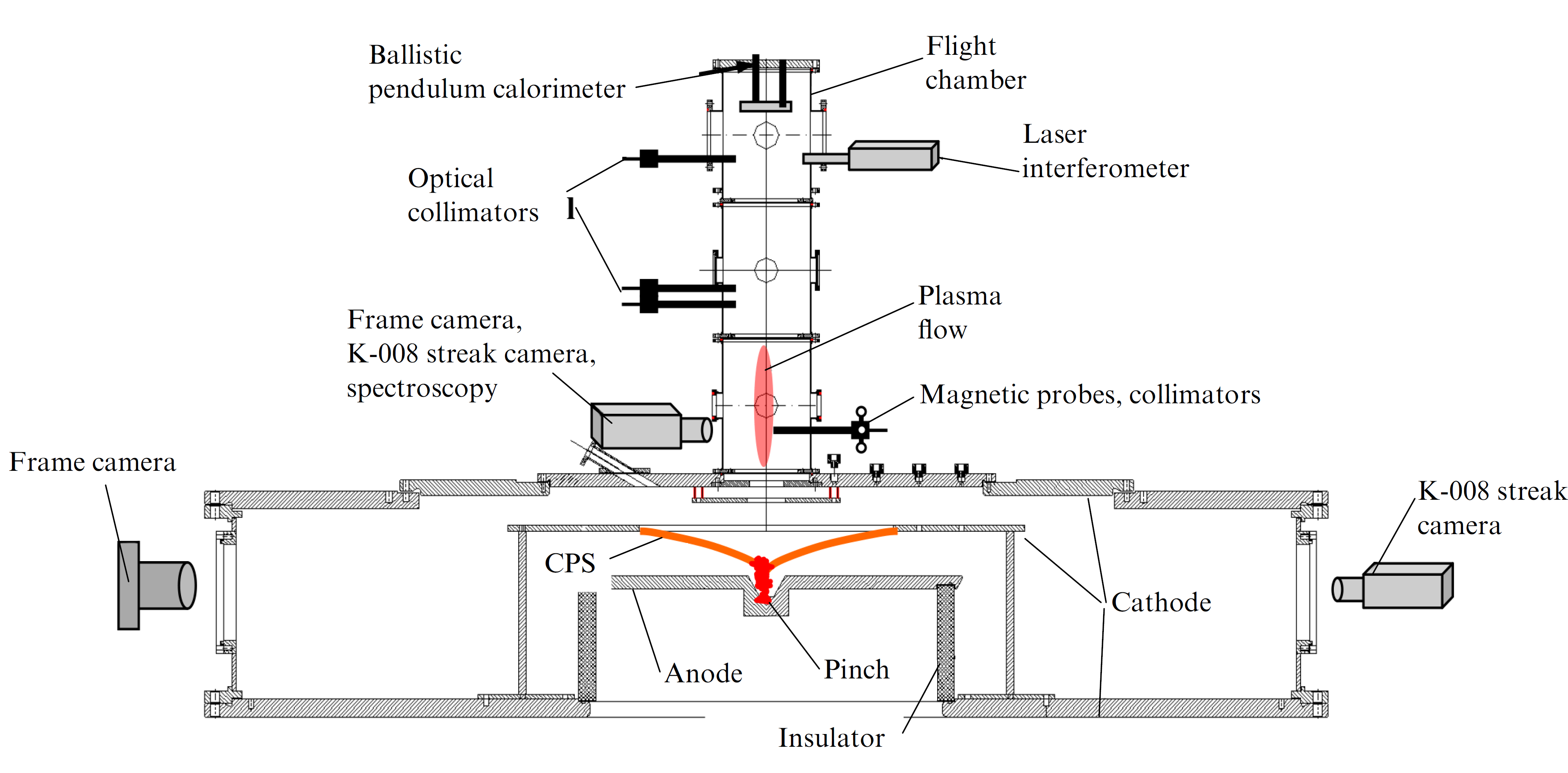}
\caption{\small Experimental setup at the PF-3 facility.}  
\label{fig06}
\end{center}
\end{figure*}

\begin{figure*}
\begin{center}
\includegraphics[width=500pt]{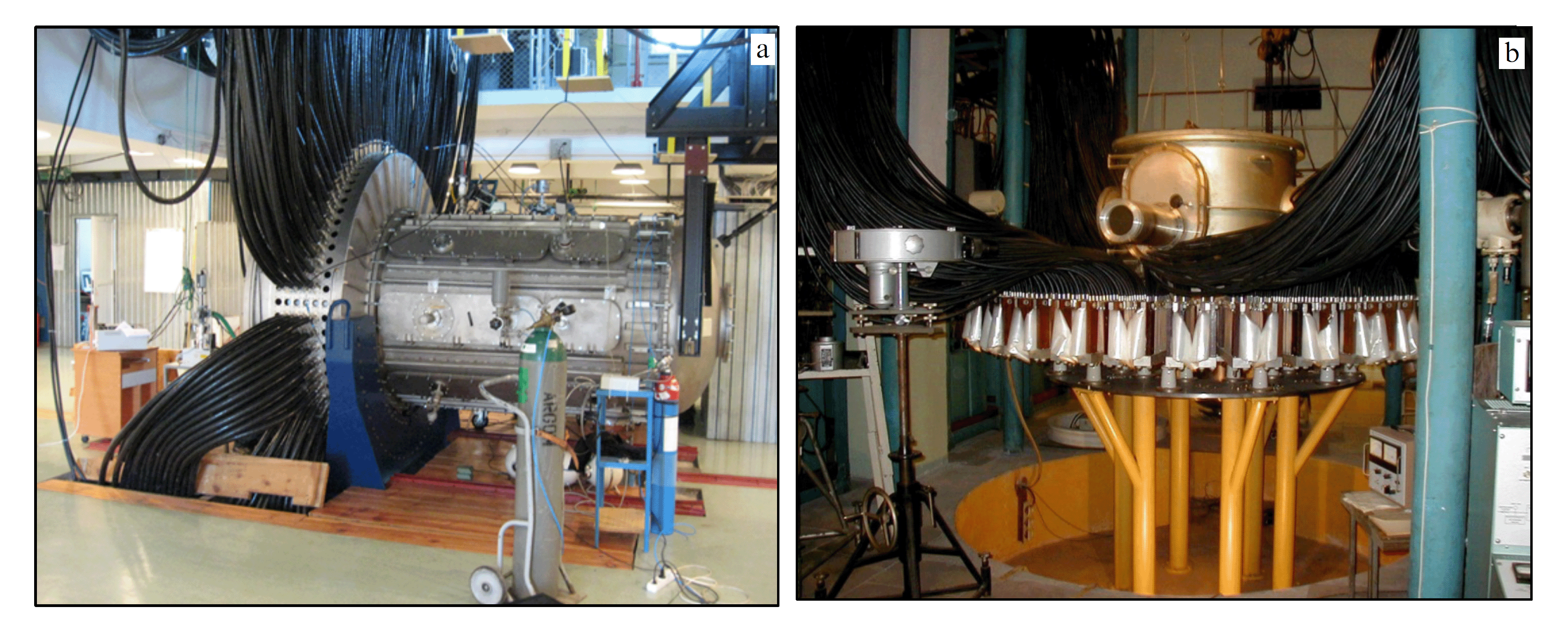}
\caption{\small PF-1000 (a) and KPF-4 (b) facilities.}  
\label{fig07}
\end{center}
\end{figure*}

Later, experiments were extended to the facilities PF-1000 (Institute of Plasma Physics and Laser Microfusion) (Warsaw, Poland) and KPF-4 Phoenix of the State Unitary Enterprise Sukhum Physical Technical Institute (SUE SPTI) (Sukhum, Abkhazia). Both setups are devices with a Mather-type discharge system. Another difference from the PF-3 facility is the option of pulsed injection of the working gas into the discharge chamber.

In the PF-1000 facility (Fig. 7a), the inner electrode (anode) is a copper cylinder 230 mm in diameter and 460 mm in length; the outer electrode (cathode) consists of twelve 80-mm thin-wall stainless steel pipes, evenly spaced around the circumference with a radius of 200 mm. The anode and cathode are separated by a ceramic insulator 230 mm in diameter and 85 mm long. The electrode system was located
in a fairly large vacuum chamber (140 cm in diameter and 250 cm long), which enabled studying the dynamics of the flow propagation over fairly long distances. At the end of the anode, there is a 50-mm axial hole with a quick-acting valve nozzle for a gas inlet. Gas (deuterium, helium, neon, and their mixtures) was injected into the chamber 1.5--2 ms before the discharge was initiated. Since the outflow of gas into the vacuum occurs during hydrodynamic times, in the process of the discharge, which lasts several tens of microseconds, the gas distribution can be considered constant. Thus, a profiled initial gas distribution was created with the density increased in the region of pinch formation and plasma flow generation. The flow propagates further in the region of low density,
which corresponds to the exit of the young star jet beyond the boundaries of the parent protostellar cloud [47]. More details about the PF-1000 device can be found, for example, in [188, 189].

The KPF-4 Phoenix device ($W_{\rm max} = 1,8$ MJ, $V_{\rm max} = 50$ kV) [190, 191], shown in Fig. 7b, is also a PF with the Mather geometry of the electrode system. However, KPF-4 used a different method of gas injection [192]. The KPF-4 electrode system consists of two coaxial electrodes: a copper cylindrical anode with a diameter of 18.2 cm and a length of the working part with an insulator of 32.6 cm, and an external cathode electrode of the 'squirrel cage' type, consisting of thirty-six 10-millimeter copper pins located at a diameter of 30 cm. The device insulator is made of aluminum oxide (alundum). The outer diameter of the insulator is the same as that of the anode electrode; its working length is up to 10 cm. The discharge system is in a vacuum chamber 109 cm in diameter and a distance between the anode end and the upper flange of 66 cm. This design made is possible to use it as a flight chamber to study the propagation of a plasma flow over considerable distances. The experiments were carried out at charging voltage \mbox{$U_{0} = 18$--20 kV,} power source energy $W = 230$--290 kJ, a discharge current of $\sim 1.5$ MA, and a current rise time to a maximum value of $\sim 7$ ms.

To inject gas into the interelectrode gap, storage is employed, from which, using 36 nozzles, gas is injected onto the anode end and into the space between the cathode and anode. The orientation of the nozzles can be changed, which makes it possible to configure the injection area. This injection method enables the creation of a required gas distribution profile. The profile features the optimal density in the insulator region and the interelectrode gap necessary for a high-quality breakdown and subsequent CPS dynamics, and a reduced density in the drift space (in the gap from the anode end to the upper flange of the vacuum chamber). This made it possible to carry out experiments with pulsed argon injection, which are equivalent to experiments with a stationary pressure of $\sim$ 1 Torr.

In the case of pulsed gas injection, the chamber is first evacuated to a pressure of the order of $10^{-2}$ Torr, and then, after the valve is opened, it is filled with gas at hydrodynamic velocities. The characteristic times of change in the gas distribution are in this case of the order of 1 ms. Given that the duration of the main stages of the plasma-focus discharge is about 10 $\mu$s, the gas distribution in the chamber during the discharge can be considered stationary, and, by adjusting the delay between the moment the valve opens and the moment the discharge is initiated, various gas distribution profiles can be created.

\subsection{Formation of plasma flows}

As noted in Section 3.1, the presence of plasma flows in PF discharges was established long ago. The well-known ability of plasma flows to propagate over considerable distances is actively used in various practical applications. However, the mechanism of formation of plasma flows in PF discharges is still a matter of discussion. First of all, we note that, despite the outward similarity of PF facilities, especially Mather ones, with coaxial plasma accelerators, the physics of the processes that lead to the generation of flows in PF facilities is closer to that of high-power electric systems. Due to the noncylindrical nature of the CPS at the stage of convergence to the axis, processes in the PF during pinch formation are similar to those occurring during compression of conical wire arrays [163]. However, there are significant differences here as well. In the PF, the return conductor is a movable plasma sheath,
and the pinch height is not set by the size of the wire liner, but grows as the current sheath expands under the action of magnetic pressure forces. As noted in [20], in the PF-3 facility, the sheath first converges in the anode cavity below the anode plane, and then the pinch starts growing upward along the axis with a phase velocity of $10^{8}$ cm s$^{-1}$ as a result of sequential convergence of different sheath segments ('zipper effect'), but no mass transfer occurs in the axial direction. This process is similar to the `magnetic tower' model implemented in experiments with radial wires, in which the pinch height increases with the formation of a 'magnetic bubble,' which is interpreted as plasma jet propagation [157]. However, it should be recalled that we are primarily interested in the mechanism of formation of supersonic plasma objects propagating over distances more than an order of magnitude greater than the dimensions of the pinch itself.

As noted, the plasma focus is a source of various types of radiation, including flows of accelerated ions and X-rays. They generate an ionization wave, which hinders identification of actual plasma flows. One of the first explanations for the formation of plasma flows was the 'cumulative model' in which, by analogy with shaped artillery charges, the flow is generated as early as the stage of CPS convergence due to its noncylindrical shape [193--197]. Such an analogy should undoubtedly be treated very carefully, since plasma is a
compressible liquid. Nevertheless, one of the first experimental events was the discovery of a quasi-spherically expanding bubble above the CPS (Fig. 8), observed using both laser diagnostics and optical frame recorders [180, 198, 199]. In [198], this bubble was treated as an ionization wave excited by accelerated ions.

In experiments with the PF-3 facility, the flow generation was studied using high-speed optical recorders: EOCs operating in the frame mode with a frame exposure of 12 ns and K-008 streak camera. The images obtained using the EOC (Fig. 9) clearly show compact radiating formations located above the current sheath. The formation of the flow, which occurs at the final stages of the development of the
discharge, can be associated with the development of MHD instabilities in the pinch. These formations are fairly well structured, and several separate formations are observed simultaneously.

An analysis of images obtained with a delay of 150 ns yielded an estimated velocity of plasma flows of \mbox{$> 10^{7}$ cm s$^{-1}$,} which is several times higher than the velocity with which the current sheath rises under the action of a pressure of the magnetic field of the discharge current of $\sim 3 \times 10^{6}$ cm s$^{-1}$ ('magnetic tower'). On slit scans obtained at a distance of 85 mm above the anode level, a glow is observed even before the CPS rises to the indicated height, which we associate with the emergence of a plasma flow in this region [196, 200]. Having escaped from the pinch, the plasma flows propagate over distances of more than 100 cm, maintaining their compact nature. It is these plasma formations that were the subject of our research.

\subsection{Justification of the applicability of plasma focus installations for laboratory simulation}

\begin{figure}
\begin{center}
\includegraphics[width=230pt]{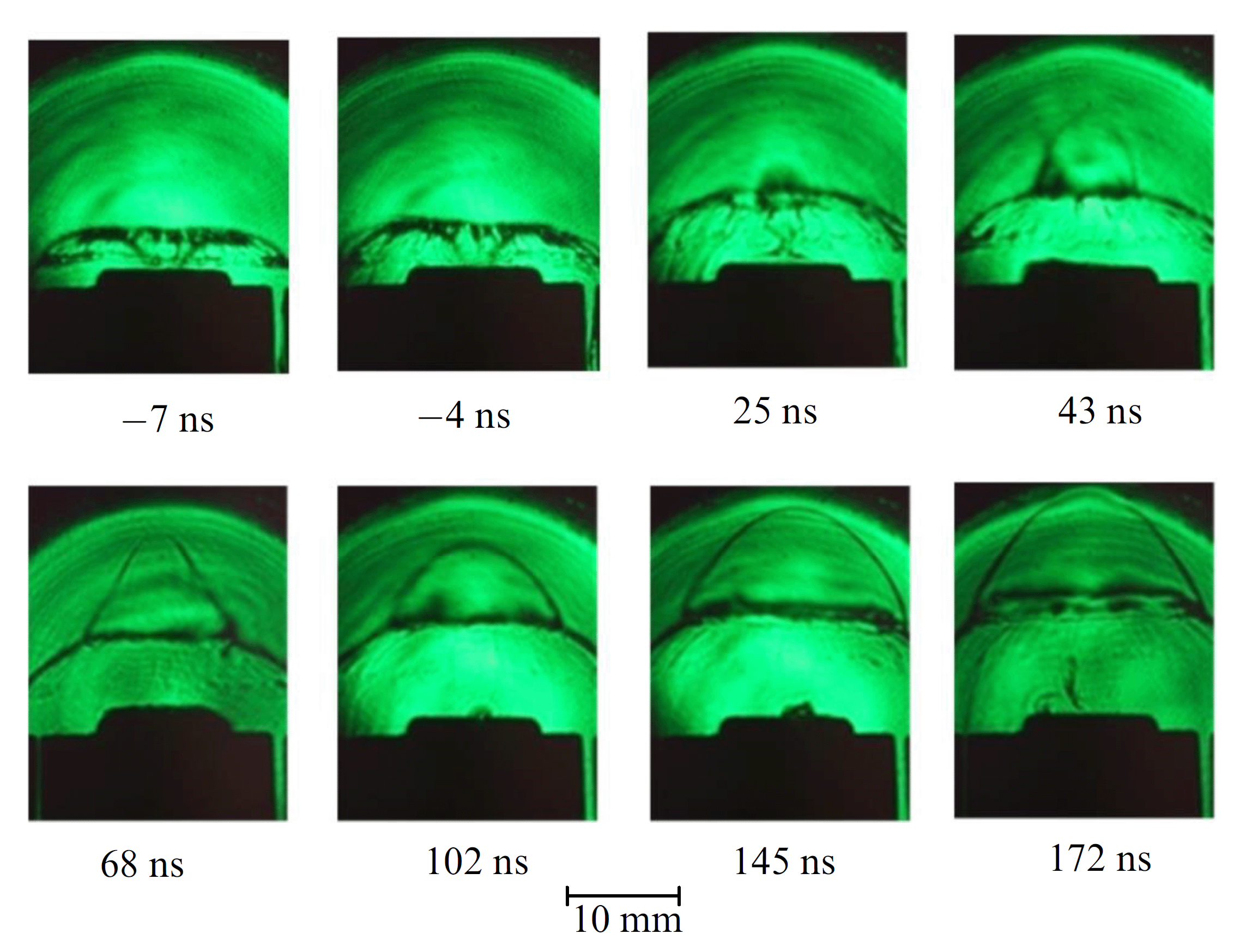}
\caption{\small Sequence of schlieren images representing the plasma dynamics
($t = 0$ corresponds to the time of the minimum value of the current
derivative, i.e., close to the pinch time). First two images correspond to
the pinch formation stage. Subsequent images correspond to the dynamics
after pinching [180].}  
\label{fig08}
\end{center}
\end{figure}

\begin{figure}
\begin{center}
\includegraphics[width=230pt]{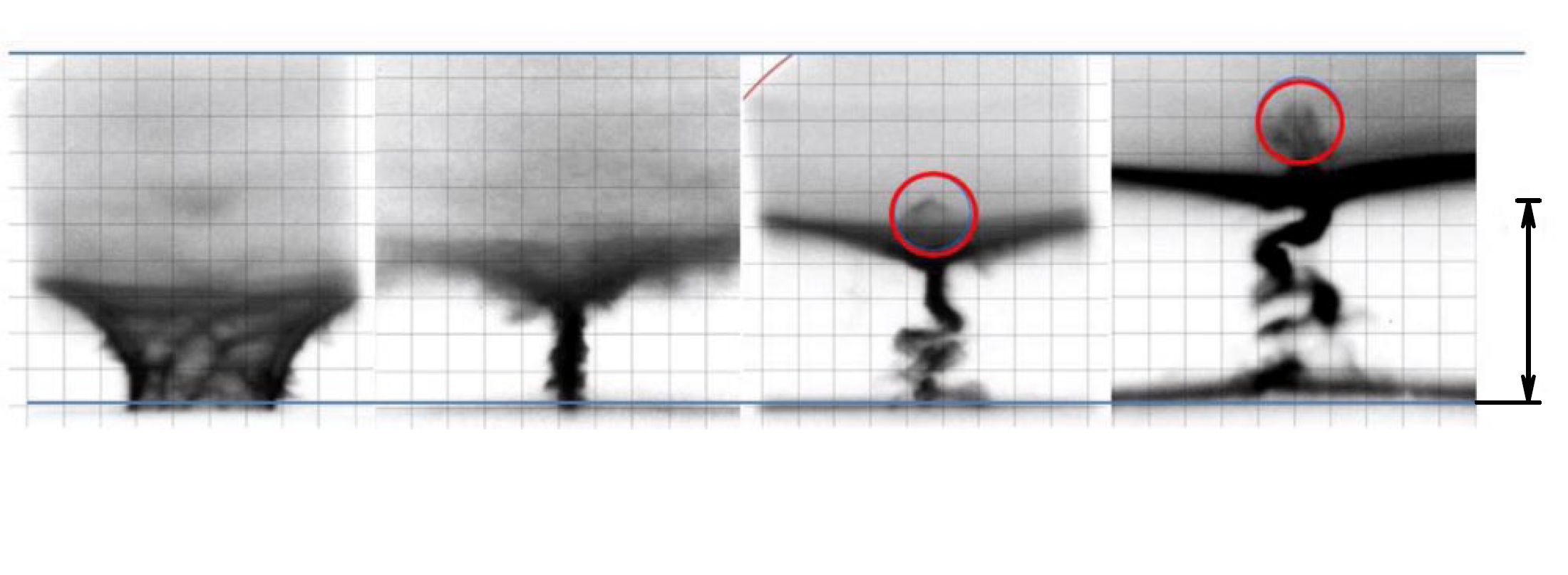}
\caption{\small Images of pinching and the formation of a plasma flow.
Exposure: 12 ns; scale: 1 cm. Time interval between two frames is 150 ns.
Formations are seen above the CPS, which are elongated along the facility
axis in the direction away from the anode (shown by red circles) [196].}  
\label{fig09}
\end{center}
\end{figure}

\begin{figure}
\begin{center}
\includegraphics[width=230pt]{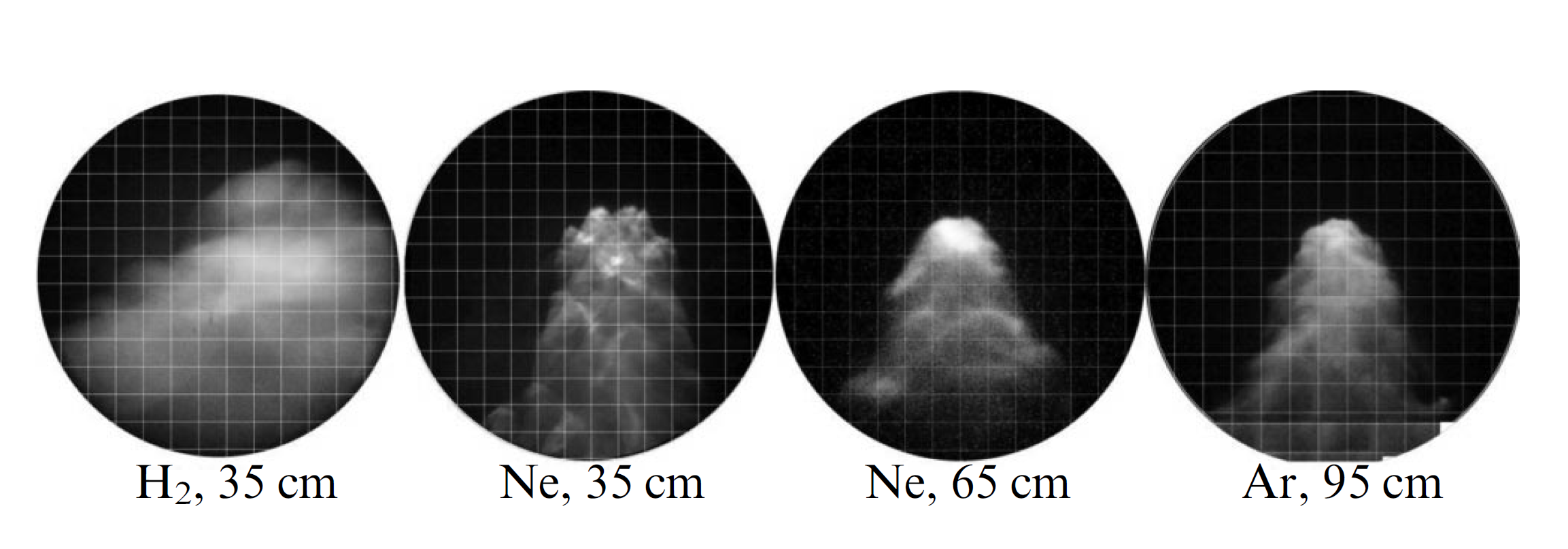}
\caption{\small Images of the plasma flow front at various distances from the
anode during a discharge in hydrogen, neon, and argon. Cell scale is 1 cm
[201].}  
\label{fig10}
\end{center}
\end{figure}

\begin{figure}
\begin{center}
\includegraphics[width=230pt]{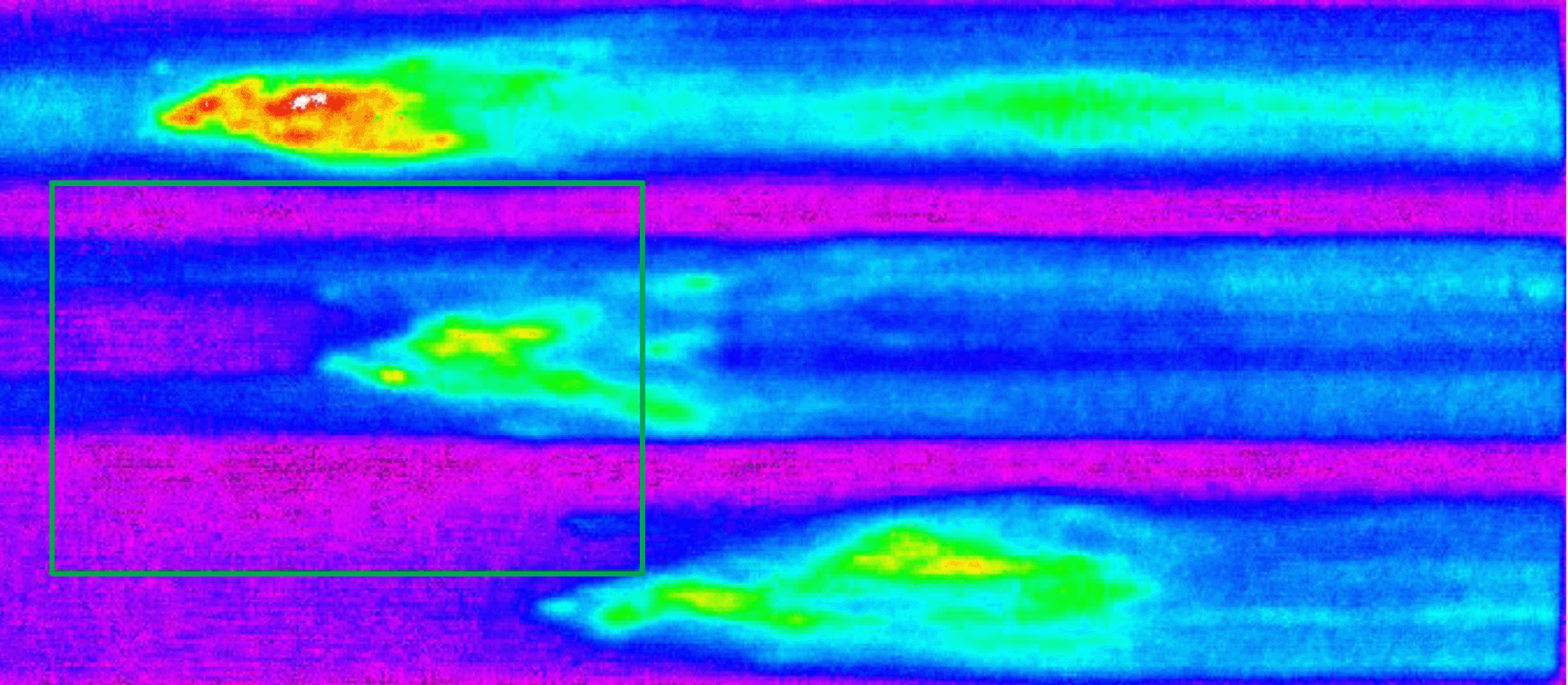}
\caption{\small Slit sweeps of a plasma flow. Three slits record radiation at a
height of 31, 35, and 39 cm. The scan duration is 6 $\mu$s.}  
\label{fig11}
\end{center}
\end{figure}

{\bf 3.4.1 Main parameters of flows in a PF.} It is clear that modeling the jets of young stars is only meaningful provided the parameters of the laboratory experiment correspond to real astrophysical conditions. Presented below are the results of studies of the key parameters of plasma flows generated in PF discharges, such as flow velocity, density and temperature (both of the flow itself and the environment), geometric dimensions, and the magnitudes of trapped magnetic fields. The data obtained made it possible to estimate the main dimensionless parameters, such as hydrodynamic and magnetic Reynolds numbers $Re$ and $Re_{\rm m}$, Peclet number $Pe$, and Mach number ${\cal M}$. A distinctive feature of the PF experiment is the ability to simulate the propagation of a plasma flow over considerable distances from the generation region. Therefore, we were primarily interested in the correspondence of the indicated values to the scaling requirements at large distances from the anode. As stated in Section 3.2, the PF-3 facility can measure plasma flow parameters in three regions: at distances of 35$\pm$5 cm, 65$\pm$5 cm, and 95$\pm$5 cm. The cross section at the level of 35 cm was chosen as the reference for evaluation of these options.

Figure 10 shows images of the plasma flow for various gases used. The effect of the gas type on the flow profile is discussed in Section 3.5.2. For now, we note that the transverse dimensions of the flow in strongly radiating gases are several centimeters, and the compactness of the flow persists until the moment it reaches the end of the flight chamber.

Figure 11 shows the time sweeps of the flow at heights of 31, 35, and 39 cm above the anode level obtained by the K-008 streak camera. It can be seen that the plasma bunch is also quite compact in the longitudinal direction. Having determined the speed of the flow from the delay in the appearance of the image at different heights, one can estimate the spatial dimensions of the compaction. In most discharges, the longitudinal size is 5--10 cm.

The conclusion about the compactness of the plasma flow is also confirmed by the results of the detection of plasma glow using optical collimators. Collimators detect radiation from a fairly small solid angle along the chamber diameter (the size of the registration area is $\sim$ 0.5 cm on the chamber axis), which is transmitted via a light guide to the input of photomultiplier tubes (PMTs). The arrival of the flow into the region of observation is accompanied by a sharp increase in the signal front (Fig. 12). The use of two collimators separated by a small distance ($\sim$ 1.5 cm) makes it possible to determine the flow velocity with high accuracy. As can be seen, an intensely emitting object is quite compact: the pulse duration at half-height is $\sim$ 1 ${\mu}$s, which at a speed of $\sim 5 \times 10^{6}$ cm s$^{-1}$ corresponds to an object length of $\sim$ 5 cm.

Flow velocity is one of the key parameters of laboratory simulations. The jets from young stellar objects are known to be nonrelativistic; their speed is $\sim 10^{7}$ cm s$^{-1}$. In a project, flow velocities were measured at three facilities, PF-3 (National Research Center Kurchatov Institute), KPF-4 (SPTI), and PF-1000 (Warsaw), using various gases and a wide set of complementary diagnostic tools: streak cameras, frame cameras with time delay between frames, spaced magnetic probes, and optical collimators of various designs.

Measurements of the plasma flow speed using various methods yielded similar results. However, the main method for studying the evolution of the flow velocity during its propagation in the background gas was the double collimator method, which is due primarily to its simplicity. This method makes it possible not only to determine the flow velocity at a given observation point but also to estimate the flow deceleration decrement as it propagates in the background gas. As an example, Figure 13 shows the dependence of the
flow velocity on the distance from the anode when neon is used as the working gas. A simple exponential approximation of this dependence yields the initial velocity of $ 10^{7}$ cm s$^{-1}$, which is in good agreement with the initial velocity determined using high-speed optical recorders. Similar results were also obtained at the KPF-4 [203], PF-1000 [204], and Flora [205] facilities. An important result was that the flow velocity weakly depends on the type of working gas and ranges from a few times $10^{7}$ cm s$^{-1}$ to $10^{6}$ cm s$^{-1}$ at different stages of motion, corresponding well to the velocities of jets from young stellar objects.

To determine the main dimensionless parameters, the parameters of both the plasma of the jet itself and the ambient gas are required. The plasma density and temperature were measured using optical spectroscopy. To this end, a diagnostic unit was used at the PF-3 setup, which included an STE-1 spectrograph with a wide spectral view in combination with a chronographic electron-optical camera [206]. The plasma concentration was determined in an experiment with helium as the working gas using the Stark broadening of lines,
while the electron temperature was determined from the ratio of the intensities of two lines pertaining to a neutral atom (HeI) and an ion (HeII) of helium [207]. The ambient plasma density did not really change from the beginning of the line glow at the moment when the X-ray pulse was emitted from the plasma focus, which ionized the background gas, up to the moment the jet arrived; it was $n_{\rm e} = 2 \times 10^{16}$ cm$^{-3}$. The jet plasma was characterized by great inhomogeneity, its maximum density was $2 \times 10^{17}$ cm$^{-3}$, and the minimum density was $10^{15}$ cm$^{-3}$. The ionization temperature of the jet plasma was 4--8 eV [208].

In the same geometry, experiments were carried out using a mixture of two gases: neon and helium [209]. Neon, which prevailed in the mixture, set the jet dynamics, and helium was a diagnostic additive (10--20\%) to determine plasma concentration and temperature. In most experiments, the ambient plasma density was \mbox{$n_{\rm e} = 4 \times 10^{16}$ cm$^{-3}$.} The maximum density in the jet plasma, determined from the Stark linewidth of helium HeI 587.6 nm when adding helium to neon as well as in experiments with pure helium, was \mbox{(2--4) $\times 10^{17}$ cm$^{-3}$} at the electron temperature of the jet plasma \mbox{$T_{\rm e} \approx$ 2--3 eV.} At a distance of 65 cm from the focus, the concentration of the ambient neon plasma was outside the detection limits of the spectral equipment: \mbox{$n_{\rm i} \ll 10^{16}$ cm$^{-3}$.} The maximum density of electrons in the jet was \mbox{(0.5--2) $\times 10^{17}$ cm$^{-3}$} at an electron temperature of the jet plasma of $T_{\rm e} \approx$ 1 eV.

Similar results were also obtained at the PF-1000U facility [204, 210]. The time sweeps of the spectra were recorded with a Mechelle$^{\circledR}$900 spectrometer in the wavelength range from 300 to 1100 nm with an exposure time adjustable from 100 ns to 50 ms. The electron density estimated based on the line broadening was \mbox{(0.4--7) $\times 10^{17}$ cm$^{-3}$,} depending on the initial gas distribution, the detection area, and the time interval of registration relative to the moment of jet generation. The ambient plasma density decreased from $\sim 10^{16}$ cm$^{-3}$ at a distance of 27 cm to $1.5 \times 10^{15}$ cm$^{-3}$ at a distance of 57 cm. Based on the ratio of the intensities of the HeII (468.6 nm) and HeI (587.6 nm) lines, the flow temperature was estimated to be 3--7 eV. As can be seen, the results obtained with two devices are also easily repeatable in this case, despite significant differences in the geometry of the discharge system.

The total energy and momentum carried by the plasma flow were determined using a ballistic pendulum, which could simultaneously be used in the calorimeter mode [211]. The plasma flow energy was estimated in two ways: by the heat absorbed in the calorimeter mode and by calculations from the measured momentum of the plasma flow and its velocity. There is a difference between the measurement results,
because, on the one hand, the calorimeter is not an absolutely black body (lower estimate), and, on the other hand, the ballistic pendulum gives an overestimated value, since the jet impact is not absolutely inelastic. For example, for experiments with argon, the maximum energy determined from the absorbed heat in the second section (65 cm from the anode plane) was 460 J, while the kinetic energy calculated from the momentum and velocity of the plasma flow was 710 J. However, these results make it possible to determine
the flow energy with fairly good accuracy. It has been shown that in experiments with argon the energy density of the incident flow at a distance of \mbox{65 cm} is as large as $\geqslant$ 10 J cm$^{-2}$.

\begin{figure}
\begin{center}
\includegraphics[width=230pt]{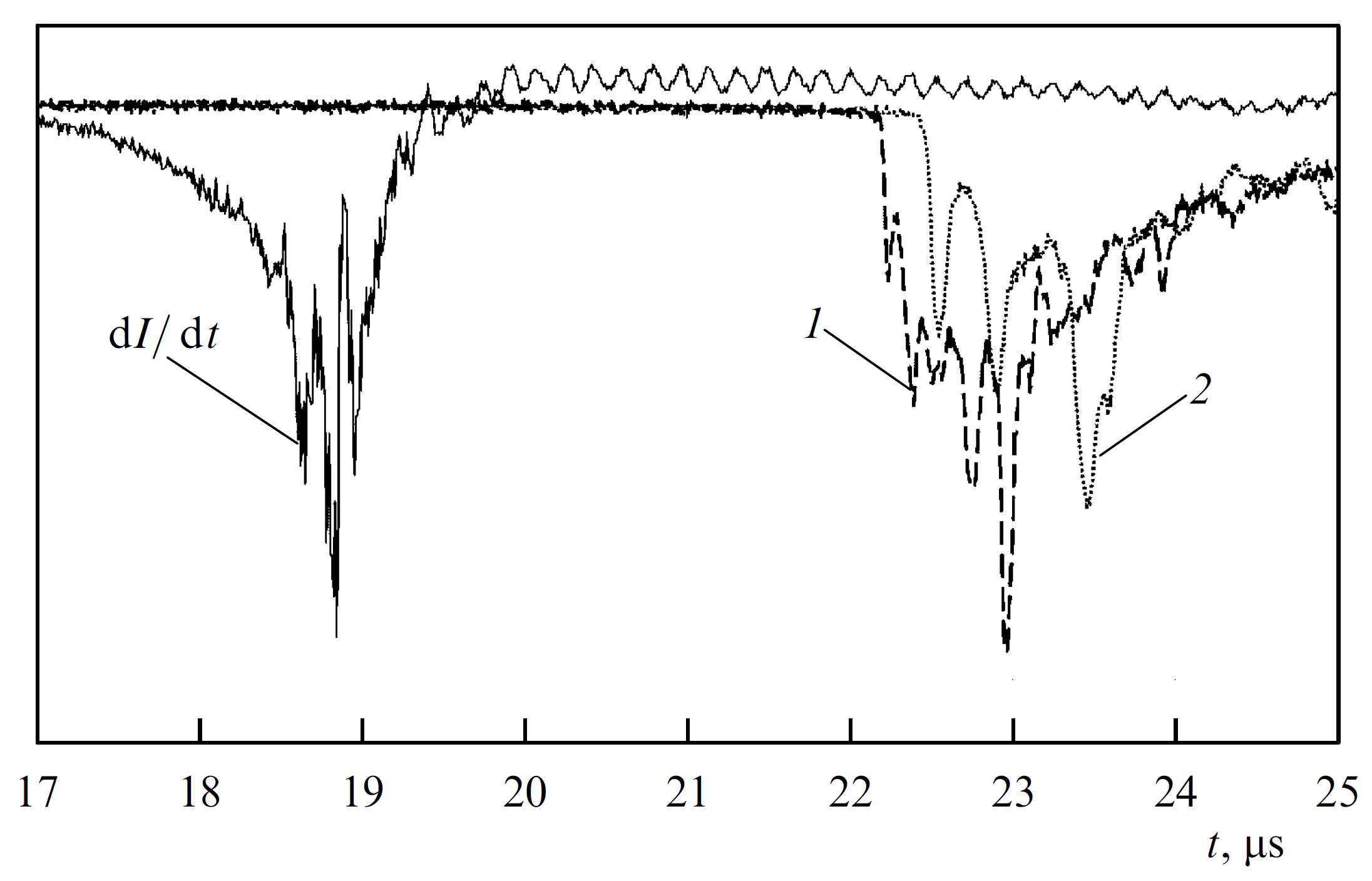}
\caption{\small Oscillograms of discharge current derivative  ${\rm d}I/{\rm d}t$ 
and signals from the double optical collimator (1, 2) at the PF-3 facility. 
Distance between collimator axes is 16 mm. Discharge occurs in neon. Distance 
to the anode plane is 35 cm [21].}  
\label{fig12}
\end{center}
\end{figure}

\begin{figure}
\begin{center}
\includegraphics[width=230pt]{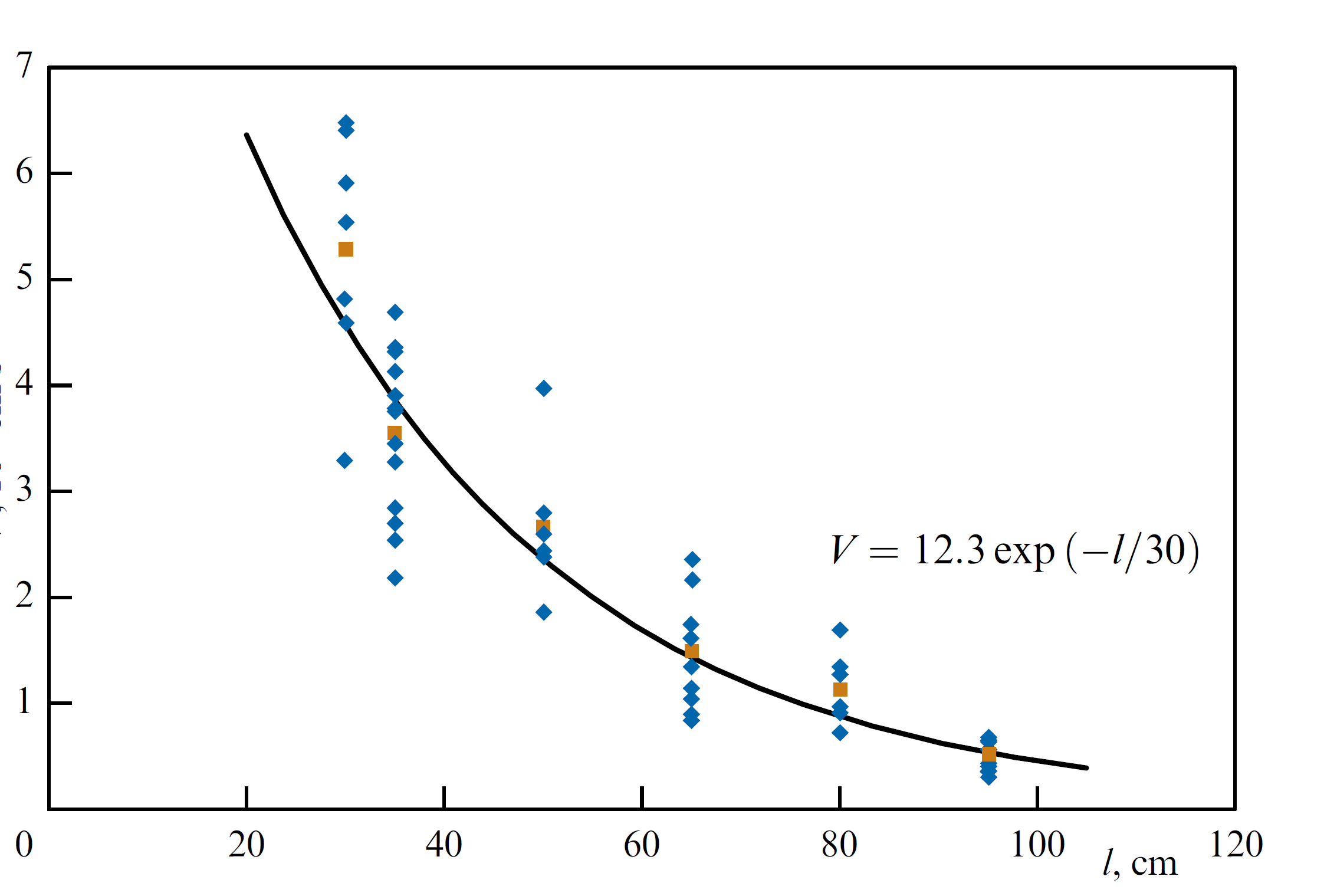}
\caption{\small Dependence of flow velocity on distance $l$ from the anode
obtained using experimental series at the PF-3 facility for invariable initial
conditions of the discharge in neon: diamonds show experimental results;
squares show average values, and the solid curve is the approximation of
experimental points [202].}  
\label{fig13}
\end{center}
\end{figure}

\begin{figure*}
\begin{center}
\includegraphics[width=480pt]{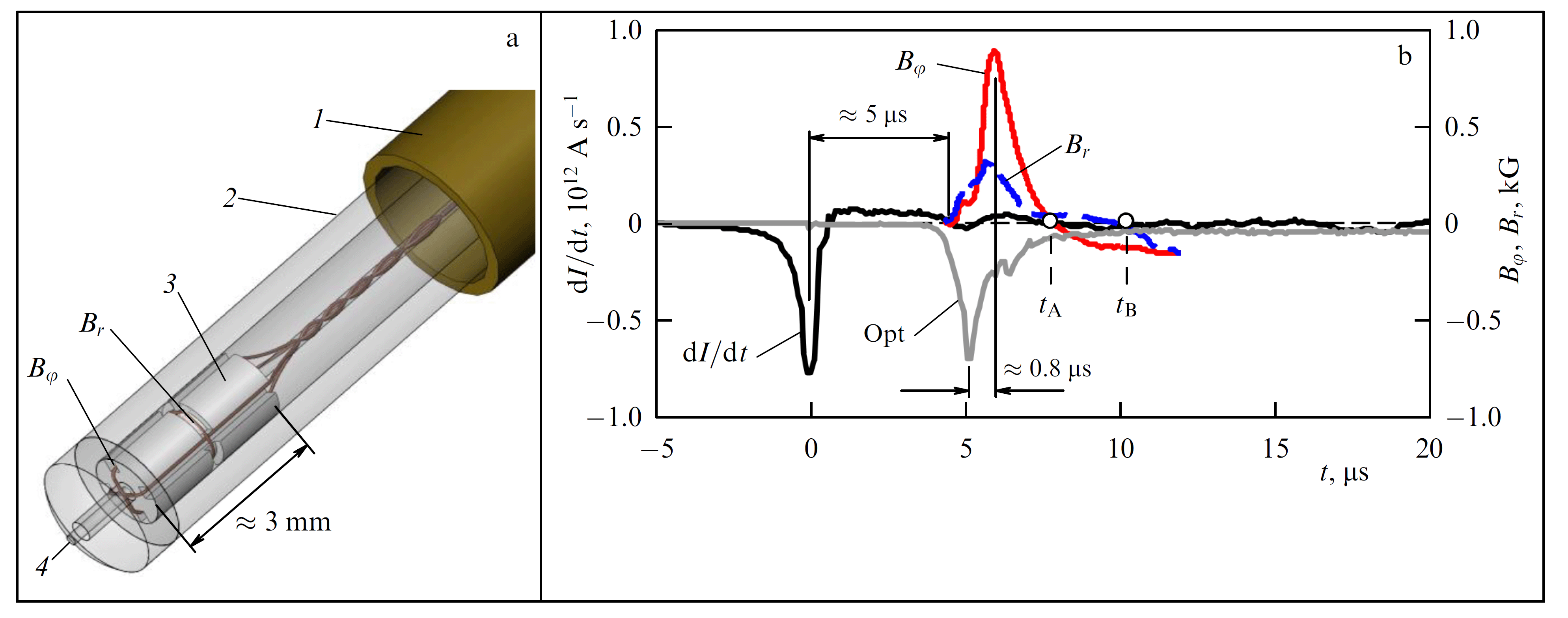}
\caption{\small (a) Sensitive element for measuring two components of the magnetic 
field ($B_{r}$,  $B_{\varphi}$) with an additional channel for detecting optical 
radiation of plasma: $1$ -- metal casing of the probe; $2$ -- glass tube 3 mm
in diameter; $3$ -- frame setting the direction of loops; $4$ -- optical fiber 
0.38 mm in diameter. $B_{r}$ and $B_{\varphi}$ are probe loops that measure 
corresponding components of the magnetic field. (b) Results of the measurements 
of the magnetic field in a plasma flow at the PF-3 facility (gas Ne, $P_{0} =$ 1.5 Torr, 
$U_{0} =$ 9 kV, $W_{0} =$ 373 kJ); ${\rm d}I/{\rm d}t$ is the derivative of 
the total current, $B_{r}$ and $B_{\varphi}$ are the induction of the radial 
and azimuthal components of the magnetic field measured at radius $r = 1.31$ cm 
at a height of 35 cm above the anode level; Opt is a signal from the
optical channel of the probe.
}
\label{fig14}
\end{center}
\end{figure*}

Finally, another important parameter necessary to estimate dimensionless scaling parameters is the magnitude of the magnetic field. One of the advantages of the PF-based experimental scheme is the fairly large size of the investigated flow (several centimeters), which makes it possible to apply the magnetic probe technique.

In experiments, specially designed probes of various designs were used, which are described in detail below in presenting specific results. Although the magnetic probe method is a contact technique and, in principle, can lead to perturbation of the parameters of the object under study, today it is the most accessible tool that allows obtaining extensive information about the magnitude, distribution, and dynamics of magnetic fields. Moreover, as was shown, the perturbation introduced by magnetic probes is within
acceptable limits, and, subject to certain requirements for the probe design, can be minimized. The parameters of the probes used in our experiments are described in detail in [212].

The first studies [21, 213] have already shown that the plasma flow propagates with its own trapped magnetic field. A complex spatial configuration of the magnetic field inside the plasma flow, which changes with time, is found. Figure 14 shows the design of a probe for measuring two components of the magnetic field ($B_{r}$, $B_{\varphi}$) at a given point in space with an additional channel for detecting the optical radiation of a plasma jet on an SNFT-3 PMT (Fig. 14a) and the results of measuring magnetic fields using this multicomponent probe (Fig. 14b).

The data displayed in Fig. 14b enable making a number of important conclusions. First of all, attention is drawn to the finite duration of the magnetic field signal, which coincides with the duration of the signal from the optical channel of the probe, implying that the magnetic field is concentrated inside the radiating plasma bunch, which has finite dimensions. Based on the data presented in Fig. 13 and assuming that the flow velocity at a distance of 35 cm from the anode is $\sim 5 \times 10^{6}$ cm s$^{-1}$, for a signal duration of 1-2 $\mu$s, we obtain the longitudinal size of the region with a magnetic field of $\sim 5$--10 cm. The optical signal appears ahead of the magnetic field signal, which, apparently, is associated with the glow of the SW front. Finally, the determined magnitude of the magnetic field was on the order of several kG, which made it possible to estimate the main dimensionless scaling parameters.

{\bf 3.4.2 Correspondence to astrophysical sources.} The most important requirement for successful modeling is, of course, the correspondence between laboratory experiment and real astrophysical sources. As noted above, this becomes possible due to the absence of intrinsic scales in the equations of ideal magnetic hydrodynamics, which enables studying identical processes that feature completely different time and space scales.

It should be kept in mind that, historically, the following three scaling parameters{\footnote{Study [141] is devoted to the analysis of the possibility of laboratory simulation of the remnants of supernova explosions.}} were used in [141]:
\begin{equation}
a_{\rm L} = \frac{L_{\rm lab}}{L_{\rm ast}}, \qquad 
a_{\rho} = \frac{\rho_{\rm lab}}{\rho_{\rm ast}}, \qquad 
a_{\rm P} = \frac{P_{\rm lab}}{P_{\rm ast}},
\label{abc}
\end{equation}
which correspond to length $L$, density $\rho$, and pressure $P$. As a result, we obtain, according to Table 1,
\begin{equation}
a_{\rm L} \sim 10^{-14}-10^{-15}, 
a_{\rho} \sim 10^{14}-10^{16}, 
a_{\rm P} \sim  10^{14}-10^{16}.   
\end{equation}
It is clear that the introduction of three scaling parameters (corresponding to the centimeter--gram--second triad) is only a choice of the corresponding rulers. Processes can be considered similar only provided all other quantities measured by such rulers turn out to be of the same order. The remarkable fact is that this feature does hold to fairly good accuracy [214].

\begin{table}
\caption{Main dimensionless parameters.}
{\footnotesize
\vspace{0.2cm}
{\begin{tabular}{|c|c|c|c|} 
 \hline
 &  &  & PF-3 \\
Parameter & YSO & Note & 35 cm from \\
& &  & 
the anode \\
\hline
Mach number   & $10-50$ & $> 1$, super- & $> 10$  \\
$ {\cal M} = V/c_{\rm s}$ &  &  sonic
flow &  (Ne, Ar) \\
\hline
Alfv{\' e}n Mach
  &  &  &   \\
number   & $ \sim 10-40$ & --- & $> 1$  \\
$ {\cal M}_{\rm A} = V/V_{\rm A}$ &  &  &  \\
\hline
  & $\ll 1$ near &  &   \\
$\beta$  &  the source & --- & $\sim 0.1$  \\
($4 \pi P/B^2$) & $\sim 1$ (100 au)  &  &  (Ne, Ar) \\
\hline
Reynolds &  & $\gg 1$, visco- &  \\
 number & $10^{6}$--$10^{8}$ & sity is of no & $10^{5}-10^{6}$ \\
${\cal R}e = LV/\eta$ &  & importance &  \\
%&  &  & \\
\hline
Magnetic &  & $> 1$,  & \\
Reynolds number & $\sim 10^{15}$ &  frozen-in & $\sim 100$ \\
${\cal R}e_{\rm m} = LV/\eta_{\rm m}$ &  & field &  \\
\hline
Peclet  &  &  $ > 1$, convec-  &   \\
number & $ \sim 10^{6} $ & tive heat  & $> 10^{4}$  \\
${\cal P}e = LV/\chi$  &  &  transfer &  \\
\hline
Lundquist  & & $> 1$, field  &  \\
 number & $\sim 10^{13}$ &diffusion & $\sim 10$ \\
${\cal L}u = LV_{\rm A}/\eta_{\rm m}$ &  & is small &  \\
\hline
Density   &   &  &  \\ 
contrast   & $>1$  &  --- & $1-10$\\ 
($n_{\rm jet}/n_{\rm amb}$) &  & & \\ 
 \hline
 \end{tabular} 
\label{ta1}
}
}
\end{table}

Indeed, we note first of all that the ratio $a_{P}/a_{\rho} \propto P/\rho$, which scales the speed squared, is close to unity. Therefore, to satisfy the similarity condition, the Mach numbers
\begin{equation}
{\cal M} = \frac{V}{(P/\rho)^{1/2}}
\label{Euler}   
\end{equation}
for astrophysical and laboratory plasma should be of the same order. As shown in Table 2, this requirement is indeed satisfied, since the flow velocities in the laboratory experiment turn out to be close to the flow velocity in astrophysical jets. Consequently, the corresponding temperatures turn out to be close with almost any device.

Furthermore, magnetohydrodynamics requires for complete similarity the magnetic field to scale as $a_{P}^{1/2}$. In other words, the plasma parameters $\beta = 4 \pi P/B^2$ of laboratory and astrophysical plasmas should also be of the same order. Consequently, the values of the Alfv{\' e}n Mach number ${\cal M}_{\rm A} = V/V_{\rm A} = \beta^{1/2}{\cal M}$ should be close. Table 2 shows that for the PF-3 installation such a similarity also holds in general.

The absence of intrinsic scales noted above only takes place in an ideal MHD. The smallness of the dissipative processes is due then to the large values of the dimensionless parameters associated with the corresponding transport processes. The list of these parameters includes: Reynolds number ${\cal R}e = LV/\nu$  (condition ${\cal R}e \gg 1$ implies that processes associated with viscous energy release can be neglected, $\nu$ is kinematic viscosity), magnetic Reynolds number ${\cal R}e_{\rm m} = LV/\nu_{\rm m}$ (condition ${\cal R}e_{\rm m} \gg 1$ implies that magnetic diffusion can be neglected, $\nu_{\rm m}$ is the magnetic viscosity), and the Peclet numbers${\cal P}e = LV/\chi$ (disregarding thermal conductivity, $\chi$  is thermal diffusivity), and Lundquist numbers ${\cal L}u = LV_{\rm A}/\nu_{\rm m}$ (disregarding magnetic diffusion for static configurations). Here, by implication, as the characteristic size $L$ and the characteristic velocity $V$, their smallest values should be used. Table 2 compares the main dimensionless parameters noted above. It can be seen that for almost all parameters the necessary conditions are met with a good
margin.

It is important to note that the transport coefficients{\footnote{The displayed formulas refer to hydrogen plasma ($Z = 1$).}}
\begin{eqnarray}
\nu & = & \frac{T^{5/2}}{4 \pi \alpha^2 e^4 n  A^{1/2} m_{\rm p}^{1/2}\Lambda}, 
\label{defnu} \\
\nu_{\rm m} & = & \frac{e^2 c^2 A^{1/2} m_{\rm e}^{1/2}\Lambda}{T^{3/2}}, 
\label{defnum} \\
\chi & = & \frac{T^{5/2}}{4 \pi \alpha e^4 n A^{1/2} m_{\rm e}^{1/2}\Lambda}
\label{defchi} 
\end{eqnarray}
are easily scaled not only in density $n = \rho/m_{\rm i}$ and temperature $T = P/n$ but also in the quantities $a_{\rho}$ and $a_{P}$, and, in addition, they depend on ionization degree $\alpha$. In turn, the ratios of $n$ and $T$ depend on atomic weight $A$:
\begin{eqnarray}
\frac{n_{\rm lab}}{n_{\rm ast}} & = & a_{\rho} \frac{A_{\rm ast}}{A_{\rm lab}}, \\
\frac{T_{\rm lab}}{T_{\rm ast}} & = & \frac{a_{\rm P}}{a_{\rho}}\,\frac{A_{\rm lab}}{A_{\rm ast}}.
\end{eqnarray}
Eventually, for dimensionless parameters (with the Coulomb logarithm $\Lambda$ and the difference between the velocities $V$ included in the definitions of corresponding parameters being neglected), we arrive at
\begin{eqnarray}
\frac{{\cal R}e_{\rm lab}}{{\cal R}e_{\rm ast}}  & = &  
a_{\rm L}  a_{\rho}^{7/2} a_{\rm P}^{-5/2}\left(\frac{A_{\rm ast}}{A_{\rm lab}}\right)^3
\left(\frac{\alpha_{\rm lab}}{\alpha_{\rm ast}}\right)^2, \\
\frac{{\cal R}e_{\rm m,lab}}{{\cal R}e_{\rm m, ast}}   & = & 
a_{\rm L}  a_{\rho}^{-3/2} a_{\rm P}^{3/2}\left(\frac{A_{\rm lab}}{A_{\rm ast}}\right)^{3/2}, \\
\frac{{\cal P}e_{\rm lab}}{{\cal P}e_{\rm ast}}  & = & 
a_{\rm L}  a_{\rho}^{7/2} a_{\rm P}^{-5/2}\left(\frac{A_{\rm ast}}{A_{\rm lab}}\right)^{7/2}
\left(\frac{\alpha_{\rm lab}}{\alpha_{\rm ast}}\right), \\
\frac{{\cal L}{\rm u}_{\rm lab}}{{\cal L}{\rm u}_{\rm ast}} & =  & 
a_{\rm L}  a_{\rho}^{-2} a_{\rm P}^{3/2}\left(\frac{A_{\rm lab}}{A_{\rm ast}}\right)^{3/2}
\left(\frac{B_{\rm lab}}{B_{\rm ast}}\right).
\end{eqnarray}

\begin{figure*}
\begin{center}
\includegraphics[width=480pt]{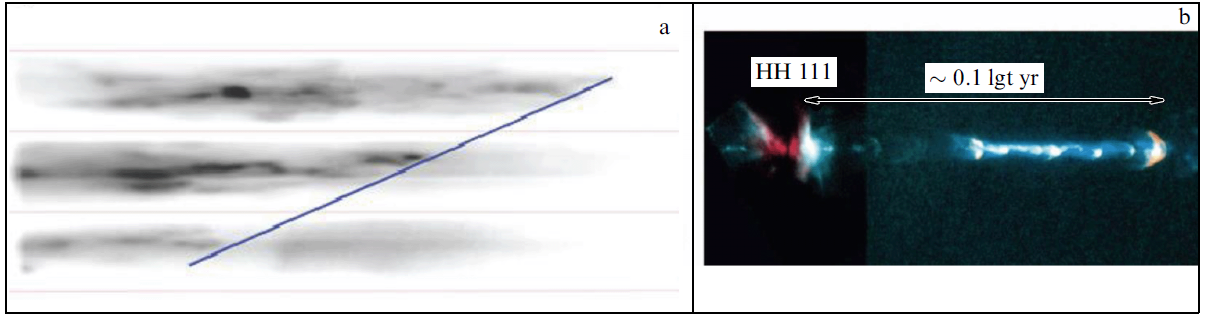}
\caption{\small (a) Time sweep of the neon plasma glow of a flow through three 
transverse slits at a height of 31, 35, and 39 cm. Scan duration is 2.5 ms, 
and the vision field in the object is 10 cm. Delay in the emergence of 
glow at different slits (blue tilted line) was used to determine the flow 
velocity $5,7 \times 10^6$ cm s$^{-1}$ [195]. (b) Optical image of the 
HH 111 object [31]. Scale segment that approximately corresponds to 0.1 
of light year (lgt yr) is also displayed.}
\label{fig15}
\end{center}
\end{figure*}

Finally, we note that scaling coefficients (62)--(67) do not depend on the degree of plasma magnetization: as was noted, the formulas for the dimensionless variables should contain the largest values of transport coefficients (59)--(61). However, under conditions of strong magnetization $\omega_{\rm B} \tau \gg 1$, where $\omega_{\rm B} = eB/m_{\rm e}c$ is the gyrofrequency and $\tau$ is the collision time, the external magnetic field can only reduce the corresponding coefficients of the transfer perpendicular to magnetic field lines (i.e., result in an increase in the corresponding dimensionless parameters). This only improves the accuracy of the approximation of the ideal MHD on which the theoretical analysis presented in Sections 2.2. is based. Since, as can be seen, in laboratory experiments with PF equipment this approximation is fulfilled with a good accuracy, one can confidently assert that both the internal structure and other features of laboratory plasma jets should correspond to jets from young stars.

In completing the analysis of the scaling factor, we come to the following conclusion: the laboratory experiment designed to test the models based on an ideal MHD makes it possible to reproduce three astrophysical parameters. Above, the conventionally used quantities were employed as such parameters: transverse jet size $L$, density $\rho$, and pressure $P$. This choice automatically ensures matching in the values of thermal velocity$c_{\rm s} \approx (P/\rho)^{1/2}$ and, in addition, with an accuracy up to the ratio of atomic weights $A^{1/2}$, and temperature $T$. The hydrodynamic velocity obtained by scaling $V = {\cal M}_{\rm lab}V_{\rm lab}$ also turned out to be close to real velocities observed in astrophysical jets. On the other hand, time scaling $t = a_{\rm t}t_{\rm lab}$, where
\begin{equation}
a_{t} = a_{\rm L} \left(\frac{a_{\rho}}{a_{\rm P}}\right)^{1/2},    
\end{equation}
corresponds to astrophysical times on the order of $10^{3}$ yr, a value which is also quite satisfactory. Finally, the magnetic field that should scale like $a_{\rm P}^{1/2}$ would correspond to values on the order of $10^{-2}$--$10^{-3}$ G, which is somewhat larger than the estimated magnetic fields in astrophysical jets. However, since the Alfv{\' e}n Mach number is in both cases larger than one, such a difference within an ideal MHD is of no significant importance.

\subsection{Main results of studies}

{\bf 3.5.1 Description of the simulation problem.} The differences between the scheme of PF-based experiment and that of experiments based on Z-pinches or laser systems enable setting new tasks for laboratory simulation. Distinguished among the differences are the propagation of the head part of the flow to significant distances from the place of its generation, no less than 100 cm, a value that exceeds by two orders of magnitude initial transverse dimensions of the generated plasma flow, and the presence of a background
environment with adjustable parameters. These features enable studying the interaction with the environment, which is accompanied by the formation of a shock wave, development of hydrodynamic instabilities resulting in `fragmentation' of the flow, which is characteristic of jets from young stellar objects, and a number of other phenomena. The possibility of carrying out experiments with a broad variety of working gases enables studying the effect of radiation cooling on plasma flow collimation. Finally, the dimensions
of the flows generated in the PF are fairly large (several centimeters), enabling the application of magnetic probe techniques; this provides options for detailed exploration of the distribution of magnetic fields and an analysis of their role in plasma flow collimation and stability. These and some other problems have been studied in the experiments described in Sections 3.5.2--3.5.6.

\begin{figure*}
\begin{center}
\includegraphics[width=500pt]{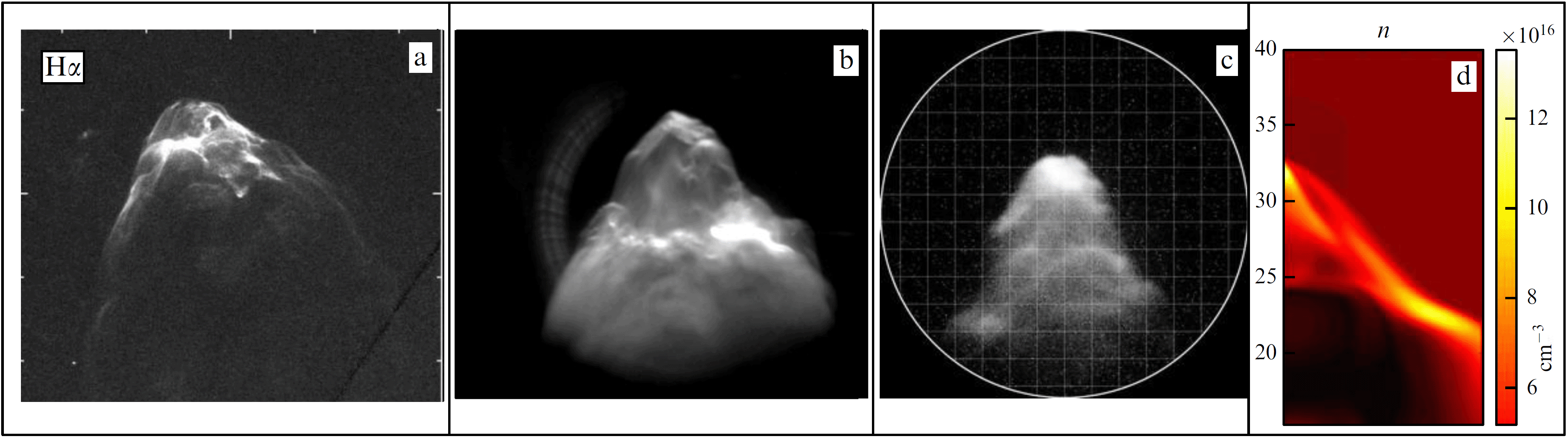}
\caption{\small (a)HH 34 object [91]. Plasma flows at facilities 
(b) PF-1000U [215, Fig. 7a] and (c) PF-3 [214, Fig. 3a] 
during a discharge in neon. (d) Results of numerical simulation 
of the plasma flow propagation at the PF-3 facility [216, Fig. 3c].}  
\label{fig16}
\end{center}
\end{figure*}

\begin{figure}
\begin{center}
\includegraphics[width=230pt]{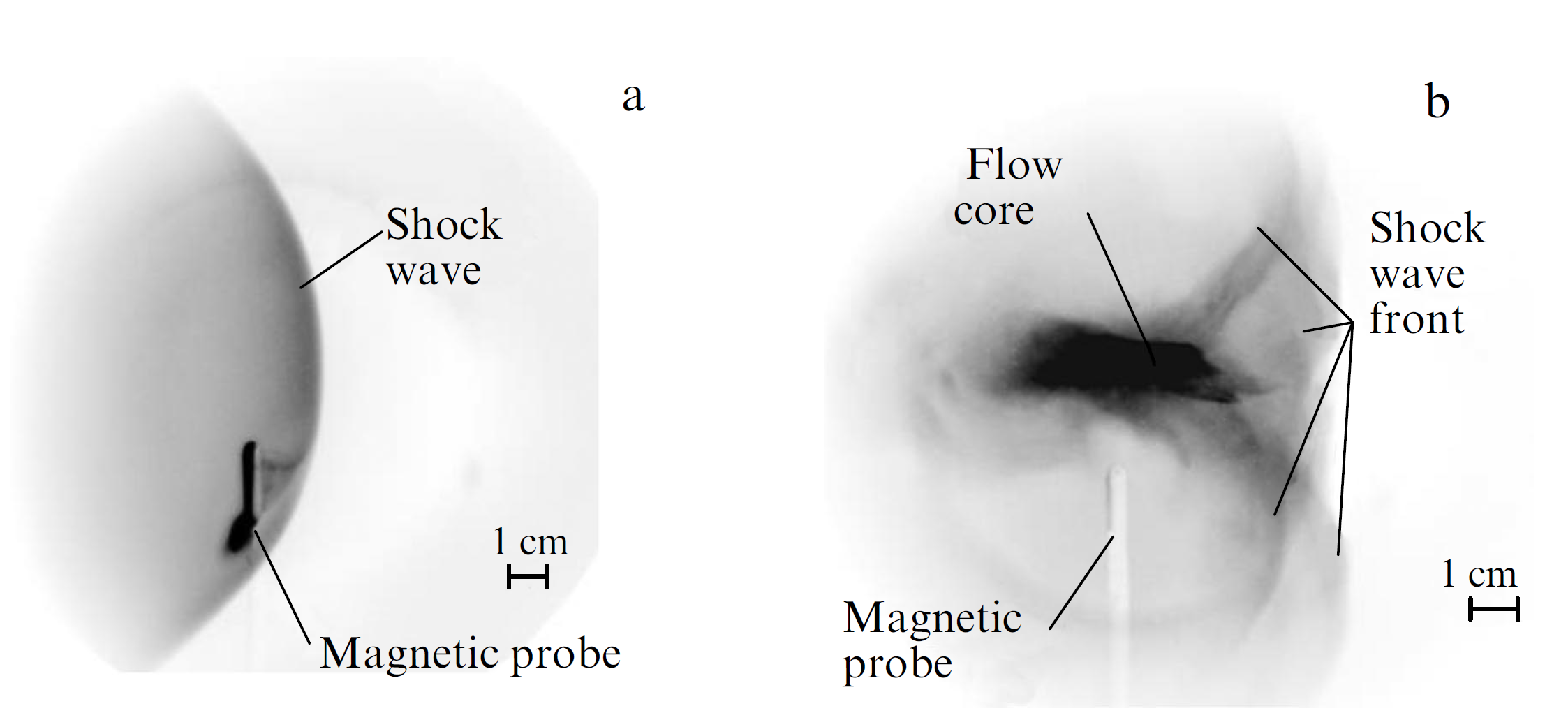}
\caption{\small Images of a plasma flow at a distance of 40 cm from the anode
face end: (a) stationary inflow of D$_{2}$ (1.2 hPa); (b) stationary 
inflow of D$_{2}$ (1.2 hPa) and additional injection of a mixture of 
deuterium (75\%) and neon (25\%). Magnetic probe that detects the 
toroidal magnetic field is seen. (Adapted Fig. 3 from [219].)}  
\label{fig17}
\end{center}
\end{figure}

\begin{figure}
\begin{center}
\includegraphics[width=230pt]{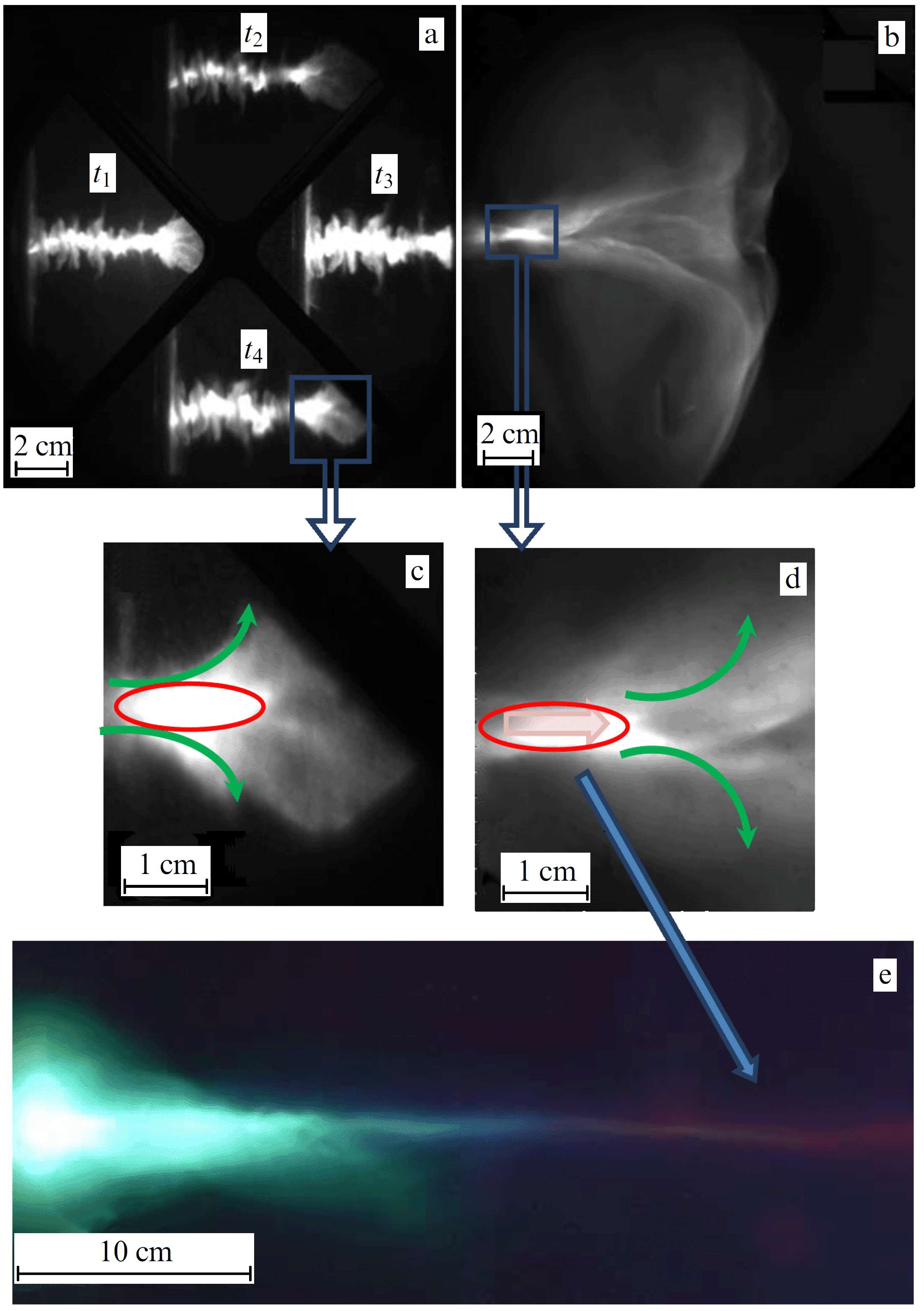}
\caption{\small Sequential images of the pinching region in the soft X-ray range
(a); time interval between frames of 20$\pm$5 ns and the region of plasma
flow generation on an enlarged scale (c). Images of plasma flow in the
visible range at a distance of 40 cm from the anode plane (b) and its central
part (d) on an enlarged scale. (e) Time-integral image of plasma flow in the
visible range. Arrows show the directions of the axial current and reverse
closing currents [215].}  
\label{fig18}
\end{center}
\end{figure}

{\bf 3.5.2 Plasma flow morphology.} The structure of plasma flow was studied using optical high-speed cameras. Already, the first studies have shown that the plasma flow in heavy gases is compact and structured [195]. Good agreement with astrophysical jets is observed. For example, the slit scanning displayed in Fig. 15, which was obtained at a height of 35 cm in a discharge in neon, is very similar to the image of the HH 111 object.

Figure 16 displays images of a real jet observed in the HH 34 object (Fig. 16a), and frame-by-frame images of the head part of the plasma flow in an optical range obtained by the PF-1000 (Fig. 16b) and PF-3 (Fig. 16c) facilities using frame cameras. A high degree of similarity between the objects under study is clearly seen. Quite likely, the outer boundary of the observed object is detected, or more specifically, the SW that is formed when the supersonic plasma flow propagates in a finite-density medium, which is confirmed by the numerical simulation carried out for the parameters of the PF-3 facility [216--218].

Experimental results show that radiation cooling significantly affects plasma flow morphology. Collimation effects are virtually unobservable in experiments with a weakly radiating gas (hydrogen, helium), while, in experiments with heavy strongly radiating gases, the flow profile acquires a conical shape (see Fig. 10).

The collimation effect was exhibited in a most pronounced way in experiments at the PF-1000U facility [215, 219]. In discharges with preliminary filling of the discharge chamber with deuterium (the main working gas in the PF-1000U facility), a quasi-spherical profile of the SW is observed (Fig. 17a), unlike the conical profile when the chamber is filled with neon (Fig. 16b). An even more
interesting result is obtained in the case of the additional pulsed injection of neon into the near-axis region of a
chamber filled with deuterium (Fig. 17b). The development of the discharge under these conditions is presented in Fig. 18. As a result of CPS formation on the axis and the development of instabilities, a plasma object (plasmoid) is formed, containing a large amount of strongly radiating neon, which is pushed away along the chamber axis. The process of the formation of such a plasmoid is shown in Figs 18a, c. The already formed plasmoid takes away part of the magnetic flux.

Figures 18b, d display images of the flow in the optical range with a 3-ns exposure of the shot, which were made at a distance of the order of 30±40 cm from the place of generation. It can be seen that, even at such a distance, a brightly shining core with a diameter of $\sim$ 1 cm persists. Moreover, the path of the motion of the brightly shining core along the chamber is clearly seen in a time integral photography made using a digital camera (Fig. 18e) and even at larger distances. The transverse dimension of the path is
virtually constant, but the radiation spectrum changes from blue, which is characteristic of NeII, to red, characteristic of neutral NeI. Since, in the setup with gas-puff, at the moment of discharge, there should be no neon at large distances, this observation indicates primarily that the injected gas is entrained in a transfer by the plasma flow and the gradual cooling and recombination of flow ions as the distance from the anode increases. As the brightly shining core propagates in a background plasma, an SW emerges to which part of the mass outflows. It creates a medium for the flow of reverse closing currents, and these currents `spread' not in a uniform way but as separate current channels. In the presence of a strongly radiating substance (under pulsed injection), the reverse closing currents are also well exhibited on the background of weakly radiating hydrogen.

\begin{figure}
\begin{center}
\includegraphics[width=230pt]{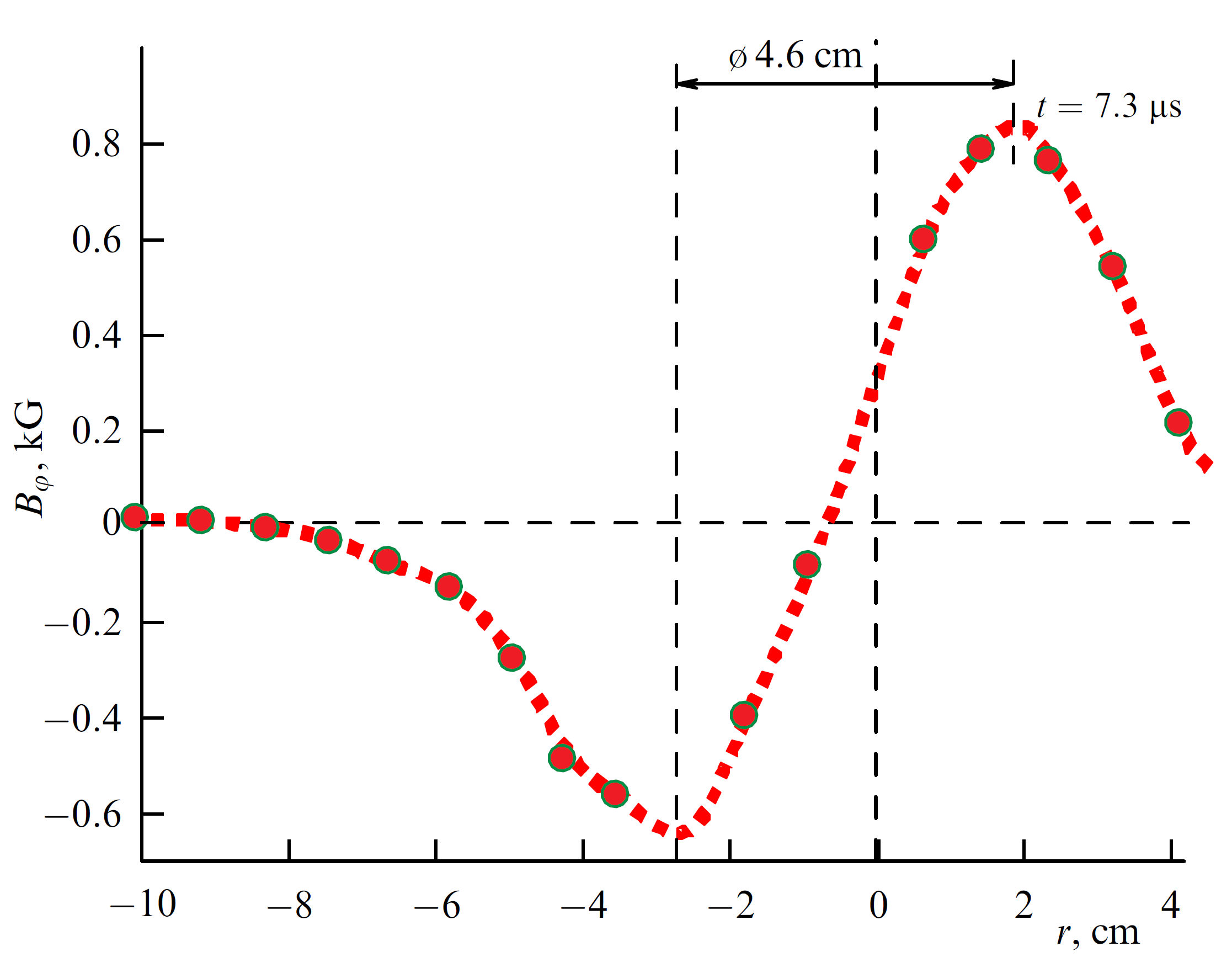}
\caption{\small Radial distribution of a toroidal magnetic field at a distance of
35 cm from the anode at time $t = 7.3$ $\mu$s relative to the flow generation
moment (the singularity of the derivative of the current). A minor shift of
the flow axis relative to the chamber axis is seen [221].}  
\label{fig19}
\end{center}
\end{figure}

{\bf 3.5.3 Distribution of magnetic fields in a plasma flow.} As was noted, currently, the most advanced model is the electro-magnetic model of jets [95], according to which the decisive role in propagation and collimation of jets is played by electromagnetic forces that emerge due to rotation of the `central engine.' In this model, an electric current flows near the jet axis and closes on the jet periphery. The transverse size of the jet is determined by the balance of the internal pressure, which depends on the strength of the magnetic field, and the pressure from an external medium (which, in turn, depends on the distance from the 'central engine') [220].

Naturally, one of the main tasks of experiments is to check the extent to which this picture can be applied in analyzing the plasma jets created by laboratory facilities. It should be noted that the nature of the emergence and the structure of fields under astrophysical and laboratory conditions are significantly different. While in astrophysics the main 'seed' field is a poloidal one, which actually sets the jet direction, and the toroidal field emerges due to jet rotation, in a PF experiment, the initial poloidal field in the pinch, which performs as the 'central engine,' is virtually absent. The magnetic field has primarily a toroidal component due to which, in essence, plasma is pinched. Indeed, as the first experiments have shown [203, 213], in plasma flows, there is a strong (up to 10 kG) toroidal magnetic field. In this review, we abstract from the mechanisms that generate this field and concentrate on discussing its distribution and dynamics in a plasma flow.

To explore in detail the radial distribution of the toroidal field, multichannel magnetic probes were made, which can concurrently conduct measurements at several (up to 18) spatial points. Figure 19 displays an example of the radial distribution of the toroidal field. The flow may be shifted relative to the axis, but in all cases several distribution zones are clearly seen. In the flow center, the field increases as $B_{\varphi} \propto r$ in correspondence with the distribution of the field of a conductor with current. The radial distribution of the magnetic field indicates that a longitudinal current of $\sim$ 10 kA flows in a near-axis region 1--3 cm in radius at a distance of 35 cm from the anode [213, 222, 223].

The size of the region with the magnetic field along the facility axis, $L_{z}$, estimated using time dependences $B_{\varphi}(t)$ with consideration for the bunch motion speed, is also finite, 5--15 cm. Detailed measurements using 18 probes installed along the camera axis have shown that the region with the trapped magnetic field moves along the flight chamber with a speed of 2--4 $\times 10^{6}$ cm s$^{-1}$ until the moment it hits the upper flange [223]. This implies the presence of reverse currents flowing at the periphery of the plasma flow and closing directly on the plasmoid itself. Measurements at the periphery of the plasma flow showed the presence of such currents, and their distribution may differ significantly from uniform. Apparently, the reverse current can flow through channels along individual plasma filaments.

Thus, the measurements carried out enable drawing the following conclusion: from a certain moment, the plasma flow exists independently of the 'central engine,' the pinch, and the currents terminate on the plasma bunch itself. Then, the duration of the interval during which the compact form of the plasma flow persists depends on the dissipation time of the currents circulating in them. Experiments conducted at various facilities have shown that the magnetic field propagating along the flight chamber decreases by more than an order of magnitude [222--225]. Nevertheless, the plasma bunch retains its compactness: a few centimeters in the head
part with a gradual expansion in the tail part [222, 224]. Studies using high-speed photographic recorders have shown that such a structure can persist to distances of up to $\approx$ 100 cm (the limit of measurements in our experiments), apparently due to a simultaneous decrease in plasma pressure as a result of radiation cooling. This assumption can be confirmed by regimes with compact bunches being observed only in discharges with strongly radiating gases such as neon and argon.

\begin{figure*}
\begin{center}
\includegraphics[width=400pt]{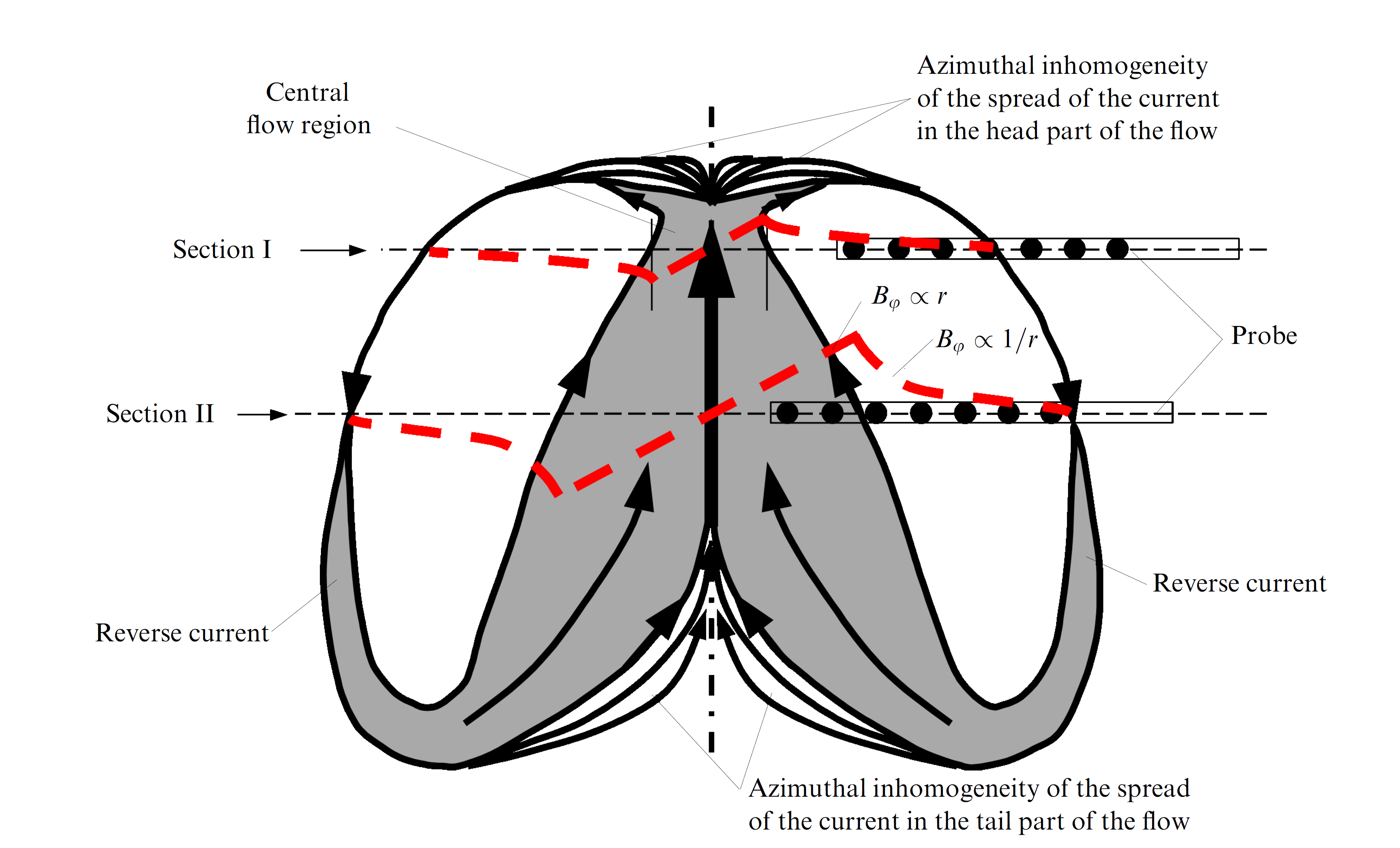}
\caption{\small Structure of the axial plasma flow. Arrows show the scheme of current
circulation; dash lines present the radial distribution of the azimuthal magnetic 
field $B_{\varphi}(r)$ in the central part of the plasma flow and on its periphery: 
two positions of the magnetic probe --- in sections I and II --- are shown [224].}  
\label{fig20}
\end{center}
\end{figure*}

The measurements carried out made it possible to develop a phenomenological model of a plasma bunch. Figure 20 shows two positions of the magnetic probe: section I corresponds to the location of the probe coils, in which some of the coils are in the magnetic field of the central current outside the zone of its flow, i.e., while $B_{\varphi}(r) \propto 1/r$, and the remaining coils are outside the zone where the magnetic field is localized in the plasma flow ($B_{\varphi}(r) \approx0$); section II corresponds to the case when some of the probe coils are located in the central current flow zone, where $B_{\varphi}(r) \propto r$, and the remaining ones are in its magnetic field $B_{\varphi}(r) \propto 1/r$. Thus, during the passage of the plasma flow through the probe, the coils located inside it can at different times be both in the zone where the central current flows and outside it, including outside the region of magnetic field localization in the plasma flow. The boundary of this region is the place where reverse currents flow. The detected change over time in the form of the radial distribution of the magnetic field from $B_{\varphi}(r) \propto r$ to $B_{\varphi}(r) \propto 1/r$ indicates a broadening of the radial size of the zone of central current flow in the direction from the leading edge of the flow to the trailing edge, as shown in Fig. 20 in gray.

Apparently, the dynamics of the parameters of both the flow plasma and the magnetic field trapped by it are largely determined by the parameters of the medium in which the flow propagates. The conditions of experiments on PF-type installations make it possible to change the parameters of the external environment in a wide range, including so-called contrast --- the ratio of the flow density to the density of the surrounding gas. The dependence of the size of the region with a trapped magnetic flux (in fact, the radius of the reverse current flow) on the parameters of the external medium was studied in detail at the KPF-4 Phoenix facility [224]. A set of
experiments with a stationary pre-filling of Ar gas at an initial pressure of $P_{0} =$ 2 Torr showed that the radial size of the plasma flow, in which the magnetic flux trapped by the plasma is localized, did not exceed 4--6 cm. When the background gas pressure in the drift space of the facility chamber was reduced to 1 Torr, the radial size of the plasma flow increased to $r =$ 7--8 cm. In the case of pulsed injection of argon, i.e., under conditions of flow propagation in a rarefied medium (see Section 3.2), signals from magnetic probes are reliably recorded at radiuses up to r $\approx$ 16 cm. In the latter case, the plasma flow has a significantly larger radial size than at a stationary pressure of 1--2 Torr, which is due to the low pressure of the background gas.

However, in the case of stationary filling of the chamber with hydrogen, the boundary for the reverse current flow is close to a radius of less than 13 cm, despite the relatively high pressure of the background gas (8 Torr). The small radial size in the case of argon is associated with increased losses of the internal energy of the plasma for radiation.

As noted, a characteristic feature of modern models of astrophysical jets is the presence of a longitudinal (poloidal) magnetic field that determines the jet direction. In a laboratory experiment, a longitudinal magnetic field is usually created by external sources, since in the most common simulation schemes, either the longitudinal magnetic field is absent at all (laser experiments) or the azimuthal magnetic field prevails in accordance with the geometry of the experiment. The influence of the applied field on the discharge dynamics and the parameters of the formed jets was studied on the MAGPIE setups in the configuration of radial wire
arrays [226] and COBRA with compression of radial aluminum foils [164]. In these experiments, a longitudinal magnetic field of several tens kG was created by an external solenoid or Helmholtz coils. In a laser experiment at the LULI laboratory, a poloidal magnetic field of $\sim$ 200 kG was generated using Helmholtz coils [151]. It is shown that such a field can lead to flow focusing.

In experiments on PF devices, evidences were obtained for the existence of an intrinsic poloidal field already at the stage of pinch formation [227]. One of the possible reasons for the appearance of this field may be the presence of azimuthal rotation or spiral filamentation of the current sheath at the stage when it converges to the axis [228]. The poloidal component of the magnetic field was also found in the formed flow [192, 203] at a considerable distance from the anode.However, at this stage, it is virtually impossible to vary in a controlled way the initial value $B_{z}$ field in the flow formation region, a circumstance that creates difficulties in interpreting the influence of the poloidal field on the flow parameters.

In experiments on the PF-3 setup, a solenoid located under the anode at a minimum distance from its plane was used as a source of an external magnetic field [229]. The solenoid created a magnetic field of 770 G on the setup axis near the anode surface with a magnetic flux from the entire anode surface 686 kMx. In the process of convergence of the conductive CPS to the axis, the generated magnetic flux is compressed, leading to an increase in the magnetic field in the pinching region up to $\sim$ 100 kG. Magnetic probe measure-
ments showed that the $B_{z}$ component also increases significantly directly in the plasma flow, up to 40 kG at distances of 10--15 cm from the anode; the direction of the field trapped by the flow also corresponds to that of the external applied field. A strong increase in the trapped toroidal magnetic field (by a factor of 2--3) turned out to be unexpected. It is important that the direction of the toroidal field did not depend on the direction of the applied poloidal field but was determined by the polarity of the PF setup itself (the central electrode was the anode). Apossible explanation for the increase in the toroidal field could be jet rotation (see discussion below).

In addition, it was found that in the case of an applied externalmagnetic field, the decrease in the $B_{\varphi}$ field trapped by the plasma flow during the flight is much weaker than in the case when $B_{z0}$ is zero. Presumably, such a significant difference in the attenuation of currents in a span of 10--65 cm can be explained by the fact that the application of an external $B_{z}$ field results in the formation of a spiral configuration of the total magnetic field of the plasma jet, which contributes to the formation of a compact axial plasma jet propagating in the drift space of the flight chamber. 

As a result, it was shown that the $B_{z}$ field trapped by the plasma flow has a radial distribution $B_{z}(r)$, the shape of which is close to the distribution of the magnetic field of a spheromak (spherical tokamak): in the central part of the plasma flow, where the central current flows, the maximum level of the $B_{z}$ field was detected (Fig. 21a). As the distance from the axis increases, a decrease in the $B_{z}$ field is observed, followed by its polarity reversal in the region of radii of 4--6 cm. This distance agrees with good accuracy with the boundary of the central current. The distribution profile of the magnetic field $B_{z}(r)$ does not depend on the direction of the external magnetic field of the coil; it is only the sign of the Bz field that changes: the $B_{z}$ field trapped inside the jet has the same direction as the initial axial magnetic field created by the current of the external solenoid. The results obtained are in good agreement with those reported in [230] (Fig. 21b, c), in which a new broad class of solutions of ideal MHD equations was found that describe closed axisymmetric stationary flows. Similar structures are also observed in jets from young stars.

{\bf 3.5.4 Interaction of the plasma flow with the background medium.} The interaction of the flow with the environment will lead to a change in the parameters of both the flow itself --- its deceleration (see Fig. 13) [202], cooling (Fig. 18e) [215], shock wave excitation [218] --- and the background plasma. In particular, experiments have shown that the radiation power of a plasma flow is sufficient to change the ionic composition of the environment [202]. Estimates made have shown that the laboratory flow can heat the background plasma up to several eV, the effect being observed at small distances (several cm) from the flow front. Thus, as in the case of astrophysical sources, the plasma flow actually propagates not in a neutral gas but in a weakly ionized plasma 'prepared' by the flow itself. It should be noted that the effect of the so-called preshock resulting from the heating of the nearby medium by UV radiation from the cooling zone of a shock wave is considered in many papers devoted to nonrelativistic astrophysical jets [231].

Moreover, a vacuum wake, a region with reduced density, can form behind the SW [216]. Experimental verification of this assumption is fairly challenging. The main problem is that, in the PF discharge, a single plasma bunch is predominantly formed, which is generated in the pinching stage. However, two or more can sometimes be formed as a result of repeated pinching. From an extensive database of sweeps, those discharges in which several bunches were generated were selected and analyzed [218].

The results of the experiment showed that the parameters of the second jet differ significantly from those of the first (Fig. 22). The velocity of the second flow can be less than that of the first one, which is explained by different conditions of flow generation as a result of re-compression of the current-carrying plasma shell. Nevertheless, in this case, no mush-room-shaped SW is formed, sufficient for the manifestation of the luminescence intensity in the experiment, which indicates motion through a medium of lower density. The second jet is fairly well collimated and has a transverse dimension of only about 2 cm at a distance of 30 cm from
the anode with an aspect ratio of $\geqslant$ 2. 

\begin{figure*}
\begin{center}
\includegraphics[width=500pt]{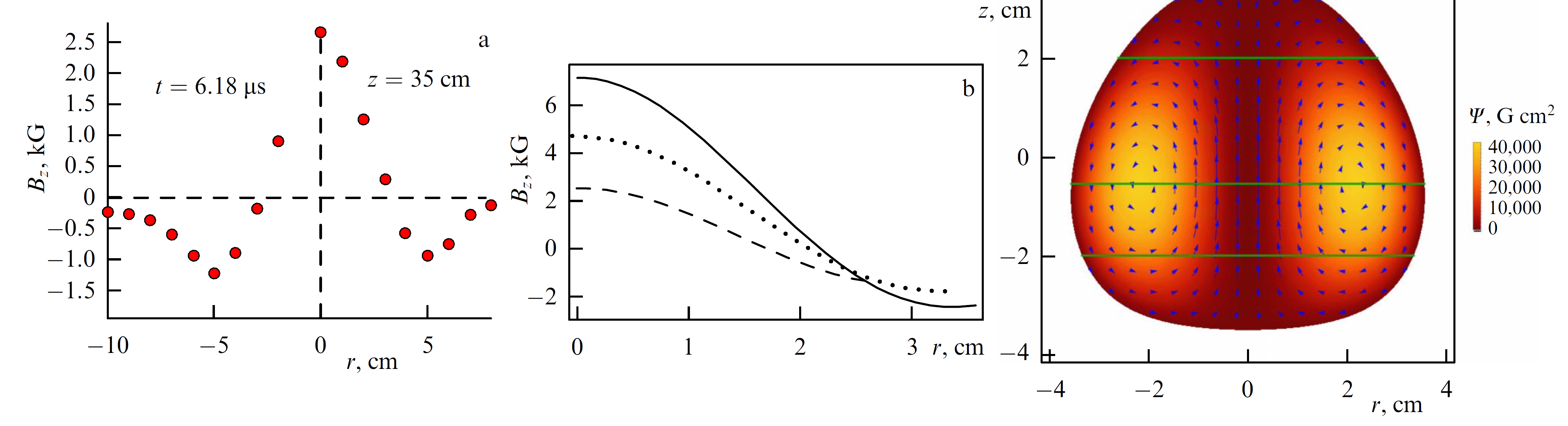}
\caption{\small (a) Radial distribution of poloidal components of 
a magnetic field at the PF-3 facility with an external magnetic field 
applied [229]. (b) Calculated radial distribution of the poloidal 
component of the magnetic field in horizontal cuts of the plasma 
flow at various heights [230]. (c) Structure of the flow within the
plasma jet. Arrows show the direction of velocities ${\bf v}$; poloidal 
magnetic field ${\bf B}$ is oriented in the same direction, and
density of electric current ${\bf j}$ is oriented in the opposite direction.}  
\label{fig21}
\end{center}
\end{figure*}

Another interesting result is the dependence of the SW profile on the composition of the working gas. As shown in Section 3.5.2, when the flow propagates in hydrogen, a quasi-spherical shock front is formed (Fig. 17a). A similar result was obtained using helium as the working gas. A different picture is observed for discharges in neon. Figure 23 shows sections of the leading edge of flows in various gases which were obtained at the PF-3 facility. Attention is drawn to the well-pronounced 'laces' of the flow front in neon.

We examined in [232] the reasons that lead to the difference in the spatial structure of plasma jets during a discharge in gases of various chemical compositions: neon, helium, and helium with an admixture of neon. As we noted, the flow is the most structured in the case of pure neon: the leading edge of the jet consists of numerous compactions, making it very similar in appearance to Herbig-Haro objects. The jet looks least structured in the case of pure helium; however, should only 1\% neon be added to helium, the shape
of the jet head changes significantly: a small-scale structure becomes noticeable in it. Estimates show that these features can be associated with a difference in the efficiency of cooling of the studied gases both in the plasma jet itself and in the shock wave that emerges when it moves through the background gas. It appears that the main reason for the emergence of inhomogeneities in a plasma bunch, as in the case of Herbig-Haro objects, is various types of instabilities that develop during effective radiative cooling.

In addition, it was found that in some cases the plasma jet can consist of several almost parallel-flying bunches that appear already at the stage of plasma pinching. The collision of shock waves generated by each of the bunches leads to the appearance of knots, which also contributes to the formation of the lacy structure of the plasma ejection. In recent years, there have been grounds to assume that the jets of young stars also consist of individual bunches flying in parallel at different velocities [93]. We believe that the results of our experiments can be considered an argument in favor of this hypothesis.

{\bf 3.5.5 Plasma flow rotation.} As mentioned in Section 2.1.2, the discovery of jet rotation in astrophysical jets [47, 233] has
become one of the most striking recent observations. The presence of plasma rotation inside the jet is of importance, among other things, because it can lead to the generation of a toroidal magnetic field and, hence, a longitudinal current [234], which, in turn, can be considered as a jet collimation mechanism [235].

As noted in Section 2.3, the flow rotation in a laboratory experiment was studied using several devices [162--164, 236]. Some indications of the possibility of plasma flow rotation in PF facilities have already been obtained in early works on magnetic probe measurements [192, 213] and in the study of plasma flows using a streak camera [195].

It is clear that the detection of rotation in a laboratory experiment could be a serious argument in favor of the very applicability of PF systems for laboratory simulation of astrophysical jets. However, observing the plasma jet rotation in a laboratory plasma, similar to a real astrophysical object, is a rather difficult task. In particular, estimates have shown that, with an expected azimuthal rotation velocity of the order of $10^{6}$ cm s$^{-1}$, Doppler broadening at the parameters of the PF-3 facility experiment will be too small compared to the Stark broadening of lines to have rotation detected and, moreover, its velocity measured. Therefore, we focused on optical methods for studying the flow dynamics, namely, on the registration of the flow with high-speed optical cameras, as well as on the magnetic-probe technique [221].

An important factor facilitating the detection of rotation can be the presence of pronounced structures: it is apparent that it is quite difficult to detect the rotation of a homogeneous object by optical methods. However, as shown in Sections 3.5.2 and 3.5.4, during a discharge in neon, the plasma flow is strongly structured, which can be associated both with the generation of several separate bunches and with the development of instabilities during the propagation in the environment. This circumstance helped us to carry out appropriate measurements.

\begin{figure}
\begin{center}
\includegraphics[width=250pt]{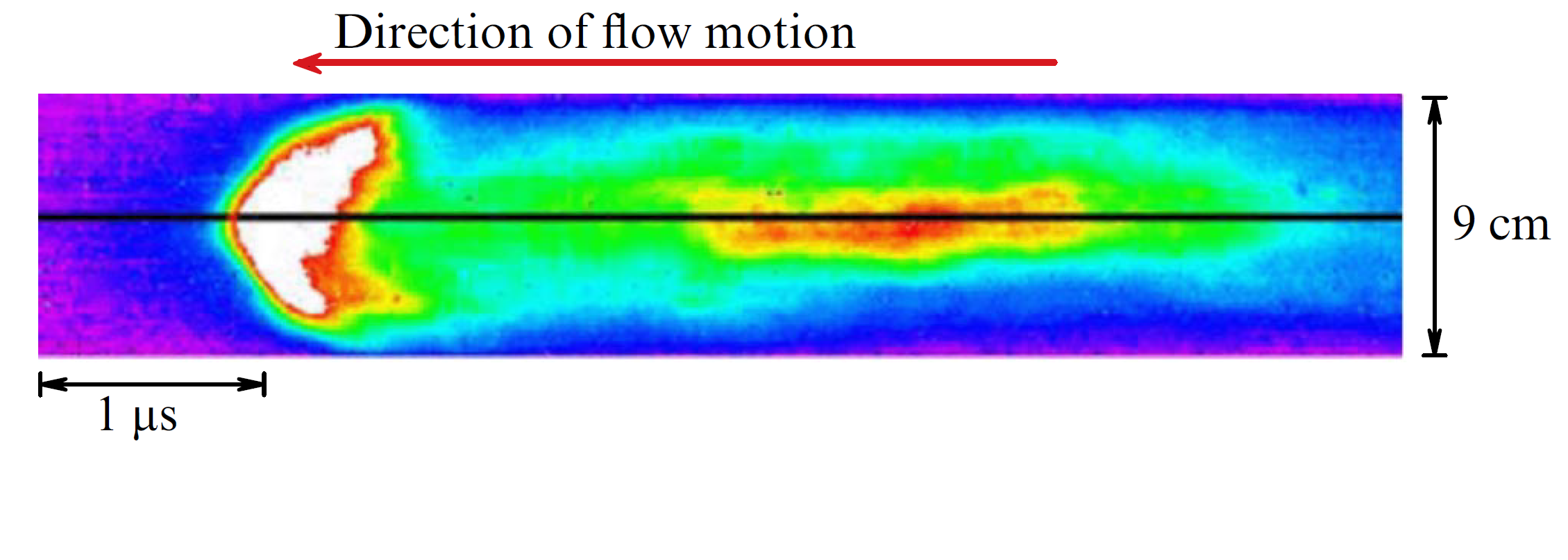}
\caption{\small Time sweep of two sequential flows at the PF-3 facility for a
discharge in helium.}  
\label{fig22}
\end{center}
\end{figure}

\begin{figure}
\begin{center}
\includegraphics[width=250pt]{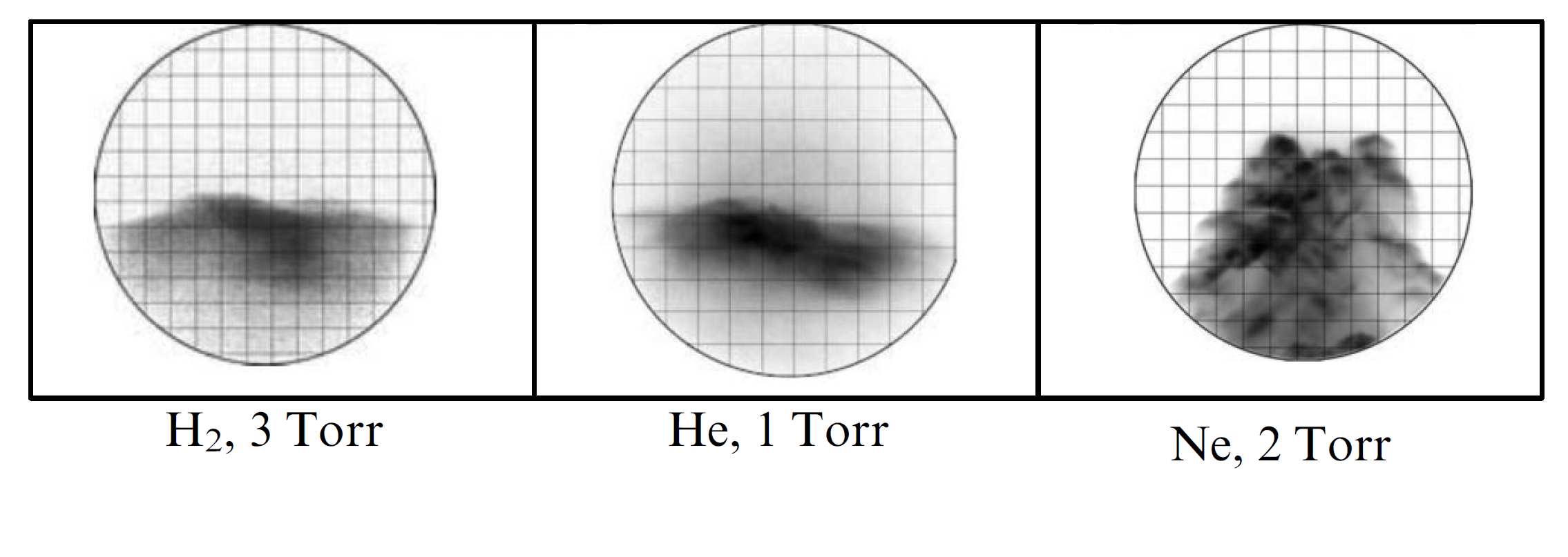}
\caption{\small Head front of a plasma flow in various gases at the PF-3 facility.
Cell scale is 1 cm.}  
\label{fig23}
\end{center}
\end{figure}

\begin{figure}
\begin{center}
\includegraphics[width=250pt]{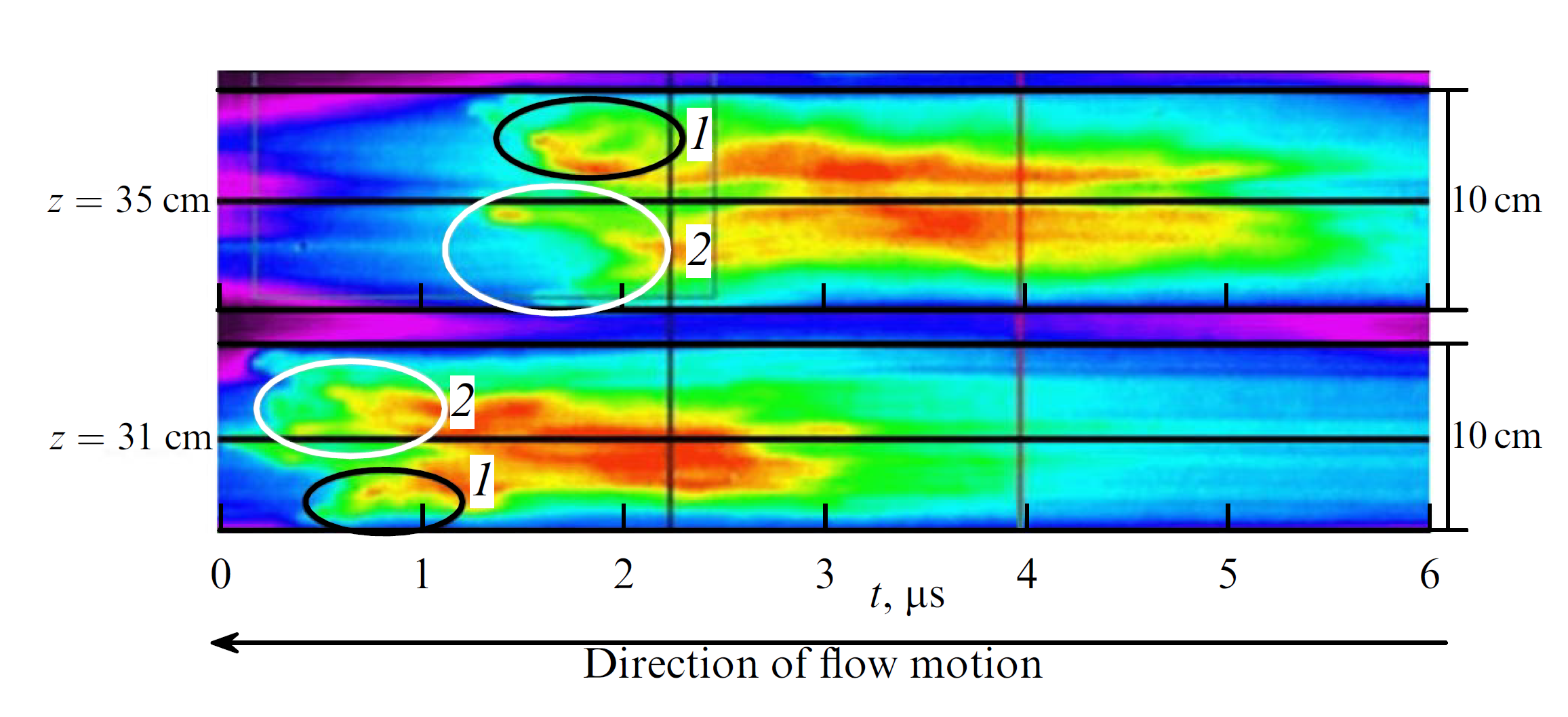}
\caption{\small Time sweeps of a plasma flow in a discharge in neon at distances
of 31 and 35 cm from the anode. Scan duration is 6 ms, and the width of
detection across the chamber diameters is 10 cm. Flow moves from right to
left; $1$ and $2$ are separated fractions of the flow.}  
\label{fig24}
\end{center}
\end{figure}

\begin{figure*}
\begin{center}
\includegraphics[width=500pt]{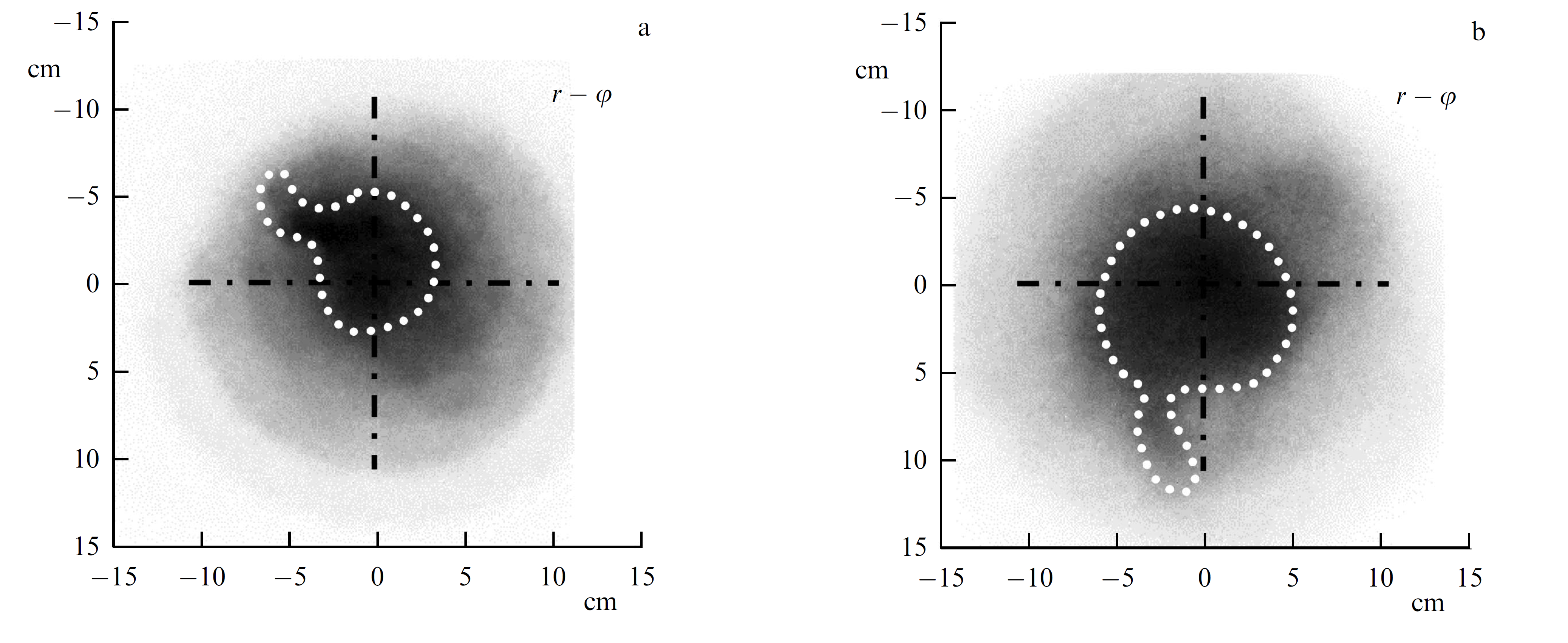}
\caption{\small Plasma rotation in an axial plasma jet based on data from 
the frame cameras in the ($r$--$\varphi$) plane when plasma passes various 
$z$ positions: (a) $z =$ 30 cm at $t = 6.4$ $\mu$s; (b) $z =$  50 cm at $t = 7.9$ $\mu$s.}  
\label{fig25}
\end{center}
\end{figure*}

\begin{figure*}
\begin{center}
\includegraphics[width=500pt]{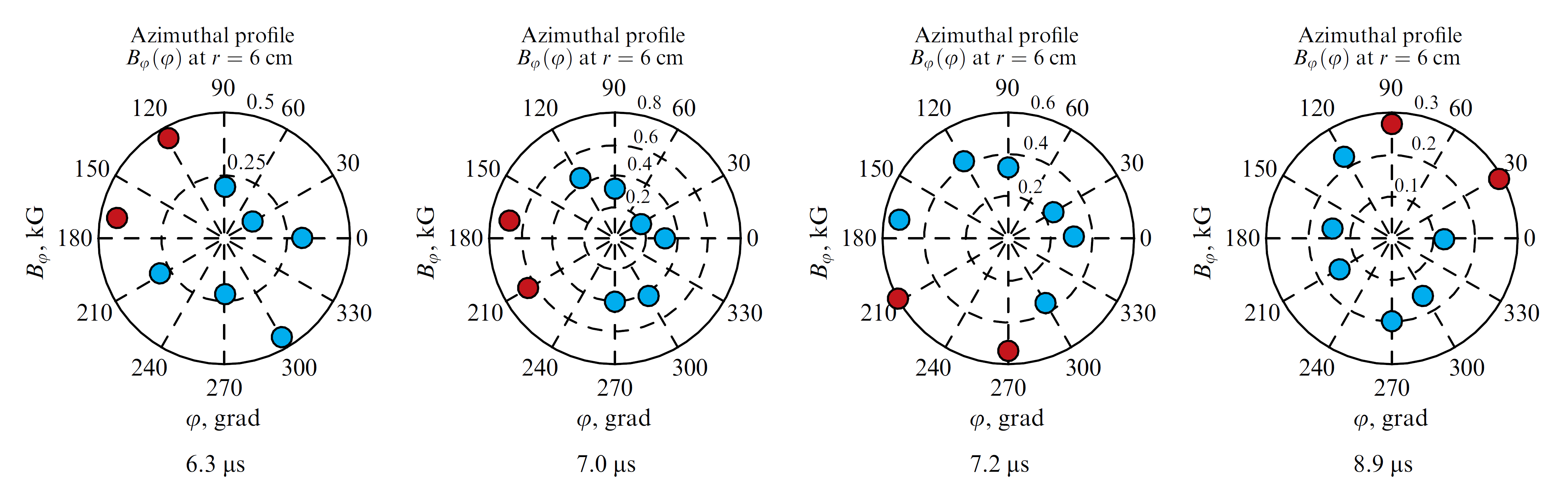}
\caption{\small Measurements of azimuthal magnetic fields in a plasma 
jet at height $z =$ 30 cm from the anode surface at the PF-3 facility 
(Ne, $P_{0} =$ 2.0 Torr, $U_{0} =$ 9 kV, $W_{0} =$ 373 kJ) at various 
points of time relative to the moment of singularity of derivative 
${\rm d}I/{\rm d}t$. Azimuthal dynamics of individual structures
of the plasma flow with circulating currents is seen.}  
\label{fig26}
\end{center}
\end{figure*}

Figure 24 shows the time sweeps of the plasma flow obtained using a streak camera at heights of 31 and 35 cm from the anode surface. Two fractions of the flow marked in the figure can be clearly distinguished. After flying a distance of 4 cm along the axis, these fractions actually swap, which corresponds to a rotation by approximately 180$^{\circ}$ in a time of $\sim$ 1 $\mu$s and a rotation speed of $\sim 3 \times 10^{6}$ cm s$^{-1}$. A more accurate estimate for two-dimensional registration in the ($r, \varphi$) plane is not possible. Therefore, the plasma flow was photographed from the end of the flight chamber using frame cameras. Figure 25 shows the results of frame recording of optical plasma images in the ($r, \varphi$) plane during the motion of an axial plasma jet on the flight base from $z =$ 30 cm to $z =$ 50 cm. The delays between frames were selected in such a way that the plasma during shooting was at the indicated $z$ positions.

Optical frames display some features of the plasma flow structure, such as the axial dense (radiating) part, in which the central current flows, and the plasma flow periphery, in which reverse currents flow. In the flow periphery region, pronounced plasma formations in the form of 'outgrowths' are observed. Actually, the method of observation of the 'outgrowths' made it possible to determine the nature of their motion. The presence of such plasma formations turned out to be useful for detecting plasma rotation in an axial plasma jet, since, as noted, in the case of a homogeneous flow structure, it would be difficult to observe the effects
associated with rotation using this technique.

%%%%%%%%%%%%%%%%%%%%%%%%%%%%%%%%%%%%%%%%%%%%%%%%%%%%%%%%%%%%%%%%%%%%%%%%%%%%%%%%%%%%%%%%%%%%%%%%%%%%%%%%%%%%%%%%%%%%%%%%%%%%

Estimation of the angular velocity of rotation depends on the direction of rotation. For example, the turn registered in optical frame images for the time $\Delta t$ 1.5 $\mu$s on the azimuth angle $\Delta \varphi_{\rm opt} \approx$ 105$^{\circ}$ counterclockwise corresponds to the angular velocity $\omega_{\rm opt} \approx 1.2 \times 10^{6}$ rad s$^{-1}$, while the clock-wise turn to the angle $\Delta \varphi_{\rm opt} \approx$ 225$^{\circ}$ corresponds to $\omega_{\rm opt} \approx 3.0 \times 10^{6}$ rad s$^{-1}$, which is in good agreement with the data obtained with streak cameras.

To refine the value of the angular velocity of plasma rotation in the flow, we analyzed the distributions of the magnetic field in the ($r$--$\varphi$) plane, obtained using eight magnetic probes located along the azimuth (Fig. 26). An analysis of the complete database of probe measurements enabled determination of the space-time characteristic of the change in the $B_{\varphi}(\varphi)$ distribution, which, combined with the data from the frame cameras, indicates a complex character of plasma motion in an axial plasma jet. Two types of motion can be traced concurrently: the azimuthal rotation of individual parts of the plasma flow and the displacement of the central current axis relative to the axis of the installation's flight chamber. The angular velocity of rotation \mbox{$\omega_{\rm mag} =$ (1.4--2.6) $\times 10^{6}$ rad s$^{-1}$} estimated based on magnetic measurements is in good agreement with the results provided by frame cameras: \mbox{$\omega_{\rm opt} \approx$ (0.5--3.0) $\times 10^{6}$ rad s$^{-1}$.}

{\bf 3.5.6 Agreement with theory predictions.} It is clear that good agreement among scaling parameters could not but lead to success in laboratory modeling of astrophysical jets, despite the geometry of a localized plasma being certainly very far from the cylindrical jet flows discussed in Section 2.2. Nevertheless, as shown in Section 3.5.2, a laboratory experiment successfully reproduced all the main morphological properties predicted by the MHD jet theory. Moreover, the main characteristics of jets not only qualitatively but also quantitatively corresponded to the predictions of the theory.

First of all, a laboratory experiment can reproduce a supersonic (${\cal M} > 1$) highly magnetized ($\sigma_{\rm M} \gg 1$) flow. We provide an estimate of these quantities below. We only note here that, presumably, the successful reproduction of a supersonic strongly magnetized flow in the experiment is due to the similarity of the very mechanism that triggers a plasma jet in PF installations to the 'central engine' in astrophysical sources [237].

Indeed, the upper part of the CPS quite accurately models the accretion disk that naturally arises near young stars. Moreover, when jets are formed from young stars, the 'central engine,' as noted, is not the young star itself, but the accretion disk. The main difference is that the source of activity in young stars is the kinetic energy of rotation stored in the accretion disk, while the radial current $j_{r}$, which closes the system of currents that carries away energy, is generated due to the rotation of a well-conducting disk, immersed in an external magnetic field. In turn, in a laboratory experiment, the source of energy is an external battery that generates a radial current $j_{r}$, the action of which, in the presence of a poloidal magnetic field, leads to rotation. However, in both cases, a potential difference arises between the outer and inner regions of a well-conductive rotating medium, which becomes a battery that drives the 'central engine.'

One of the main elements of the 'central engine,' namely, a regular magnetic field penetrating the disk, can naturally emerge both due to compression of the geomagnetic field and under the action of toroidal Hall currents [237]. The existence of such a field in experiments at the PF facility is now beyond doubt [229]. In the presence of a vertical magnetic field, the diverging radial current in the upper part of the CPS will inevitably lead to the rotation of the plasma due to the Amp{\' e}re force ${\bf j} \times {\bf B}$. As already noted, the rotation caused by the action of just such a divergent electric current was detected. For example, using the COBRA facility, the rotation speed reached 20 km s$^{-1}$ [164], and, with the equipment at the Weizmann Institute (Israel), it was \mbox{50 km s$^{-1}$ [236].}

As for the plasma jet itself, a laboratory experiment has shown the presence of a narrow central core containing almost the entire outflowing electric current $I$. In this case, in the region $r < r_{\rm core}$, the toroidal magnetic field $B_{\varphi}$ should increase linearly with increasing distance $r$ from the jet axis and decrease according to the conventional law $B_{\varphi} \propto r^{-1}$ at large distances. It is this structure of the toroidal magnetic field that is observed in our experiment. As we noted in Section 3.5.3, for typical parameters of young stars (see Sections 2.1.1 and 2.1.2), the characteristic scale $r_{\rm core} \sim$ 0.3 au is beyond the resolution of modern telescopes. Therefore, the above example is an excellent illustration of the fact that it is the laboratory experiment that enables refinement of the properties of astrophysical jets. Finally, using Eqn (51), we obtain for the cocoon size (i.e., the size at which the law $B_{\varphi} \propto r^{-1}$ is no longer true)
\begin{equation}
r_{\rm coc} \sim 10 
\left(\frac{B_{\varphi}(r_{\rm core})}{10^{3} \, {\rm G}}\right)
\left(\frac{P_{\rm ext}}{1 \, {\rm Torr}}\right)^{-1/2} \, {\rm cm},
\end{equation}
in full agreement with the experimental results.

Another important feature that confirmed the adequacy of laboratory modeling was the detection of plasma jet rotation [221]. Here, too, there is good agreement with the predictions of the theory. As noted, the angular velocity of rotation was already roughly estimated from the data of the first experiments at the PF-3 facility. Indeed, using the relation $r_{\rm core} =  v_{z}/\Omega_{\rm F}$ (44) for the radius of the central core, we immediately obtain $\Omega_{\rm F} \sim (10^6$--$10^7$) s$^{-1}$  for \mbox{$r_{\rm core} \sim 1$ cm} and \mbox{$v_z \sim (10^6$--$10^7$) cm s$^{-1}$.}

The above estimate of the angular velocity can also be obtained from other considerations. The point is that, as shown in Section 2.2, the rotation in a strongly magnetized plasma is due to a radial (in the direction of the rotation axis) electric field $E_{r}$, the strength of which is
\begin{equation}
E_{r} = \frac{\Omega_{\rm F} r}{c} B_{z},
\label{3}
\end{equation}
where $B_{z}$ is the longitudinal (poloidal) magnetic field. Using the condition that the magnetic field is frozen, \mbox{${\bf E} + {\bf v} \times {\bf B}/c = 0$,} we estimate the radial field as \mbox{$E_r \sim v_z B_{\varphi}/c$.} As a result, we obtain
\begin{equation}
\Omega_{\rm F} \sim \frac{v_{z}}{B_{z}} \, \frac{{\rm d}B_{\varphi}}{{\rm d}r}.
\label{4}
\end{equation}
Thus, the main uncertainty here is related to the magnitude of the longitudinal magnetic field $B_{z}$. Introducing for convenience the quantity
\begin{equation}
B_{\rm M} = \frac{{\rm d} B_{\varphi}}{{\rm d} r} r_{\rm core},
\label{5}
\end{equation}
which corresponds to the amplitude of the toroidal magnetic field on the scale $r = r_{\rm core}$ and the ratio \mbox{$b = B_{z}/B_{\rm M}$}, we find
\begin{equation}
\Omega_{\rm F} \sim b^{-1} \, \frac{v_{z}}{r_{\rm core}}.
\label{6}
\end{equation}

The only difference between Eqn (73) and the estimate obtained from Eqn (44) is the parameter $b^{-1}$. It is natural to associate such a difference both with the uncertainty of the estimate $E_r \sim v_z B_{\varphi}/c$ and with the fact that both relations, (44) and (73), were obtained for cylindrical configurations. On the other hand, earlier measurements of the $B_{z}$ field at the PF-3 [21] and KPF-4 [203] facilities have shown that the value of $b$ can be in the range of 0.1--1, so estimates (44) and (73) are close to each other. We also note that, as follows from these estimates, at a distance $r \sim r_{\rm core}$, the toroidal velocity becomes comparable in order of magnitude to the poloidal one, in agreement with the results obtained in the LabJet experiment [162].

If Eqn (73) depends on the parameter $b$, another important value can be obtained independently of the measurement of the poloidal magnetic field. Indeed, combining Eqn (70) and the freezing-in condition \mbox{$E_r \sim v_{z} B_{\varphi}/c$} in the region $r < r_{\rm core }$, we finally obtain  \mbox{$B_{\varphi} = (\Omega_{\rm F} r/v_{z})B_{z}$.} Determining now the longitudinal current $I$ from this toroidal magnetic field $B_{\varphi}$ and normalizing it again to the Goldreich-Julian current $I_{\rm GJ} = \pi r^2 c \rho_{\rm GJ}$, we immediately find $i_{0} = I/I_{\rm GJ} = c/v_{z}$, in full agreement with Eqn (35). As shown in Section 2.2.3, this condition exactly coincides with the one for the current in the central regions of nonrelativistic supersonic jets.

For a quantitative comparison with the theoretical estimate $\omega \approx \sigma_{\rm n}^{-1/3}{\cal M}^{-2} \, \Omega_{\rm F}$ (52), we need to determine the values of ${\cal M}^{2}$ and $\sigma_{\rm n}$. For the square of the Alfv{\' e}n Mach number ${\cal M}^{2}$ (20), we obtain the following estimate:
\begin{eqnarray}
{\cal M}^2 \approx 10 \, 
\left(\frac{b}{0.1}\right)^{-2}
\left(\frac{\rho}{10^{-7} \, {\rm g} \, {\rm cm}^{-3}}\right)
\nonumber \\
\times \left(\frac{v_{z}}{30 \, {\rm km} \, {\rm s}^{-1}}\right)^{2} 
\left(\frac{B_{\rm M}}{10^{4} \, {\rm G}}\right)^{-2}.
\end{eqnarray}
Consequently, for the magnetization parameter $\sigma_{\rm n}$ (26), we have $\sigma_{\rm n} \sim 10^2$. This estimate is rough, because the definition of $\sigma_{\rm n}$ includes the total magnetic flux in the 'central engine' $\Psi_{\rm tot}$, while in (26) we only considered the flux within the current core $\pi r_{\rm core}^2 B_{z}$. Thus, this value is a kind of lower estimate. As a result, we obtain
\begin{eqnarray}
\omega(r_{\rm core}) \sim 10^6 \,{\rm rad} \, {\rm s}^{-1}
\left(\frac{{\cal M}}{10}\right)^{-2}
\left(\frac{\sigma_{\rm n}}{10^2}\right)^{-1/3}
\nonumber \\
\times \left(\frac{v_{z}}{10^7 \, {\rm cm} \, {\rm s}^{-1}}\right)
\left(\frac{r_{\rm core}}{1 \, {\rm cm}}\right) ,
\end{eqnarray}
in good agreement with the angular velocity of plasma rotation $\omega$ determined using laboratory data.

Finally, we return to the problem that the shape of the laboratory jet significantly differs from cylindrical, for which many of the results presented in Section 2.2 have been obtained. Nevertheless, it turns out that some important properties persist for plasma jets studied in the laboratory. As shown in Fig. 21, the internal structure of the plasma jet, obtained as a solution to Eqn (23), reproduces well its main morphological features. These include both a narrow current channel near the jet axis and the increase in the width of the rear part of the jet observed in the KPF-4 Phoenix experiment (see Fig. 20) as well as the presence of a characteristic 'funnel' in its head.

%%%%%%%%%%%%%%%%%%%%%%%%%%%%%%%%%%%%%%%%%%%%%%%%%%%%%%%%%%%%%%%%%%%%%%%%%%%%%%%%%%%%%%%%%%%%%%%%%%%%%%%%%%%%%%%%%%%%%%%%%%%%%

\section{Conclusion}

Laboratory modeling of jets at plasma focus facilities really made it possible to reproduce almost all the main features predicted by analytical theory (the presence of a central region containing almost the entire current, rotation, matching of the transverse size of the current core and the angular velocity, and the magnitudes of the longitudinal current and longitudinal flow velocity). This agreement is due to the fact that both astrophysical jets and plasma in laboratory experiments are described with good accuracy by the laws of ideal MHD.

In particular, the PF-3 facility, which enables direct measurement of the main flow parameters, makes it possible to directly study the propagation, transverse structure, and stability of nonrelativistic jets. Moreover, as shown in this review, many key parameters determined in the experiment, such as magnetization parameter $\sigma_{\rm n}$, Alfv{\' e}n Mach number ${\cal M}$, and angular velocity $\Omega$, are in good agreement with the conclusions of MHD models for real jets from young stars.

Of course, in a laboratory experiment, where the dimensionless parameters are not much larger than unity, dissipative processes can play a certain role. Many of them are described in detail in our review --- such processes include, for example, a decrease in the magnetic field and plasma jet temperature with increasing distance from the generation region. It is clear that such effects cannot be described within the ideal MHD. An analysis of such dissipative processes, however, is beyond the scope of our review, which is devoted
to testing the adequacy of the 'zero' model developed within the ideal MHD. As we have shown, all the key properties of such a model fully agree with the predictions of the theory.

In conclusion, we formulate once again the main astrophysical problems that can be solved by laboratory simulation of jets. First of all, we are talking about the study of the internal structure of the jet. The hydrodynamic and electromagnetic forces acting on the plasma depend on the specific arrangement of the flow of matter in the jet. Of importance for this problem is the relationship among the motion of the plasma, the generation of electric currents by moving charges, the structure of the magnetic field, and the formation of bulk forces, which again cause the motion of the plasma. We note that in a highly conductive plasma, the magnetic field is entrained by the matter, so if plasma undergoes complex motion, the formation of a complex structure of the magnetic field can be expected. It is also interesting to study the processes that occur when the system deviates from equilibrium, which can have an oscillatory character and, in particular, lead to the formation of observable radiation. It is also important to determine the factors that set the rate of approach to equilibrium, since they define the regime in which the jet propagates and the extent to which the regime is close to equilibrium or a stationary state. Thus, the problem of the stability of jets and, in general, their stationarity deserves the closest attention.

The study of plasma jets in laboratory experiments makes it possible to elucidate the structure and the mechanism for the collimation and stability of jets, despite the fact that they exist for a limited time. We especially emphasize the need for research based on the close relationship among astrophysical observations, physical theory, and laboratory experiment. Astrophysical observations make it possible to fix the jet structure at some point in time, but cannot trace the entire process from the moment the jet originates. However, an active space experiment can not be conducted, since it is not possible to change the state of the medium and other
parameters that characterize the astrophysical plasma jet. Although the cosmic scales of the jet length are not available in a laboratory experiment, it is possible not only to observe the jet from the moment of its origin to the moment of disappearance but also to change the experimental conditions and thereby explore the response of the system to external influence. The repeatability and reproducibility of a laboratory experiment are also important, especially in relation to the problem of stability and stationarity of jets.

The authors are grateful to E P Velikhov, who initiated the experiments at the PF-3 facility, and to V V Myalton, K N Mitrofanov, V P Vinogradov, A M Kharrasov, I V Ilyichev, \fbox{Yu G Kalinin}, S A Dan'ko, S S Ananiev, and Yu V Vinogradova, who carried out the experiments. We also thank M Paduch and the PF-1000 installation team and D A Voitenko and the KPF-4 installation team who participated in the experiments using this equipment. Finally, we are grateful to A V Dodin, \fbox{Ya N Istomin}, I Yu Kalashnikov, A M Kiselev, V I Pariev, and D N Sob'yanin for numerous discussions that clarified many essential issues. The review was prepared with the support of the Russian Foundation for Basic Research (project 18-29-21006) and in the frame of Program 10, Experimental Laboratory Astrophysics and Geophysics of the National Center for Physics and Mathematics.

\section*{References}

\vspace{0.3cm}

{\footnotesize

1. Weinfurtner S et al. {\it Phys. Rev. Lett.} {\bf 106} 021302 (2011)

2. Bernatowicz T J et al. {\it Astrophys. J.} {\bf 631} 988 (2005)

3. Sabri T et al. {\it Astron. Astrophys.} {\bf 575} A76 (2015)

4. Losseva T V, Popel S I, Golub' A P {\it Plasma Phys. Rep.} {\bf 46} 1089 (2020); {\it Fiz. Plazmy} {\bf 46} 1007 (2020)

5. Nakamura Y, Bailung H, Shukla P K {\it Phys. Rev. Lett.} {\bf 83} 1602 (1999)

6. Burdonskii I N et al. {\it Tech. Phys. Lett.} {\bf 46} 1041 (2020); {\it Pis'ma Zh. Tekh. Fiz.} {\bf 46} (20) 47 (2020)

7. Frank A G {\it Phys. Usp.} {\bf 53} 941 (2010); {\it Usp. Fiz. Nauk} {\bf 180} 982 (2010)

8. Frank A G, Artemyev A V, Zelenyi L M {\it J. Exp. Theor. Phys.} {\bf 123} 699 (2016); {\it Zh. Eksp. Teor. Fiz.} {\bf 150} 807 (2016)

9. Lindl J D et al. {\it Phys. Plasmas} {\bf 11} 339 (2004)

10. Ryutov D D, Derzon M S, Matzen M K {\it Rev. Mod. Phys.} {\bf 72} 167 (2000)

11. Bulanov S V et al. {\it Phys. Usp.} {\bf 56} 429 (2013); {\it Usp. Fiz. Nauk} {\bf 183} 449 (2013)

12. Bulanov S V et al. {\it Plasma Phys. Rep.} {\bf 41} 1 (2015); {\it Fiz. Plazmy} {\bf 41} 3 (2015)
 
13. Zakharov Yu P et al. {\it Quantum Electron.} {\bf 46} 399 (2016); Kvantovaya Elektron.{\bf 46} 399 (2016)

14. Zakharov Yu P et al.{\it  Quantum Electron.} {\bf 49} 181 (2019); {\it Kvantovaya Elektron.} {\bf 49} 181 (2019)

15. Kurbatov E P et al. {\it Astron. Rep.} {\bf 62} 483 (2018); {\it Astron. Zh.} {\bf 95} 509 (2018)

16. Soloviev A A et al. {\it Radiophys. Quantum Electron.} {\bf 63} 876 (2020); {\it Izv. Vyssh. Uchebn. Zaved. Radiofiz.} {\bf 63} 973 (2020); {\it Quantum Electron.} {\bf 50} 1115 (2020); {\it Kvantovaya Elektron.} {\bf 50} 1115 (2020)

17. Remington B A, Drake R P, Ryutov D D {\it . Mod. Phys.} {\bf 78} 755 (2006)

18. Lebedev S V, Frank A, Ryutov D D {\it Rev. Mod. Phys.} {\bf 91} 025002 (2019)

19. Filippov N V et al. Phys. Lett. A {\bf 211} 168 (1996)

20. Krauz V I et al. Plasma Phys. Rep. {\bf 36} 937 (2010); Fiz. Plazmy {\bf 36} 997 (2010) 

21. Krauz V et al., in 42nd EPS Conf. on Plasma Physics, 22--26 June 2015, Lisbon, Portugal (Europhysics Conf. Abstracts, Vol. 39E, Eds
R Bingham et al.) (Lisbon: European Physical Society, 2015) P4.401; http://ocs.ciemat.es/EPS2015PAP/pdf/P4.401.pdf

22. Surdin V G {\it Rozhdenie Zvezd} (Birth of Stars) (Moscow: URSS, 2001)

23. Bodenheimer P H Principles of Star Formation (Berlin: Springer, 2011)

24. Li Z-Y et al., in {\it Protostars and Planets VI} (Eds H Beuther et al.) (Tucson, AZ: Univ. of Arizona Press, 2014) p. 173

25. Nakamura F et al. {\it Publ. Astron. Soc. Jpn.} {\bf 71} 117 (2019) 

26. Kuhi L V {\it Astrophys. J.} {\bf 140} 1409 (1964)

27. Bisnovatyi-Kogan G S, Lamzin S A {\it Sov. Astron.} {\bf 21} 720 (1977); {\it Astron. Zh.} {\bf 54} 1268 (1977)

28. Herbig G H {\it Astrophys. J.} {\bf 111} 11 (1950)

29. Haro G {\it Astron. J.} {\bf 55} 72 (1950)

30. Herbig G H, Jones B F {\it Astron. J.} {\bf 86} 1232 (1981)

31. Reipurth B, Bally J {\it Annu. Rev. Astron. Astrophys.} {\bf 39} 403 (2001)

32. Mundt R, Fried J W {\it Astrophys. J.} {\bf 274} L83 (1983)

33. Dopita M A, Schwartz R D, Evans I {\it Astrophys. J.} {\bf 263} L73 (1982)

34. Reipurth B ``VizieR Online Data Catalog: General Catalogue of Herbig-Haro Objects'', VizieR On-line Data Catalog: V/104 (Boulder, CO: Center for Astrophysics, Univ. Colorado, 2000)

35. Winston E et al. {\it Astrophys. J.} {\bf 743} 166 (2011)

36. McGroarty F, Ray T P {\it Astron. Astrophys.} {\bf 420} 975 (2004)

37. Melnikov S Yu et al. {\it Astron. Astrophys.} {\bf 506} 763 (2009)

38. Hartigan P et al.{\it Astrophys. J.} {\bf 736} 29 (2011)

39. Berdnikov L N et al. {\it Astrophys. Bull.} {\bf 72} 277 (2017); {\it Astrofiz. Byull.} {\bf 72} 304 (2017)

40. Mundt R, Eisl{offel J {\it Astron. J.} {\bf 116} 860 (1998)

41. Bally J Annu. Rev. {\it Astron. Astrophys.} {\bf 54} 491 (2016)

42. Bacciotti F et al. {\it Astrophys. J.} {\bf 576} 222 (2002)

43. Coffey D et al. {\it Astrophys. J.} {\bf 663} 350 (2007)

44. Launhardt R et al. {\it Astron. Astrophys.} {\bf 494} 147 (2009)

45. Coffey D et al.{\it Astrophys. J.} {\bf 749} 139 (2012)

46. De Colle F, Cerqueira A H, Riera A{\it Astrophys. J.} {\bf 832} 152 (2016)

47. Pudritz R E, Ray T P {\it Front. Astron. Space Sci.} {\bf 6} 54 (2019)

48. Snell R L, Loren R B, Plambeck R L {\it Astrophys. J.} {\bf 239} L17 (1980)

49. Frank A et al., in {\it Protostars and Planets VI} (Eds H Beuther et al.) (Tucson, AZ: Univ. of Arizona Press, 2014) p. 451

50. Cabrit C, in {\it Jets from Young Stars: Models and Constraints} (Lectures Notes in Physics, Vol. 723, Eds J Ferreira, C Dougados, E Whelan) (Berlin: Springer, 2007) p. 21

51. Tafalla M {\it Mem. Soc. Astron. Ital.} {\bf 88} 619 (2017)

52. Bouvier J et al. {\it Astron. Astrophys.} {\bf 272} 176 (1993)

53. Johns-Krull C M, Valenti J A, Koresko C {\it Astrophys. J.} {\bf 516} 900 (1999)

54. Johns-Krull C M {\it Astrophys. J.} {\bf 664} 975 (2007)

55. Jardine M, Collier Cameron A, Donati J-F {\it Mon. Not. R. Astron. Soc.} {\bf 333} 339 (2002)

56. Donati J-F et al. {\it Mon. Not. R. Astron. Soc.} {\bf 390} 545 (2008)

57. J{\" a}rvinen S P et al. {\it Mon. Not. R. Astron. Soc.} {\bf 489} 886 (2019)

58. Reiter M et al. {\it Astrophys. J.} {\bf 852} 5 (2018)

59. Johns-Krull C M et al. {\it Astrophys. J.} {\bf 700} 1440 (2009)

60. Dullemond C P, Monnier J D {\it Annu. Rev. Astron. Astrophys.} {\bf 48} 205 (2010)

61. Hartmann L, Herczeg G, Calvet N {\it Annu. Rev. Astron. Astrophys.} {\bf 54} 135 (2016)

62. Kurosawa R, Romanova M M, Harries T J {\it Mon. Not. R. Astron. Soc.} {\bf 385} 1931 (2008)

63. Dodin A V, Lamzin S A, Sitnova T M {\it Astron. Lett.} {\bf 39} 315 (2013); {\it Pis'ma Astron. Zh.} {\bf 39} 353 (2013)

64. Dodin A {\it Mon. Not. R. Astron. Soc.} {\bf 475} 4367 (2018)

65. Kwan J, Tademaru E {\it Astrophys. J.} {\bf 454} 382 (1995)

66. Blandford R D, Payne D G {\it Mon. Not. R. Astron. Soc.} {\bf 199} 883 (1982)

67. Ercolano B, Owen J E {\it Mon. Not. R. Astron. Soc.} {\bf 460} 3472 (2016)

68. Hartmann L, Bae J{\it Mon. Not. R. Astron. Soc.} {\bf 474} 88 (2018)

69. Hartigan P, Edwards S, Pierson R {\it Astrophys. J.} {\bf 609} 261 (2004) 

70. Matt S, Pudritz R E, in {\it Star-Disk Interaction in Young Stars}, Symp. S243 (Proc. IAU Symp., Vol. 3, Eds J Bouvier, I Appenzeller)(Cambridge: Cambridge Univ. Press, 2007) p. 299, https://doi.org/10.1017/S1743921307009659

71. Cranmer S R {\it Astrophys. J} {\bf 689} 316 (2008)

72. Errico L, Lamzin S A, Vittone A A {\it Astron. Astrophys.} {\bf 377} 557 (2001)

73. Pikelner S B {\it Astrophys. Lett.} {\bf 2} 97 (1968)

74. Hartigan P {\it Astrophys. J.} {\bf 339} 987 (1989)

75. Hollenbach D, in {\it Herbig-Haro Flows and the Birth of Stars} Proc. (IAU Symp. No. 182, Eds B Reipurth, C Bertout) (Dordrecht: Kluwer Acad. Publ., 1997) p. 181

76. Smith M D, Khanzadyan T, Davis C J {\it Mon. Not. R. Astron. Soc.} {\bf 339} 524 (2003)

77. Raga A C et al. {\it Astrophys. J.} {\bf 364} 601 (1990)

78. Dopita M A, Sutherland R S {\it Astrophys. J. Suppl.} {\bf 229} 35 (2017)

79. Hartigan P et al. {\it Astrophys. J.} {\bf 661} 910 (2007)

80. K{\" o}nigl A {\it Astrophys. J.} {\bf 261} 115 (1982)

81. Li J Z et al. {\it Astrophys. J.} {\bf 549} L89 (2001)

82. Krist J E et al. {\it Astron. J.} {\bf 136} 1980 (2008)

83. Chevalier R A, Raymond J {\it Astrophys. J.} {\bf 225} L27 (1978)

84. Heathcote S et al.{\it Astron. J.} {\bf 112} 1141 (1996)

85. Cox D P, Raymond J C {\it Astrophys. J.} {\bf 298} 651 (1985)

86. Hartigan P, Write A {\it Astrophys. J.} {\bf 811} 12 (2015)

87. Morse J A et al. {\it Astrophys. J} {\bf 339} 231 (1992)

88. Morse J A et al. {\it Astrophys. J.} {\bf 410} 764 (1993)

89. Tesileanu O et al. {\it Astron. Astrophys.} {\bf 507} 581 (2009)

90. Tesileanu O et al. {\it Astrophys. J.} {\bf 749} 96 (2012)

91. Reipurth B et al. {\it Astron. J.} {\bf 123} 362 (2002)

92. Zaninetti L {\it Int. J. Astron. Astrophys.} {\bf 9} 302 (2019)

93. Hansen E C et al. {\it Astrophys. J.} {\bf 837} 143 (2017)

94. Beskin V S {\it Phys. Usp.} {\bf 53} 1199 (2010); {\it Usp. Fiz. Nauk} {\bf 180} 1241 (2010)

95. Beskin V S {\it MHD Flows in Compact Astrophysical Objects: Accretion, Winds and Jets} (Berlin: Springer, 2010); Translated from Russian: {\it Osesimmetrichnye Statsionarnye Techeniya v Astrofizike} (Moscow: Fizmatlit, 2006)

96. Lipunov V M {\it Astrophysics of Neutron Stars} (Berlin: Springer-Verlag, 1992); Translated from Russian: {\it Astrofizika Neitronnykh Zvezd} (Moscow: Nauka, 1987)

97. Pudritz R E, Norman C A {\it Astrophys. J.} {\bf 301} 571 (1986)

98. Heyvaerts J, Norman C {\it Astrophys. J.} {\bf 347} 1055 (1989)

99. Pelletier G, Pudritz R E {\it Astrophys. J.} {\bf 394} 117 (1992)

100. Shu F et al. {\it Astrophys. J.} {\bf 429} 781 (1994)

101. Ferraro V C A {\it Mon. Not. R. Astron. Soc.} {\bf 97} 458 (1937)

102. Lery T et al. {\it Astron. Astrophys.} {\bf 337} 603 (1998)

103. Beskin V S {\it Phys. Usp.} {\bf 40} 659 (1997); {\it Usp. Fiz. Nauk} {\bf 167} 689 (1997)

104. Beskin V S, Malyshkin L M {\it Astron. Lett.} {\bf 26} 208 (2000); {\it Pis'ma Astron. Zh.} {\bf 26} 253 (2000)

105. Bogovalov S, Tsinganos K {\it Mon. Not. R. Astron. Soc.} {\bf 305} 211 (1999)

106. Beskin V S, Okamoto I {\it Mon. Not. R. Astron. Soc.} {\bf 313} 445 (2000)

107. Ferreira J {\it Astron. Astrophys.} {\bf 319} 340 (1997)

108. Heyvaerts J, in {\it Plasma Astrophysics}, Astrophysics School VII, San Miniato, Italy, 3--14 October 1994 (Lecture Notes in Physics,Vol. 468, Eds C Chiuderi, G Einaudi) (Berlin: Springer-Verlag, 1996) p. 31

109. Goldreich P, Julian W H {\it Astrophys. J.} {\bf 157} 869 (1969)

110. Pravdo S H et al. {\it Nature} {\bf 413} 708 (2001)

111. Bogovalov S V {\it Sov. Astron. Lett.} {\bf 18} 337 (1992); {\it Pis'ma Astron. Zh.} {\bf 18} 832 (1992)

112. Beskin V S, Kuznetsova I V, Rafikov R R {\it Mon. Not. R. Astron. Soc.} {\bf 299} 341 (1998)

113. Lynden-Bell D {\it Mon. Not. R. Astron. Soc.} {\bf 279} 389 (1996)

114. Beskin V S, Nokhrina E E {\it Mon. Not. R. Astron. Soc.} {\bf 397} 1486 (2009)

115. Beskin V S, Okamoto I {\it Mon. Not. R. Astron. Soc.} {\bf 313} 445 (2000)

116. Balbus S A, Hawley J F {\it Astrophys. J.} {\bf 376} }214 (1991)
 
117. Romanova M M et al.{\it Mon. Not. R. Astron. Soc.} {\bf 399} 1802 (2009)

118. Bai X-N, Stone J M {\it Astrophys. J.} {\bf 769} 76 (2013)

119. Lesur G, Ferreira J, Ogilvie G I {\it Astron. Astrophys.} {\bf 550} A61 (2013)

120. Hawley J F, Gammie Ch F, Balbus S A {\it Astrophys. J.} {\bf 440} 742 (1995)

121. Brandenburg A et al. {\it Astrophys. J.} {\bf 446} 741 (1995)

122. Stone J M et al. {\it Astrophys. J.} {\bf 463} 656 (1996)

123. Goodson A P, Winglee R M, B{\" o}hm K-H {\it Astrophys. J.} {\bf 489} 199 (1997)

124. Armitage P J {\it Astrophys. J.} {\bf 501} L189 (1998)

125. Long M, Romanova M M, Lovelace R V E {\it Astrophys. J.} {\bf 634} 121 (2005)

126. Zanni C, Ferreira J {\it Astron. Astrophys.} {\bf 550} A99 (2013)

127. Bacciotti F, Eisl{\" o}ffel J, Ray T P {\it Astron. Astrophys.} {\bf 350} 917 (1999)

128. Dougados C et al. {\it Astron. Astrophys.} {\bf 357} L61 (2000)

129. Komissarov S S et al. {\it Mon. Not. R. Astron. Soc.} {\bf 380} 51 (2007)

130. Tchekhovskoy A, McKinney J C, Narayan R {\it Astrophys. J.} {\bf 699} 1789 (2009)

131. Porth O et al. {\it Astrophys. J. {\bf 737} 42} (2011)

132. Norman M L et al. {\it Astron. Astrophys.} {\bf 113} 285 (1982)

133. Blondin J M, Fryxell B A, K{\" o}nigl A {\it Astrophys. J.} {\bf 360} 370 (1990)

134. Stone J M, Norman M L {\it Astrophys. J.} {\bf 413} 210 (1993)

135. Raga A C et al. {\it Astron. Astrophys.} {\bf 465} 879 (2007)

136. Stone J M, Hardee Ph E {\it Astrophys. J.} {\bf 540} 192 (2000)

137. Hansen E C, Frank A, Hartigan P {\it Astrophys. J.} {\bf 800} 41 (2015)

138. Kajdi{\u c} P, Raga A C {\it Astrophys. J.} {\bf 670} 1173 (2007)

139. Ciardi A, in {\it Jets from Young Stars IV} (Lecture Notes in Physics, Vol. 793, Eds P J Valente Garcia, J M Ferreira) (Berlin: Springer-Verlag, 2010) p. 31

140. Bocchi Met al. {\it Astrophys. J.} {\bf 767} 84 (2013)

141. Ryutov D et al. {\it Astrophys. J.} {\bf 518} 821 (1999)

142. Logory L M, Miller P E, Stry P E {\it Astrophys. J. Suppl.} {\bf 127} 423 (2000)

143. Stone J Met al. {\it Astrophys. J. Suppl.} {\bf 127} 497 (2000)

144. Farley D R et al. {\it Phys. Rev. Lett.} {\bf 83} 1982 (1999)

145. Li C K et al. {\it Nat. Commun.} {\bf 7} 13081 (2016)

146. Gregory C D et al. {\it Astrophys. J.} {\bf 676} 420 (2008)

147. Foster J M {\it Bull. Am. Astron. Soc.} {\bf 41} 849 (2010)

148. Hansen J F et al. {\it Phys. Plasmas} {\bf 18} 082702 (2011)

149. Hartigan P et al. {\it Astrophys. J.} {\bf 705} 1073 (2009)

150. Yirak K et al. {\it Astrophys. J.} {\bf 746} 133 (2012)

151. Albertazzi B et al. {\it Science} {\bf 346} 325 (2014)

152. Belyaev V S et al. {\it Astron. Rep.} {\bf 62} 162 (2018); {\it Astron. Zh.} {\bf 95} 171 (2018)

153. Sasorov P V, Oleinik G M, in {\it Entsiklopediya Nizkotemperaturnoi Plazmy} (Encyclopedia of Low-Temperature Plasma) Ser. B Spravochnye Prilozheniya, Bazy i Banki Dannykh (Reference Applicat ions, Databases, and Databanks) Topical Vol . IX-2. {\it Vysokoenergetichnaya Plazmodinamika} (High-Energy Plasmodynamics) (Exec. Ed. A S Kingsep) (Moscow: Yanus-K, 2007) p. 278 

154. Lebedev S V et al. {\it Astrophys. J.} {\bf 564} 113 (2002)

155. Ampleford D J et al. {\it Astrophys. Space Sci.} {\bf 298} 241 (2005)

156. Lebedev S V et al. {\it Astrophys. J.} {\bf 616} 988 (2004)

157. Lebedev S V et al. {\it Mon. Not. R. Astron. Soc.} {\bf 361} 97 (2005)

158. Ciardi A et al. {\it Astrophys. J.} {\bf 678} 968 (2008)

159. Ciardi A et al. {\it Phys. Plasmas} {\bf 14} 056501 (2007)

160. Bellan P M {\it J. Plasma Phys.} {\bf 84} 755840501 (2018)

161. Hsu S C, Bellan P M {\it Mon. Not. R. Astron. Soc.} {\bf 334} 257 (2002)

162. Lavine E S, You S {\it Phys. Rev. Lett.} {\bf 123} 145002 (2019)

163. Ampleford D J et al. {\it Phys. Rev. Lett.} {\bf 100} 035001 (2008)

164. Byvank T et al. {\it Phys. Plasmas} {\bf 24} 122701 (2017)

165. NicolaõÈ Ph et al. {\it Phys. Plasmas} {\bf 15} 082701 (2008)

166. Suzuki-Vidal F et al. {\it Phys. Plasmas} {\bf 19} 022708 (2012)

167. Suzuki-Vidal F et al. {\it High Energy Density Phys.} {\bf 9} 141 (2013)

168. Ciardi A et al. {\it Astrophys. J.} {\bf 691} L147 (2009)

169. Agra-Amboage V et al. {\it Astron. Astrophys.} {\bf 532} A59 (2011)

170. Huarte-EspinosaMet al. {\it Astrophys. J.} {\bf 757} 66 (2012)

171. Petrov D P et al., in {\it Plasma Physics and Controlled Thermonuclear Reactions} Vol. 4 (Ed. M A Leontovich) (New York: Pergamon Press, 1961); Translated from Russian: in {\it Fizika Plazmy i Problemy Upravlyaemykh Termoyadernykh Reaktsii} Vol. 4 (Ed.MA Leontovich) (Moscow: Izd. AN SSSR, 1958) p. 170

172. Filippov N V, Filippova T I, Vinogradov V P {\it Nucl. Fusion Suppl.} {\bf 2} 577 (1962)

173. Mather J W {\it Phys. Fluids} {\bf 8} 366 (1965)

174. Burtsev V A, Gribkov V A, Filippova T I, in {\it Itogi Nauki i Tekhniki. Fizika Plazmy} (Results of Science and Technology. Plasma Physics) Vol. 2 (Moscow: VINITI, 1981) p. 80

175. Filippov N V {\it Sov. J. Plasma Phys.} {\bf 9} 14 (1983); {\it Fiz. Plazmy} {\bf 9} 25 (1983)

176. Bernard A et al. {\it J. Moscow Phys. Soc.} {\bf 8} 93 (1998)

177. Krauz V I ``Plazmennyi fokus'' (``Plasma focus''), in {\it Entsiklopediya Nizkotemperaturnoi Plazmy} (Encyclopedia of Low-Temperature
Plasma) Ser. B Spravochnye Prilozheniya, Bazy i Banki Dannykh (Reference Applications, Databases, and Databanks) Topical Vol. IX-2 {\it Vysokoenergetichnaya Plazmodinamika} (High-Energy Plasmodynamics) (Exec. Ed. A S Kingsep) (Moscow: Yanus-K, 2007) p. 152

178. Sadowski M J et al. {\it Phys. Scr.} {\bf 2006} (T123) 66 (2006)

179. Rawat R S {\it Nanosci. Nanotechnol. Let.} {\bf 4} 251 (2012)

180. Soto L et al. {\it Phys. Plasmas} {\bf 21} 122703 (2014)

181. Borovitskaya I V et al. {\it Russ. Metall.} {\bf 2018} (3) 266 (2018); {\it Metally} (2) 54 (2018)

182.Lerner E J {\it Laser Part. Beams} {\bf 4} 193 (1986)

183. Pouzo J O , Milanese M M {\it IEEE Trans. Plasma Sci.} {\bf 31} 1237 (2003)

184. Filippov N V et al. {\it Czech. J. Phys.} {\bf 50} (Suppl. 3) 127 (2000)

185. Filippov N V et al. {\it Nukleonika} {\bf 46} 35 (2001)

186. Mourenas D et al. {\it Phys. Plasmas} {\bf 10} 605 (2003)

187. Krauz V et al. {\it Phys. Scr.} {\bf 161} (T161) 014036 (2014)

188. Scholz M et al. {\it Nukleonika} {\bf 45} 155 (2000)

189. Mitrofanov K N et al. {\it Plasma Phys. Rep.} {\bf 41} 379 (2015); {\it Fiz. Plazmy} {\bf 41} 413 (2015)

190. Andreeshchev E A et al. {\it Plasma Phys. Rep.} {\bf 33} 218 (2007); {\it Fiz. Plazmy} {\bf 33} 247 (2007)

191. Krauz V I et al. {\it Plasma Phys. Rep.} {\bf 39} 888 (2013); {\it Fiz. Plazmy} {\bf 39} 993 (2013)

192. Voitenko D A et al. {\it Plasma Phys. Rep.} {\bf 43} 1132 (2017); {\it Fiz. Plazmy} {\bf 43} 967 (2017)

193. Kvartskhava I F, NinidzeML, Khautiev E Iu {\it Sov. J. Plasma Phys.} {\bf 2} 22 (1976); {\it Fiz. Plazmy} {\bf 2} 40 (1976)

194. Filippov N V et al. {\it Prikl. Fiz.} {\bf 5} 43 (1999)

195. Ananyev S S et al. {\it Vopr. Atom. Nauki Tekh. Ser. Termoyad. Sintez} {\bf 36} 102 (2013)

196. Ananyev S S et al. {\it Vopr. Atom. Nauki Tekh. Ser. Termoyad. Sintez} {\bf 40} 21 (2017)

197. Polukhin S N et al. {\it Plasma Phys. Rep.} {\bf 46} 127 (2020); {\it Fiz. Plazmy} {\bf 46} 99 (2020)

198. Bernard A et al. {\it Phys. Fluids} {\bf 18} 180 (1975)

199. Baronova E Oet al. {\it Plasma Phys. Rep.} {\bf 38} 751 (2012); {\it Fiz. Plazmy} {\bf 38} 815 (2012)

200. Il'ichev I et al. {\it Eur. Phys. J. Plus} {\bf 136} 557 (2021)

201. Krauz VI, Beskin VS, Velikhov E P {\it Int. J. Mod. Phys.} D{\bf 27} 1844009 (2018)

202. Il'ichev I V et al. {\it Plasma Phys. Rep.} {\bf 46} 506 (2020); {\it Fiz. Plazmy} {\bf 46} 419 (2020)

203. Krauz V I et al. {\it Vopr. Atom. Nauki Tekh. Ser. Termoyad. Sintez} {\bf 38} 19 (2015)

204. Skladnik-Sadowska E et al. {\it Phys. Plasmas} {\bf 25} 082715 (2018)

205. Polukhin S N et al. {\it Plasma Phys. Rep.} {\bf 42} 1127 (2016); {\it Fiz. Plazmy} {\bf 42} 1080 (2016)

206. Ananyev S S, Dan'ko S A, Kalinin Yu G {\it Instrum. Exp. Tech.} {\bf 59} 810 (2016); {\it Prib. Tekh. Eksp.} (6) 37 (2016)

207. Ananyev S S et al. {\it Plasma Phys. Rep.} {\bf 42} 269 (2016); {\it Fiz. Plazmy} {\bf 42} 282 (2016)

208. Ananyev S S et al. {\it Vopr. Atom. Nauki Tekh. Ser. Termoyad. Sintez} {\bf 39} 58 (2016)

209. Dan'ko S A et al. {\it Plasma Phys. Control. Fusion} {\bf 59} 0450003 (2017)

210. Skladnik-Sadowska E et al. {\it Phys. Plasmas} {\bf 23} 122902 (2016)

211. Krauz V I et al. {\it Vopr. Atom. Nauki Tekh. Ser. Termoyad. Sintez} {\bf 41} 48 (2018)

212. Mitrofanov K N et al. {\it Instrum. Exp. Tech.} {\bf 61} 239 (2018); {\it Prib. Tekh. Eksp.} (2) 78 (2018)

213. Mitrofanov K N et al. {\it J. Exp. Theor. Phys.} {\bf 119} 910 (2014); {\it Zh. Eksp. Teor. Fiz.} {\bf 146} 1035 (2014)

214. Beskin V S et al. {\it Radiophys. Quantum Electron.} {\bf 59} 900 (2017); {\it Izv. Vyssh. Uchebn. Zaved. Radiofiz.} {\bf 59} 1004 (2016)

215. Krauz V I et al. {\it Europhys. Lett.} {\bf 129} 15003 (2020)

216. Kalashnikov I et al. {\it Phys. Plasmas} {\bf 25} 06290 1 (2018)

217. Shatalov N A et al. {\it J. Phys. Conf. Ser.} {\bf 1390} 012069 (2019)

218. Kalashnikov I Yu et al. {\it Astron. Rep.} {\bf 65} 447 (2021); {\it Astron. Zh.} {\bf 98} 476 (2021)

219. Krauz V I et al. {\it J. Phys. Conf. Ser.} {\bf 907} 012026 (2017)

220. Beskin V S et al. {\it Mon. Not. R. Astron. Soc.} {\bf 472} 3971 (2017)

221. Krauz V I et al. {\it Astron. Rep.} {\bf 65} 26 (2021); {\it Astron. Zh.} {\bf 98} 29 (2021)

222. Mitrofanov K N et al. {\it Astron. Rep.} {\bf 61} 138 (2017); {\it Astron. Zh.} {\bf 94} 152 (2017)

223. Krauz V I et al. {\it Plasma Phys. Rep.} {\bf 47} 912 (2021); {\it Fiz. Plazmy} {\bf 47} 829 (2021)

224. Krauz V I et al. {\it Astron. Rep.} {\bf 63} 146 (2019); {\it Astron. Zh.} {\bf 96} 156 (2019)

225. Krauz V I et al. {\it J. Plasma Phys.} {\bf 86} 905860607 (2020)

226. Suzuki-Vidal F et al. {\it IEEE Trans. Plasma Sci.} {\bf 38} 581 (2010)

227. Krauz V I et al. {\it Europhys. Lett.} {\bf 98} 45001 (2012)

228. Vinogradov VP et al. {\it Plasma Phys. Rep.} {\bf 42} 1079 (2016); {\it Fiz. Plazmy} {\bf 42} 1033 (2016)

229. Krauz V I et al. {\it Astron. Rep.} {\bf 67} 15 (2023); {\it Astron. Zh.} {\bf 100} 19 (2023)

230. Beskin V S, Kalashnikov I Yu {\it Astron. Lett.} {\bf 46} 462 (2020); {\it Pis'ma Astron. Zh.} {\bf 46} 494 (2020)

231. Sutherland R S, Dopita M A {\it Astrophys. J. Suppl.} {\bf 229} 34 (2017)

232. Krauz V I et al. {\it Plasma Phys. Rep.} {\bf 48} 606 (2022); {\it Fiz. Plazmy} {\bf 48} 506 (2022)

233. Mertens F et al. {\it Astron. Astrophys.} {\bf 595} A54 (2015)

234. Bisnovatyi-Kogan G S {\it Mon. Not. R. Astron. Soc.} {\bf 376} 457 (2007)

235. Bisnovatyi-Kogan G S, Komberg B V, Fridman A M {\it Astron. Rep.} {\bf 13} 369 (1969); {\it Astron. Zh.} {\bf 46} 465 (1969)

236. Cveji{\' c} M et al. {\it Phys. Rev. Lett.} {\bf 128} 015001 (2022)

237. Beskin V S {\it Astron. Rep.} {\bf 67} 27 (2023); {\it Astron. Zh.} {\bf 100} 32 (2023)
}

\end{document}